\documentclass[journal]{IEEEtran}
\usepackage{amsmath,amsfonts}
\usepackage{amssymb}
\usepackage{amsthm}
\usepackage{cite}
\usepackage{amsmath,amssymb,amsfonts}
\usepackage{algorithmic}
\usepackage{algorithm}
\usepackage{array}
\usepackage{textcomp}
\usepackage{url}
\usepackage{multirow}
\usepackage{verbatim}
\usepackage{graphicx}
\usepackage{soul, color, xcolor}
\usepackage{caption}
\usepackage{subcaption}
\usepackage{booktabs}
\usepackage{adjustbox} 
\usepackage{pifont}
\usepackage{xcolor}
\newcommand{\cmark}{\textcolor{black}{\ding{51}}}
\newcommand{\xmark}{\textcolor{black}{\ding{55}}}
\usepackage{tikz}
\usepackage{threeparttable}
\usepackage{makecell}
\usepackage{longtable}
\newcolumntype{L}[1]{>{\raggedright\let\newline\\\arraybackslash\hspace{0pt}}m{#1}}
\graphicspath{{img/},{pics/}}
\def\BibTeX{{\rm B\kern-.05em{\sc i\kern-.025em b}\kern-.08em
    T\kern-.1667em\lower.7ex\hbox{E}\kern-.125emX}}
\usepackage{balance}

\usepackage{csquotes}
\ifCLASSINFOpdf

\else
 
\fi

\begin{document}
%
\title{A Survey on Large Language Models for Communication, Network, and Service Management: Application Insights, Challenges, and Future Directions}
%
%
%

\author{Gordon Owusu Boateng,~\IEEEmembership{Member,~IEEE,}~Hani Sami,~Ahmed Alagha, Hanae Elmekki, Ahmad Hammoud, Rabeb Mizouni, Azzam Mourad,~\IEEEmembership{Senior~Member,~IEEE,} Hadi Otrok,~\IEEEmembership{Senior~Member,~IEEE,} Jamal Bentahar, Sami Muhaidat,~\IEEEmembership{Senior~Member,~IEEE,} Chamseddine Talhi,~Zbigniew Dziong,~\IEEEmembership{Senior~Member,~IEEE,}~and~Mohsen Guizani,~\IEEEmembership{Fellow,~IEEE}
\thanks{This work is supported by .......). (Corresponding author: Azzam Mourad. email: azzam.mourad@ku.ac.ae).}
\thanks{G. O. Boateng is with KU 6G Research Center, Department of CS, Khalifa University, UAE (email: gordon.boateng@ku.ac.ae)}
\thanks{H. Sami is with Department of Software and IT Engineering, Ecole de Technologie Superieure, Montreal, Canada, and also with Artificial Intelligence \& Cyber Systems Research Center, Department of CSM, Lebanese American
University, Beirut, Lebanon (email: hani.sami@etsmtl.ca)}
\thanks{A. Alagha and H. Elmekki are with Concordia Institute for Information Systems Engineering, Concordia University, Montreal, QC, Canada (email: ahmed.alagha@mail.concordia.ca, hanae.elmekki@mail.concordia.ca)}
\thanks{A. Hammoud is with Department of Electrical Engineering, Ecole de Technologie Superieure, Montreal, Canada, and also with Artificial Intelligence \& Cyber Systems Research Center, Department of CSM, Lebanese American
University, Beirut, Lebanon (email: ahmad.hammoud.1@ens.etsmtl.ca)}
\thanks{R. Mizouni and H. Otrok are with Center of Cyber-Physical Systems (C2PS), Department of CS, Khalifa University, Abu Dhabi, UAE (email: rabeb.mizouni@ku.ac.ae, Hadi.Otrok@ku.ac.ae}
\thanks{A. Mourad is with KU 6G Research Center, Department of CS, Khalifa University, UAE, and also with Artificial Intelligence \& Cyber Systems Research Center, Department of CSM, Lebanese American
University, Beirut, Lebanon (email: azzam.mourad@ku.ac.ae)}
\thanks{J. Bentahar is with KU 6G Research Center, Department of CS, Khalifa University, UAE, and also with Concordia Institute for Information Systems Engineering, Concordia University, Montreal, QC, Canada (email: jamal.bentahar@ku.ac.ae)}
\thanks{S. Muhaidat  is with KU 6G Research Center, Department
of Electrical Engineering and Computer Science, Khalifa University,
Abu Dhabi, UAE, and also with Department of Systems and Computer
Engineering, Carleton University, Ottawa, ON K1S 5B6, Canada (email: sami.muhaidat@ku.ac.ae)}
\thanks{C. Talhi is with Department of Software and IT Engineering, Ecole de Technologie Superieure, Montreal, Canada (email: Chamseddine.Talhi@etsmtl.ca)}
\thanks{Z. Dziong is with Department of Electrical Engineering, Ecole de Technologie Superieure, Montreal, Canada (email: Zbigniew.Dziong@etsmtl.ca)}
\thanks{M. Guizani is with Department of ML, Mohamed Bin Zayed University of Artificial Intelligence, Abu Dhabi, UAE (email: Mohsen.Guizani@mbzuai.ac.ae)}}


\markboth{Journal of \LaTeX\ Class Files,~Vol.~14, No.~8, August~2015}%
{Shell \MakeLowercase{\textit{et al.}}: Bare Demo of IEEEtran.cls for IEEE Journals}
%



\maketitle

\begin{abstract}
The rapid evolution of communication networks in recent decades has intensified the need for advanced Network and Service Management (NSM) strategies to address the growing demands for efficiency, scalability, enhanced performance, and reliability of these networks. Large Language Models (LLMs) have received tremendous attention due to their unparalleled capabilities in various Natural Language Processing (NLP) tasks and generating context-aware insights, offering transformative potential for automating diverse communication NSM tasks. Contrasting existing surveys that consider a single network domain, this survey investigates the integration of LLMs across different communication network domains, including mobile networks and related technologies, vehicular networks, cloud-based networks, and fog/edge-based networks. First, the survey provides foundational knowledge of LLMs, explicitly detailing the generic transformer architecture, general-purpose and domain-specific LLMs, LLM model pre-training and fine-tuning, and their relation to communication NSM. Under a novel taxonomy of network monitoring and reporting, AI-powered network planning, network deployment and distribution, and continuous network support, we extensively categorize LLM applications for NSM tasks in each of the different network domains, exploring existing literature and their contributions thus far. Then, we identify existing challenges and open issues, as well as future research directions for LLM-driven communication NSM, emphasizing the need for scalable, adaptable, and resource-efficient solutions that align with the dynamic landscape of communication networks. We envision that this survey serves as a holistic roadmap, providing critical insights for leveraging LLMs to enhance NSM.
\end{abstract}

\begin{IEEEkeywords}
Large language models, communication network and service management, monitoring, planning, deployment, support.
\end{IEEEkeywords}

%
\IEEEpeerreviewmaketitle

\section{Introduction}\label{sec:introduction}
%
%
%
%
\subsection{Background}
\IEEEPARstart{T}{he} rapid development of wired and wireless network technologies has strengthened the backbone of communication networks for modern ubiquitous connectivity. Notably, communication networks are an enabler for seamless network connections between devices, infrastructure, and systems across varied distances and diverse applications \cite{8438489,9598915,FORTUNA20091354,7462480}. These networks span multiple domains, from traditional wired communication networks to modern wireless and mobile network infrastructures, ensuring flawless, pervasive data flow, and massive connectivity to support the communication network ecosystem, while bridging the digital divide. Among the diverse types of communication networks, mobile networks, especially the recent deployment of the Fifth-Generation (5G) and beyond technologies, facilitate high-speed data rate, low latency, high reliability, and wide coverage density applications \cite{10.5555/3307066,9019680}. For instance, the envisioned Sixth-Generation (6G) network is expected to achieve peak data rates of up to 1 Terabit per second (Tbps), user experience data rate of 1 to 10 Gigabits per second (Gbps), sub-millisecond latency (as low as 0.1 milliseconds), reliability of 99.99999\% (\textit{\textquote{seven nines}}), and about 10 million device connectivity per square kilometer \cite{10597064}. In parallel, the Internet of Things (IoT) paradigm connects billions of devices, ranging from sensors in smart homes to industrial applications such as the Industrial IoT \cite{9795904}. Vehicular networks are capable of enabling real-time data exchange among vehicles and Roadside Units (RSUs) via Dedicated Short-Range Communications (DSRC) or Cellular-Vehicle-to-Everything (C-V2X) technologies \cite{10689485}. This enhances safety and efficiency in the evolving Intelligent Transportation Systems (ITS). Additionally, cloud-based networks provide centralized, scalable, and abundant computing and storage resources for processing and delivering services in communication networks \cite{9882121}, while fog/edge-based networks bring such processing and services closer to end users by enabling localized processing \cite{10044183}. With fog/edge-based networks, computation latency and bandwidth consumption are reduced drastically.

Network and Service Management (NSM) in modern communication networks constitutes optimizing the underlying network infrastructure and critical key performance indicators such as latency, bandwidth, energy efficiency, and Quality of Service (QoS) to ensure efficient resource utilization and service continuity \cite{LIYANAGE2022103362,10504277,9915455}. These management tasks involve monitoring, reporting, planning, deploying, distributing, and continuously supporting the network for full-scale functionality and flexibility. Effective NSM ensures smooth inter-domain service provisioning and efficient utilization of resources. Though closely related, network management and service management focus on different aspects of the communication network. Network management primarily concerns the optimization and maintenance of the underlying physical infrastructure, ensuring that the physical and logical components, such as base stations and routers, operate efficiently \cite{PANEK2023109984,7543814,8613269,8533352}. It handles tasks such as base station siting, cloud data center deployment, and fog node distribution. On the other hand, service management is concerned with the quality and delivery of services provided by the network, ensuring that end users and applications enjoy their desired performance levels \cite{7448886,8573813,8703470}. This includes service provisioning management, adherence to Service Level Agreements (SLAs), and ensuring QoS. Thus, while network management caters for the technical performance of the network itself, service management enhances end user experience and the smooth operation of applications and services running on top of the network.

Traditionally, NSM approaches rely on conventional optimization techniques, e.g., rule-based \cite{9219456,5277977}, heuristic-based \cite{5708471,10138589}, and static \cite{6503647,WU201849} algorithms, which are tailored to address specific network and service challenges like traffic monitoring, network configuration, intrusion detection, root-cause analysis, and security enhancement. Rule-based algorithms use predefined rules to govern management decision-making, e.g., traffic management systems in mobile networks often rely on these rules to dynamically allocate bandwidth resources. Heuristic-based methods are designed to quickly find good enough solutions when the optimal solution is complex or computationally intensive to determine. For instance, cloud-based networks use heuristics to distribute tasks efficiently among servers to minimize latency and energy consumption. Static algorithms tend to assume a controlled environment without frequent changes in network behavior, e.g., in vehicular networks, pre-defined routing and scheduling solutions can be developed using static methods. Although effective in controlled environments, static methods often fall short when faced with highly dynamic, real-world conditions. Despite their maturity and widespread adoption to solve NSM challenges, conventional optimization methods have been proven inadequate for handling the growing complexity of emerging network technologies \cite{7414384,6842585}. The lack of flexibility to adapt to changing infrastructure and resource demand on-the-fly renders them inefficient in today's heterogeneous communication networks.

To overcome the challenges faced by conventional optimization techniques, Artificial Intelligence (AI) \cite{9424691} and Machine Learning (ML) \cite{8255757} approaches have shown promising signs of enhancing NSM tasks in communication networks. In particular, Deep Reinforcement Learning (DRL) has achieved great strides in handling dynamic decision-making in complex and non-stationary network environments. In the context of mobile networks and IoT, DRL has been utilized to optimize resource allocation in the 5G Radio Access Network (RAN) \cite{8730413} and Core Network (CN) \cite{9779742}. Similarly, in cloud-based \cite{8297294} and fog/edge-based \cite{9591490} environments, DRL has enabled more efficient task offloading, ensuring that computational tasks are handled by the most appropriate resources with minimum delay. Despite their strengths, traditional AI/ML algorithms often struggle with unstructured data such as network configuration intents, logs, and multimodal data, which are common in many real-world applications. These models lack the ability to interpret context and relationships between different sources of data, leading to suboptimal decisions in dynamic environments. Moreover, AI/ML models are often highly task-specific, which require retraining and fine-tuning for new tasks and varying network conditions. Their adaptability to evolving environments is limited, and model updates can be computationally expensive and time-consuming.

The recent rise of Large Language Models (LLMs) in both industry and academia has unraveled endless new possibilities in various research fields \cite{minaee2024largelanguagemodelssurvey,zhao2024surveylargelanguagemodels,zhou2024largelanguagemodelllm,qu2024mobileedgeintelligencelarge}. With its notable success in Natural language Processing (NLP), LLMs can process and generate human-like text, understand complex commands, and make high-level decisions based on input prompts or patterns in the data they train on. Building on their NLP tasks capabilities, LLMs can now handle huge amounts of datasets and derive actionable insights and reasoning to solve real-world problems. Unlike AI/ML algorithms, LLMs excel at processing and understanding unstructured data, particularly natural language. They can analyze data from multiple sources and extract meaningful insights, improving decision-making and troubleshooting in complex systems \cite{hu2023bliva}. Trained on large datasets, LLMs are better equipped to understand context and nuances in data for multi-dimensional decision-making. This allows them to make more informed decisions by considering not only data points but also other relationships, content, and high-level meanings. Specifically, LLMs have the potential to augment NSM by automating network operations, detecting anomalies, and performing predictive maintenance \cite{10574890,10588835}. For instance, Mekrache \textit{et al.} \cite{10574890} proposed an architecture that leverages LLMs to facilitate the entire intent Life-Cycle (LC) in next-generation networks. The primary role of LLM was to enable seamless communication between humans and machines by utilizing advanced natural language understanding and task generation capabilities, thereby automating various network management processes.
Some authors \cite{10588835} proposed ConnectGPT, a novel pipeline that leverages LLM to analyze traffic data from infrastructure sensors, automate the generation of standardized Cooperative ITS (C-ITS) safety messages, and enhance traffic management efficiency and road safety for Connected Autonomous Vehicles (CAVs).

Despite their enormous potential, the application and customization of LLMs for NSM is still in its nascent stages, especially in the context of communication networks. Current and emerging communication networks are highly multifaceted, encompassing a wide range of application scenarios across diverse domains, including mobile networks, vehicular networks, cloud-based infrastructures, and fog/edge computing environments. This diversity and complexity underscore the need for a comprehensive study on the integration of LLMs within these network domains. In response, we aim to curate an extensive survey that explores the customization and utilization of LLMs across various communication network domains, namely mobile networks and associated technologies, vehicular networks, cloud-based networks, and fog/edge-based networks. Based on existing body of literature, we provide a detailed analysis of how LLMs have been integrated into current communication network architectures, identifying trends, challenges, and opportunities for future advancements.

\subsection{Motivation}
The integration of LLMs into communication networks for NSM tasks is becoming increasingly crucial as these networks grow more complex, spanning multiple domains such as mobile, vehicular, cloud-based, and fog/edge-based networks. 
Considering that NSM plays a critical role in communication networks, it is essential to explore how LLMs can be applied to enhance NSM across the diverse network domains. While existing surveys have focused on the application of LLMs in areas like telecommunications \cite{zhou2024largelanguagemodelllm}, Autonomous Driving Systems (ADS) \cite{fourati2024xlmautonomousdrivingsystems}, the Internet of Vehicles (IoV) \cite{xu2024integrationmixtureexpertsmultimodal}, Unmanned Aerial Vehicles (UAVs) \cite{10643253},  Integrated Satellite-Aerial-Terrestrial Networks (ISATNs) \cite{javaid2024leveraginglargelanguagemodels}, Intelligent Transportation Systems (ITS) \cite{10401518}, and Mobile Edge Intelligence (MEI) \cite{qu2024mobileedgeintelligencelarge}, separately, there still remains a significant gap in a comprehensive review targeting the potential of LLMs for NSM in communication networks. The existing literature overlooks the broader application of LLMs for optimizing and managing communication networks in domains such as mobile networks and technologies, vehicular networks, cloud-based networks, and fog/edge networks. This survey aims to fill the gap by providing an in-depth review of the role of LLMs in communication NSM. Additionally, it will explore how LLMs can be leveraged to tackle the unique challenges faced in these networks, offering insights into open issues, potential solutions, and future research directions in the field of LLM-driven NSM for communication networks.

\renewcommand\arraystretch{1.4} 
\begin{table*}[!htbp]
\centering
\caption{Comparison of Related Surveys}
\label{tab-surveys}
\small 
\begin{adjustbox}{width=\textwidth}
\begin{tabular}{|p{0.9cm}|p{0.9cm}|p{1.8cm}|p{1.8cm}|p{5.7cm}|p{4cm}|p{4.8cm}|}
    \hline
    \textbf{Ref.} & \textbf{Year} & \textbf{LLM Fundamentals} & \textbf{Methodology} & \textbf{Main Focus/Scope} & \textbf{Application-specific Domain} & \textbf{Taxonomy} \\ \hline

    \cite{10433480} & 2024 & \cmark & \cmark & Provides a structured overview of LLMs, addressing their evolution, datasets utilized in LLM studies, social impact, various applications of LLMs, and challenges faced in their deployment & Biomedical and healthcare, education, social media, business, agriculture & Resources of LLMs, domain-specific applications, impact of LLMs on society, industrial applications of LLMs \\ \hline

    \cite{10401518} & 2023 & \xmark & \xmark & Investigates the impact of frontier AI, foundation models (FMs), and LLMs in ITS & ITS & Traffic prediction and management, sentiment analysis of social media data for traffic insights, emergency response and disaster management, multimodal transportation planning \\ \hline

    \cite{10495592} & 2024 & \xmark & \xmark & Investigates the integration of LLMs in autonomous driving and mapping systems, highlighting the current tools, datasets, and benchmarks available in the domain & Autonomous driving & MLLMs for autonomous driving, datasets \\ \hline

    \cite{fourati2024xlmautonomousdrivingsystems} & 2024 & \cmark & \cmark & Reviews the architectures, tools, and frameworks associated with XLMs, discusses their deployment strategies in ADS & ADS/ITS & XLMs (LLMs/VLMs) for autonomous driving (prompt engineering-based methods, fine-tuning-based methods, RLHF-based methods, LLM and GAI-based methods), datasets and simulators for ADS \\ \hline

    \cite{10643253} & 2024 & \cmark & \xmark & Evaluates LLM architectures for UAV integration, identifying innovative applications and outlining future research directions necessary for optimizing the synergy between LLMs and UAVs & UAVs & Surveillance and reconnaissance applications, emergency response and disaster management, delivery service and logistics, environmental monitoring and wildlife conservation, satellite and HAP communications \\ \hline

    \cite{javaid2024leveraginglargelanguagemodels} & 2024 & \xmark & \xmark & Investigates the integration of LLMs into ISATNs, emphasizing their potential to enhance data flow, signal processing, and network management through enhanced AI/ML techniques & ISATNs & Network optimization, resource allocation, traffic routing \\ \hline

    \cite{xu2024integrationmixtureexpertsmultimodal} & 2024 & \xmark & \xmark & Investigates the integration of MoE and GAI to achieve AGI in IoV, aiming for full autonomy with minimal human oversight. Discusses applications in environmental monitoring, traffic management, and autonomous driving & IoV & MoE and GAI for distributed perception and monitoring, cooperative decision-making and planning, generative modeling for simulation \\ \hline

    \cite{zhou2024largelanguagemodelllm} & 2024 & \cmark & \xmark & Provides a comprehensive survey of LLM fundamentals while exploring its applications in generation, classification, optimization, and prediction tasks relevant to the telecom domain & Telecom & LLM for generation problems in telecom, LLM-enabled classification in telecom, LLM-enabled optimization techniques, LLM-enabled prediction in telecom \\ \hline

    \cite{qu2024mobileedgeintelligencelarge} & 2024 & \cmark & \xmark & Provides a comprehensive survey on the utilization of MEI to enhance LLMs, addressing the limitations of running LLMs on resource-constrained edge devices compared to powerful cloud infrastructures & MEI & Application scenarios, edge caching and delivery for LLMs, edge training for LLMs, edge inference for LLMs \\ \hline

    \textbf{Our work} & \textbf{2024} & \cmark & \cmark & \textbf{Provides a comprehensive survey of LLMs for NSM in communication networks, exploring application scenarios of mobile network and IoT technologies, vehicular networks, cloud-based networks, and fog/edge-based networks} & \textbf{Communication networks (mobile network and IoT technologies, vehicular networks, cloud-based networks, and fog/edge-based networks)} & \textbf{LLM for network monitoring and reporting, LLM for AI-powered network planning, LLM for network deployment and distribution, LLM for continuous network support} \\ \hline
\end{tabular}
\end{adjustbox}
\end{table*}

\begin{figure*}
\centerline{\includegraphics[width=6.5in]{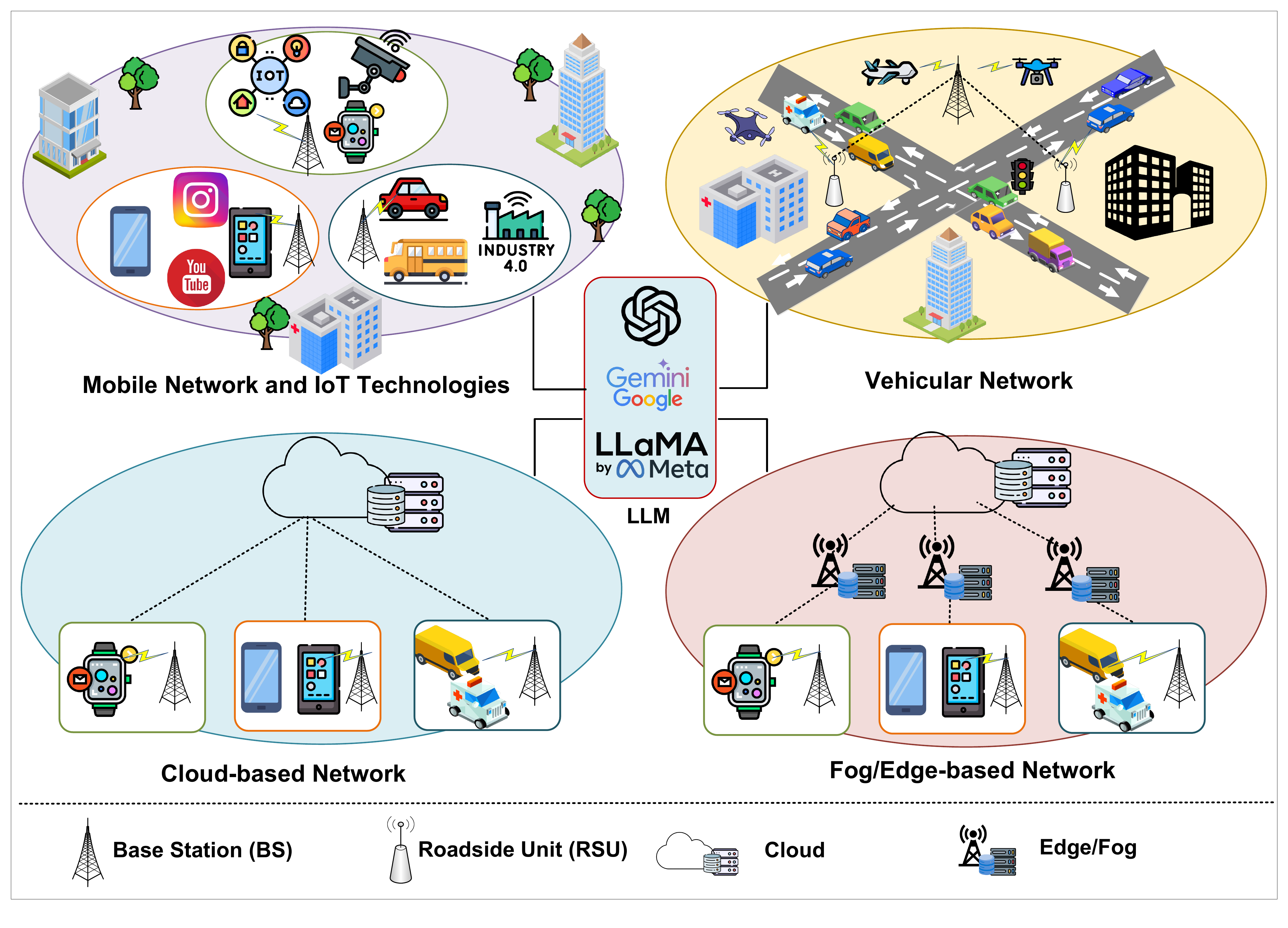}}
\caption{LLM application in communication network domains.}
\label{mainfig}
\end{figure*}
\subsection{Related Surveys}
In the recent past, Generative AI (GenAI), Foundation Models (FMs), and LLMs have received enormous attention from researchers in both industry and academia due to their benefits in text generation, task automation, and rich semantic abilities. As such, numerous studies have investigated the potential of these flagship technologies in various domains and application scenarios, including communication networks. {Specifically, researchers have surveyed the application of LLMs in specific domains such as ITS \cite{10401518}, ADS \cite{fourati2024xlmautonomousdrivingsystems}, IoV \cite{xu2024integrationmixtureexpertsmultimodal}, telecommunications \cite{zhou2024largelanguagemodelllm}, and the capabilities of techniques such as MEI to complement LLM training and model inference \cite {qu2024mobileedgeintelligencelarge}, using different taxonomies. Though informative, most of these surveys do not extensively shed light on the potential of LLMs for communication NSM, considering that communication networks require efficient optimization of network resources and appropriate infrastructure distribution and deployment to herald a sustainable network. The work in \cite{10433480} provided a comprehensive review of LLMs, addressing their foundational concepts, diverse applications, and open challenges that remain in their deployment. This work focused on LLM resources, domain-specific applications, the impact of LLMs on society, and industrial applications of LLMs in biomedical and healthcare, education, social media, business, and agriculture. However, this survey did not emphasize the application of LLMs for communication NSM.

In terms of ITS and ADS, several surveys have been conducted to showcase the potential of LLM integration \cite{10401518,10495592,fourati2024xlmautonomousdrivingsystems}. Shoaib \textit{et al.} \cite{10401518} explored the transformative impact of frontier AI, FMs, and LLMs on ITS.
The survey focused on the dynamic synergy between LLMs and ITS, investigating their roles and delving into applications in traffic management, integration into autonomous vehicles, and their role in shaping smart cities, while addressing challenges brought by frontier AI and FMs. Specific use cases such as traffic prediction and management, sentiment analysis of social media data for traffic insights, emergency response and disaster management, and multimodal transportation planning were considered, highlighting impressive research findings and how LLMs are employed to solve these problems in ITS. The work in \cite{10495592} systematically surveyed the integration of LLMs in autonomous driving and mapping systems context, highlighting the current landscape of Multimodal LLMs (MLLMs), their applications, existing tools, datasets, and benchmarks available in the domain. Some authors \cite{fourati2024xlmautonomousdrivingsystems} comprehensively reviewed the architectures, tools, and frameworks of Vision Language Models (VLMs) and MLLMs, collectively referred to as XLMs in the context of ADS. This survey discussed deployment strategies for the effective integration of XLMs into autonomous driving frameworks under the taxonomy of prompt-engineering-based, fine-tuning-based, Reinforcement Learning with Human Feedback (RLHF)-based, and LLM and GAI-based methods.

Existing research has surveyed the significant potential of LLMs for vehicular network domains such as UAVs \cite{10643253}, ISATNs \cite{javaid2024leveraginglargelanguagemodels}, and IoV \cite{xu2024integrationmixtureexpertsmultimodal}. Javaid \textit{at al.} \cite{10643253} examined the architectures of LLMs for their suitability in UAV systems, emphasizing their potential to improve data analysis, decision-making, and operational efficiency. The survey was conducted focusing on application taxonomies such as surveillance and reconnaissance, emergency response and disaster management, delivery service and logistics, environmental monitoring and wildlife conservation, and satellite and High Altitude Platform (HAP) communications. The same authors \cite{javaid2024leveraginglargelanguagemodels} extended the scope of their survey to investigate the integration of LLMs within ISATNs, emphasizing their impact on operational efficiency, resource allocation, traffic routing, and network optimization strategies in the 5G/6G technology context. Xu \textit{et al.} \cite{xu2024integrationmixtureexpertsmultimodal} curated a survey to investigate the synergistic integration of GAI and Mixture of Experts (MoE) to advance Artificial General Intelligence (AGI) in IoV. The survey highlighted the potential of GAI and MoE under the taxonomy of distributed perception and monitoring, cooperative decision-making and planning, and generative modeling for simulation. 

LLMs are expected to properly understand and generate content that aligns with real-world details and specific requirements of telecommunications applications. To realize this effort, Zhou \textit{et al.} \cite{zhou2024largelanguagemodelllm} provided a thorough overview of the application of LLMs in telecom applications. The main focus of the survey was to introduce LLM-enabled key techniques and telecom applications in terms of generation, classification, optimization, and prediction problems. In as much as LLMs have the potential to revolutionize networked systems and applications, their integration requires increased computation and storage resources, especially during model pre-training and fine-tuning. In this light, Qu \textit{et al.} \cite{qu2024mobileedgeintelligencelarge} surveyed the integration of MEI with LLMs, exploring resource-efficient techniques and applications that highlight the advantages of edge computing in various sensitive contexts. The authors structured the taxonomy of their review into application scenarios, edge caching and delivery for LLMs, edge training for LLMs, and edge inference for LLMs.

From the aforementioned related surveys, it can be observed that existing studies provide a consolidated overview of LLMs for specific application domains and under varying taxonomies. However, a few of these surveys fail to present the fundamentals of LLMs, which is a catalyst for understanding the nuances of the flagship technology and how it can be incorporated in various fields. Moreover, most of these works do not highlight their research methodology, making it challenging to judge whether these works conducted extensive reviews in their studies. Besides, the scope of each related survey is fixated on different domains ranging from ITS, ADS, UAVs, ISATNs, and IoV to telecoms and spanning unique taxonomies. From not an entirely different perspective, this work seeks to provide a comprehensive survey of LLMs for NSM in the context of communication networks. Considering that integrating LLMs into communication networks to enhance NSM is a complicated task, we first lay the foundation by introducing the fundamentals of LLMs, including but not limited to the basic transformer architecture, comparing general-purpose and domain-specific LLMs, and LLM model pre-training and fine-tuning in relation to the subject matter. Unlike existing works that focus narrowly on a single application domain like UAVs in \cite{10643253}, IoV in \cite{xu2024integrationmixtureexpertsmultimodal}, or Telecom in \cite{zhou2024largelanguagemodelllm} with limited scope, our survey adopts a broader and more inclusive perspective. We comprehensively examine the application of LLMs for NSM 
across key communication network paradigms. These paradigms include mobile network and IoT technologies, vehicular networks, cloud-based networks, and fog/edge-based network domains, thereby providing a holistic view of the potential and versatility of LLMs in addressing the multifaceted challenges of NSM. In terms of taxonomy, we explore existing studies of LLM applications in the various domains based on network monitoring and reporting, AI-powered network planning, network deployment and distribution, and continuous network support. Table \ref{tab-surveys} summarizes a comparison of related surveys and our work, and Fig. \ref{mainfig} presents an illustration of the application of LLMs in various communication network domains.

\subsection{Main Contributions}
The main contribution of this survey is to provide a comprehensive study of the potential of LLMs for NSM in current and emerging communication networks. 
Specifically, this work presents a unified perspective on LLM integration across diverse network environments, including mobile networks and IoT technologies, vehicular networks, cloud-based networks, and fog/edge-based networks. We highlight the unique capabilities of LLMs to address complex NSM tasks such as network monitoring, anomaly detection, resource optimization, network configuration, and security enhancements that are critical in supporting the scale and performance requirements of emerging network technologies. Along the line, we discuss the methodology of our survey to highlight an exhaustive literature survey that contributes to the comprehension of the prospective reader.

In an attempt to fill the void left by prior surveys, this paper first lays the foundation for understanding the fundamental concepts and aspects of LLMs and their transformative potential in NSM, detailing the basic transformer architecture, comparing general-purpose and domain-specific LLMs, and LLM pre-training and fine-tuning. Additionally, this survey proposes a novel taxonomy for LLM applications in communication NSM, covering the main areas of network monitoring and reporting, AI-powered network planning, network deployment and distribution, and continuous network support. Under each categorization, we review existing and current research papers, identify their innovative points, discuss their experimental outcomes, and outline specific methods used to address some vital challenges faced in deploying LLMs for NSM, considering each of the different communication networks involved in this survey. Finally, we discuss challenges, open issues, and future research directions associated with LLMs for NSM in communication networks. \textit{To the best of our knowledge, this is the first extensive survey conducted on LLM-enabled NSM in communication networks involving mobile networks and IoT technologies, vehicular networks, cloud-based networks, and fog/edge-based networks}. We envision this structured and exhaustive material to serve as a roadmap for future research and practical implementations of LLMs in NSM, addressing the demands of next-generation communication systems. The main contributions of this survey are outlined as follows:

\begin{table*}[ht]
    \renewcommand\arraystretch{1.2}
    \centering
    \caption{Electronic Database Search}
    \label{tab:electronic_database_search}
    \begin{threeparttable}
        \resizebox{\textwidth}{!}{
            \begin{tabular}{L{3cm} L{3.5cm} L{3cm} L{6.5cm}}
                \Xhline{1.2pt}
                \textbf{Electronic Database} & \textbf{Database Type} & \textbf{Information Used} & \textbf{URL} \\
                \Xhline{1.2pt}
                Google Scholar & Research-oriented search engine & Keyword search & \url{https://scholar.google.com/} (Accessed on Sept 6, 2024) \\
                \hline
                Semantic Scholar & Research-oriented search engine & Keyword search & \url{https://www.semanticscholar.org/} (Accessed on Sept 6, 2024) \\
                \hline
                IEEE Xplore & Digital library & Title, full text search & \url{https://ieeexplore.ieee.org/Xplore/home.jsp} (Accessed on Sept 6, 2024) \\
                \hline
                ACM Digital Library & Digital library & Title, full text search & \url{https://dl.acm.org/} (Accessed on Sept 8, 2024) \\
                \hline
                Springer Link & Digital library & Title, full text search & \url{https://link.springer.com/} (Accessed on Sept 8, 2024) \\
                \hline
                Elsevier (ScienceDirect) & Digital library & Title, full text search & \url{https://www.sciencedirect.com/} (Accessed on Sept 8, 2024) \\
                \Xhline{1.2pt}
            \end{tabular}
        }
    \end{threeparttable}
\end{table*}

\begin{figure}[!t]
    \centering
    \includegraphics[width=3.0in]{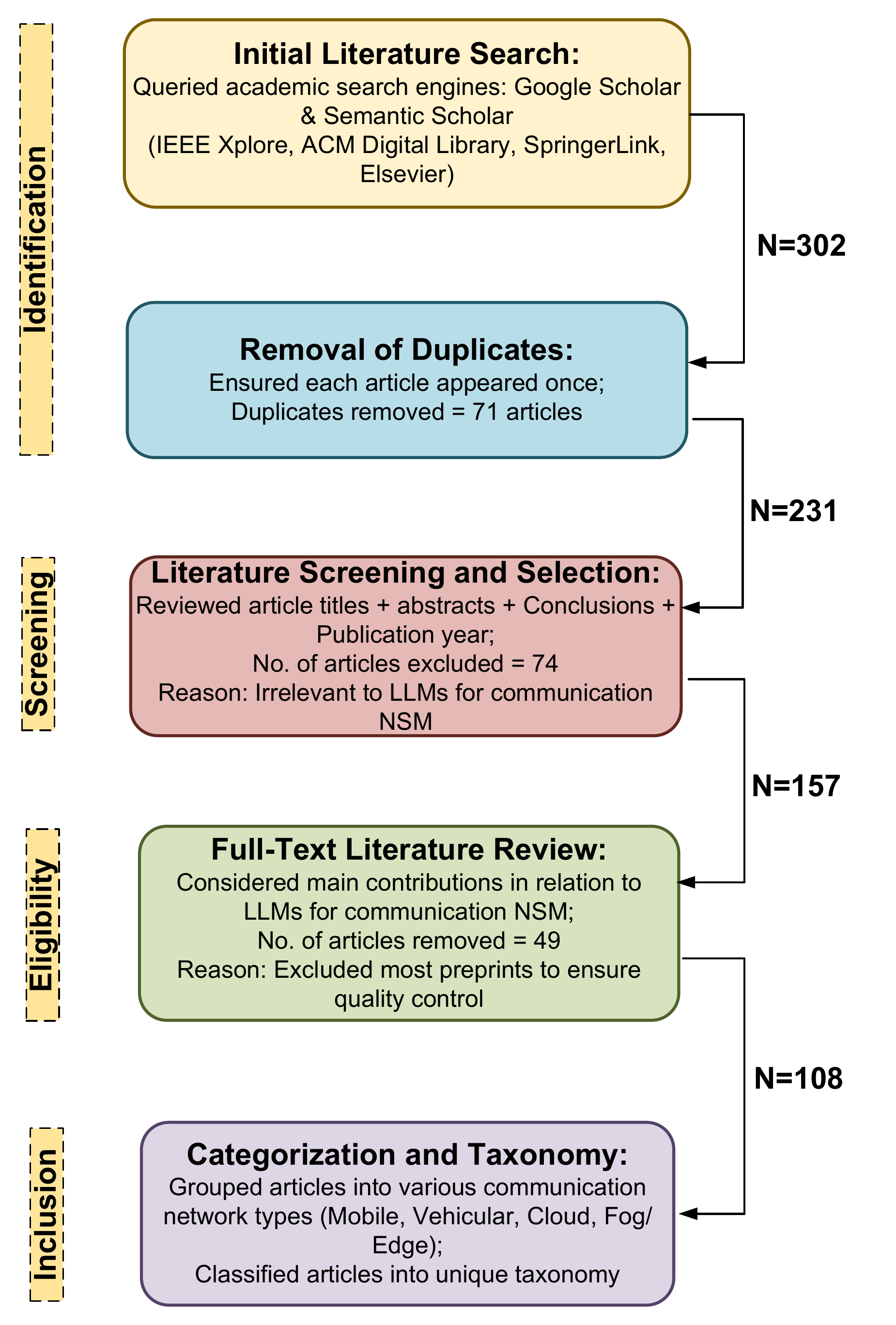}
    \caption{PRISMA flow diagram.}
    \label{prisma}
\end{figure}
\begin{itemize}
    \item We provide a well-structured comprehensive survey of the potential of LLMs for NSM across current and emerging communication networks, including mobile networks and IoT technologies, vehicular networks, cloud-based networks, and fog/edge-based networks.

    \item We present the fundamentals of LLMs in terms of the basic transformer architecture, general-purpose and domain-specific LLMs, model pre-training and fine-tuning, and their integration into communication networks for tackling NSM tasks.

    \item We curate a unique taxonomy for LLM applications in communication NSM, exploring extensive literature on the aforementioned network types and categorizing them under four main broad tasks, namely, LLM for network monitoring and reporting, LLM for AI-powered network planning, LLM for network deployment and distribution, and LLM for continuous network support.

    \item Finally, we discuss critical challenges, open issues, and potential future research directions on the application of LLMs for NSM in communication networks.

\end{itemize}

\subsection{Survey Methodology}
This survey employed the Preferred Reporting Items for Systematic Reviews and Meta-Analyses (PRISMA) framework to conduct a rigorous, thorough, and fair literature review, as it ensures a systematic approach to identifying, screening, and selecting relevant studies \cite{Pagen71}. The PRISMA framework helps to maintain a high level of quality and consistency throughout the literature review process.

\subsubsection{Initial literature search}
The initial search was conducted using Google Scholar and Semantic Scholar, which query major relevant digital libraries such as IEEE Xplore, ACM Digital Library, Springer Link, Elsevier (ScienceDirect), etc. A broad combination of keywords relevant to the study was used to capture almost all related studies as follows:
\begin{itemize}
    \item \textit{\textquote{Large Language Models}} OR \textit{\textquote{LLMs}} AND \textit{\textquote{Network}} AND/OR \textit{\textquote{Service}} AND/OR \textit{\textquote{Management}} AND \textit{\textquote{Mobile Network}}

    \item \textit{\textquote{Large Language Models}} OR \textit{\textquote{LLMs}} AND \textit{\textquote{Network}} AND/OR \textit{\textquote{Service}} AND/OR \textit{\textquote{Management}} AND \textit{\textquote{Vehicular Network}}

    \item \textit{\textquote{Large Language Models}} OR \textit{\textquote{LLMs}} AND \textit{\textquote{Network}} AND/OR \textit{\textquote{Service}} AND/OR \textit{\textquote{Management}} AND \textit{\textquote{Cloud Network}}

    \item \textit{\textquote{Large Language Models}} OR \textit{\textquote{LLMs}} AND \textit{\textquote{Network}} AND/OR \textit{\textquote{Service}} AND/OR \textit{\textquote{Management}} AND \textit{\textquote{Fog Network}} AND/OR \textit{\textquote{Edge Network}}
\end{itemize}
In all, the initial literature search resulted in 302 manuscripts, including 14 survey papers and several magazines, conferences, and journal manuscripts, of which some of them are preprints. This comprehensive literature search approach allowed for a broad collection of studies on LLMs for NSM across multiple communication network domains.

\subsubsection{Removal of duplicates}
After the initial literature search, duplicate papers were removed to streamline the literature review process and ensure that each study was only reviewed once. A total of 71 papers were duplicated, which were removed, and the remaining 231 moved to the next stage of literature screening and selection. 

\subsubsection{Literature screening and selection}
The remaining papers were screened based on their titles, abstracts, conclusions, and publication year on predefined inclusion and exclusion criteria. The criteria focused on whether an article really reflects the main content of LLM for NSM in communication networks, and whether it was published in the past five years. If yes, the article was included; otherwise, the article was excluded. This step ensured that only studies relevant to the objectives of the survey were considered. At this stage, 74 papers were excluded, and the remaining 157 papers were considered for full-text literature review. 

\subsubsection{Full-text literature review}
The remaining articles were then thoroughly reviewed to evaluate their quality and relevance to the survey. The in-depth review considered the main contributions of each study in relation to LLMs for communication NSM, specifically identifying its innovative methods or approaches. Since articles in preprint are not peer-reviewed, we removed most of them at this stage to ensure quality control. Finally, 108 manuscripts were considered for the categorization and taxonomy. 

\subsubsection{ Categorization and taxonomy}
To organize the findings and design a more accurate taxonomy, the selected studies were categorized into four primary network domains: \textit{\textquote{Mobile networks and IoT technologies}}, \textit{\textquote{Vehicular networks}}, \textit{\textquote{Cloud-based networks}}, and \textit{\textquote{Fog/Edge-based networks}}.  The categorization was tabulated with the following attributes for each paper: paper title, the year it was published, publisher, an overview of the paper, the problem it addressed, main contributions, the proposed LLM solution, datasets used in the simulation/experiments, and the evaluation metrics. Within each domain, the studies were further classified under the following taxonomies:
 \begin{itemize}
     \item LLM for Network Monitoring and Reporting

     \item LLM for AI-powered Network Planning

     \item LLM for Network Deployment and Distribution

     \item LLM for Continuous Network Support
 \end{itemize}
This classification approach provides a clear structure for analyzing the role of LLMs in communication NSM and highlights the contributions of each study within its specific network and functional category.

Fig. \ref{prisma} depicts the PRISMA flow diagram used in this survey, and Table \ref{tab:electronic_database_search} describes the electronic database search for selecting articles from various database sources.

\subsection{Outline of Survey}
The rest of the survey is organized as follows: Section \ref{sec:llmfundamentals} presents the fundamentals of LLMs, and Section \ref{sec:taxonomy} provides a clear proposed taxonomy of LLMs for NSM in communication networks. Section \ref{sec:mobile} focuses on LLMs for mobile networks and technologies-based NSM, Section \ref{sec:vehicular} focuses on LLMs for vehicular networks-based NSM, Section \ref{sec:cloud} focuses on LLMs for cloud-based NSM, and Section \ref{sec:fogedge} focuses on LLMs for fog/edge-based NSM. Section \ref{sec:challenges} identifies challenges, open issues, and future research directions. Finally, Section \ref{sec:conclusion} concludes this survey. The outline of our survey is illustrated in Fig. \ref{org}.

\begin{figure*}
\centerline{\includegraphics[width=5.5in]{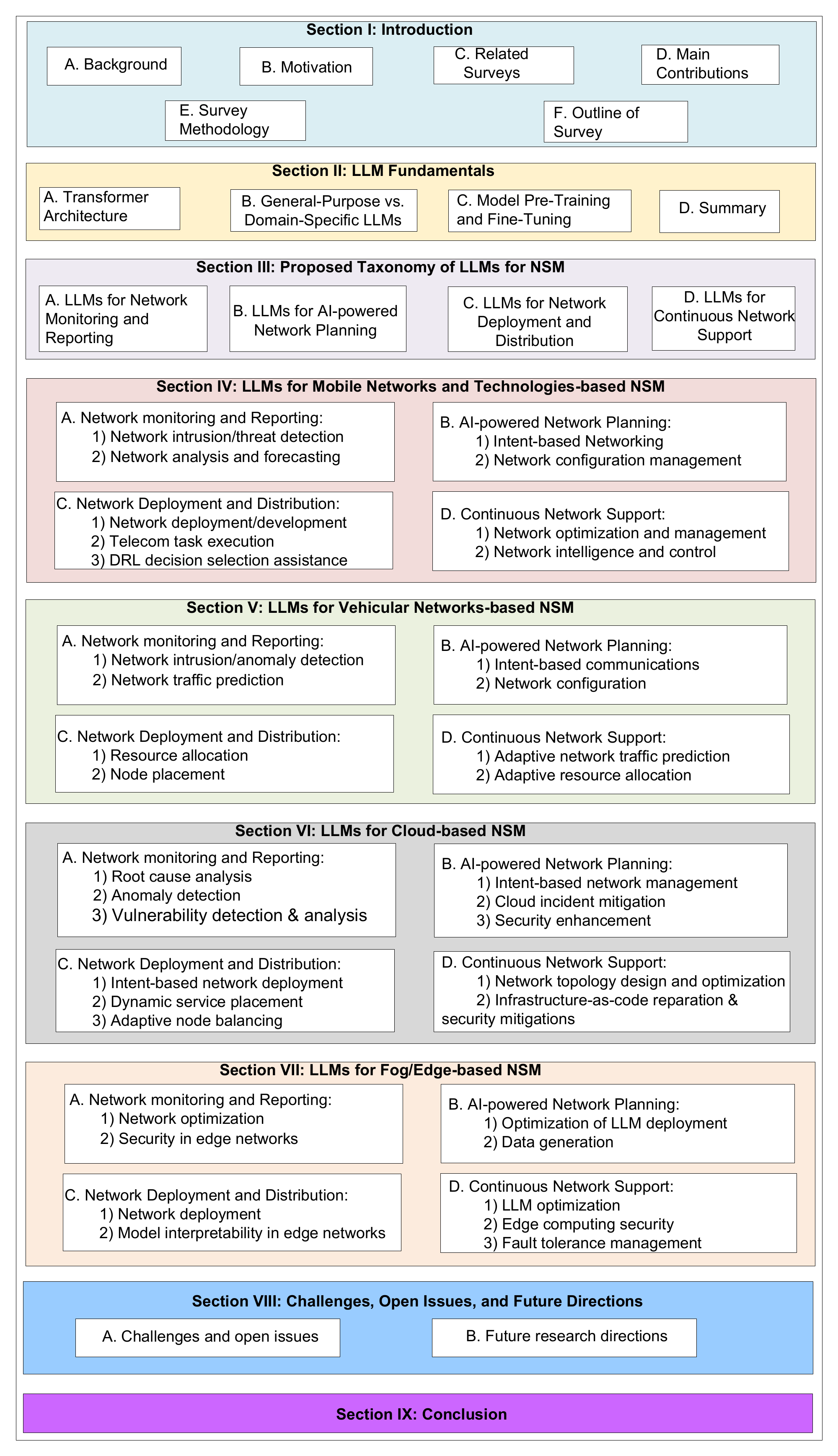}}
\caption{Outline of survey.}
\label{org}
\end{figure*} 
\section{LLM Fundamentals}\label{sec:llmfundamentals}

LLMs are a class of ML models that have exceptional capabilities in processing and generating human-like output for tasks across domains such as NLP and computer vision by capturing intricate language patterns \cite{10.1145/3641289}. Their capability to learn complex language patterns stems from their extensive training on large-scale datasets, enabling them to generalize across applications. Capitalizing on advanced neural network architectures, LLMs have demonstrated the ability to understand, interpret, and generate language with a level of contextual understanding that often surpasses traditional models such as Recurrent Neural Networks (RNNs). Notably, LLMs have demonstrated exceptional performance in a broad range of NLP tasks, e.g., text classification \cite{zhang2024pushinglimitllmcapacity}, question-answering (Q\&A) \cite{Li_Fan_Gu_Li_Duan_Dong_Liu_Wang_2024}, and sentiment analysis \cite{miah2024multimodal}. Their capabilities now go beyond NLP, expanding into other domains of computer vision, e.g., image recognition \cite{koziolek2024llm}, object detection \cite{zang2024contextual}, and multimodal applications, positioning them as versatile, scalable tools for a range of complex tasks. 

The LLM architecture is built on transformers \cite{vaswani2017attention}, which leverage attention mechanisms to capture intricate dependencies between inputs and outputs to learn real relationships between them without the constraints of sequential processing. The ability of the transformer architecture to manage and interpret extensive language information through self-attention and parallel processing sets it apart from other sequence models, such as RNNs and LSTM networks, which are constrained by sequential dependencies. A typical transformer architecture comprises two essential components, namely, the encoder and the decoder, both structured as stacks of self-attention and feed-forward layers that allow for managing substantial input data flexibly and interpreting complex language patterns. The encoder is responsible for breaking down inputs as tokens and generating rich representations that encapsulate contextual dependencies between input tokens. The decoder interprets the encoder's output representations, especially in tasks where output generation is required, such as text translation, text generation, and text summarization. The self-attention mechanism lies at the heart of the transformer and is crucial for understanding inter-dependencies.

\subsection{Transformer Architecture}
The transformer architecture consists of several interconnected components that enable it to process sequential data with high efficiency and accuracy. The core components of a transformer architecture are the input layer, embedding layers (input and output embeddings), positional encoding layers, encoder and decoder modules (containing self-attention mechanisms and normalization layers), and output layer, as illustrated in Fig. \ref{transformer}. In the following, we present the function of each component:

\begin{itemize}
    \item \textbf{Input layer:} The input layer is where raw data is parsed as tokens into the transformer. For language-related tasks, the input generally consists of a sequence of tokens, such as words, subwords, or characters, that represent the text to be processed by the model. The tokens are typically mapped from a vocabulary to numerical indices that the model can interpret.

     \item \textbf{Embedding layers:} The transformer architecture consists of the input embedding layer and output embedding layer. The input embedding layer precedes the encoder and converts tokens from the input sequence into dense vector representations. These embeddings allow the model to understand the semantic relationships between tokens, which are essential for capturing contextual information. On the other hand, the output embedding layer exists after the decoder and converts the model's internal continuous representations back into discrete tokens in the output vocabulary, typically in the form of probabilities for each token. 

     \item \textbf{Positional encoding layer:} Since transformers process tokens in parallel, they lack the ability to order tokens. The positional encoding layer solves this issue by providing the position of tokens in a sequence, preserving sequential information.

     \begin{figure}[!t]
    \centering
    \includegraphics[width=3.5in]{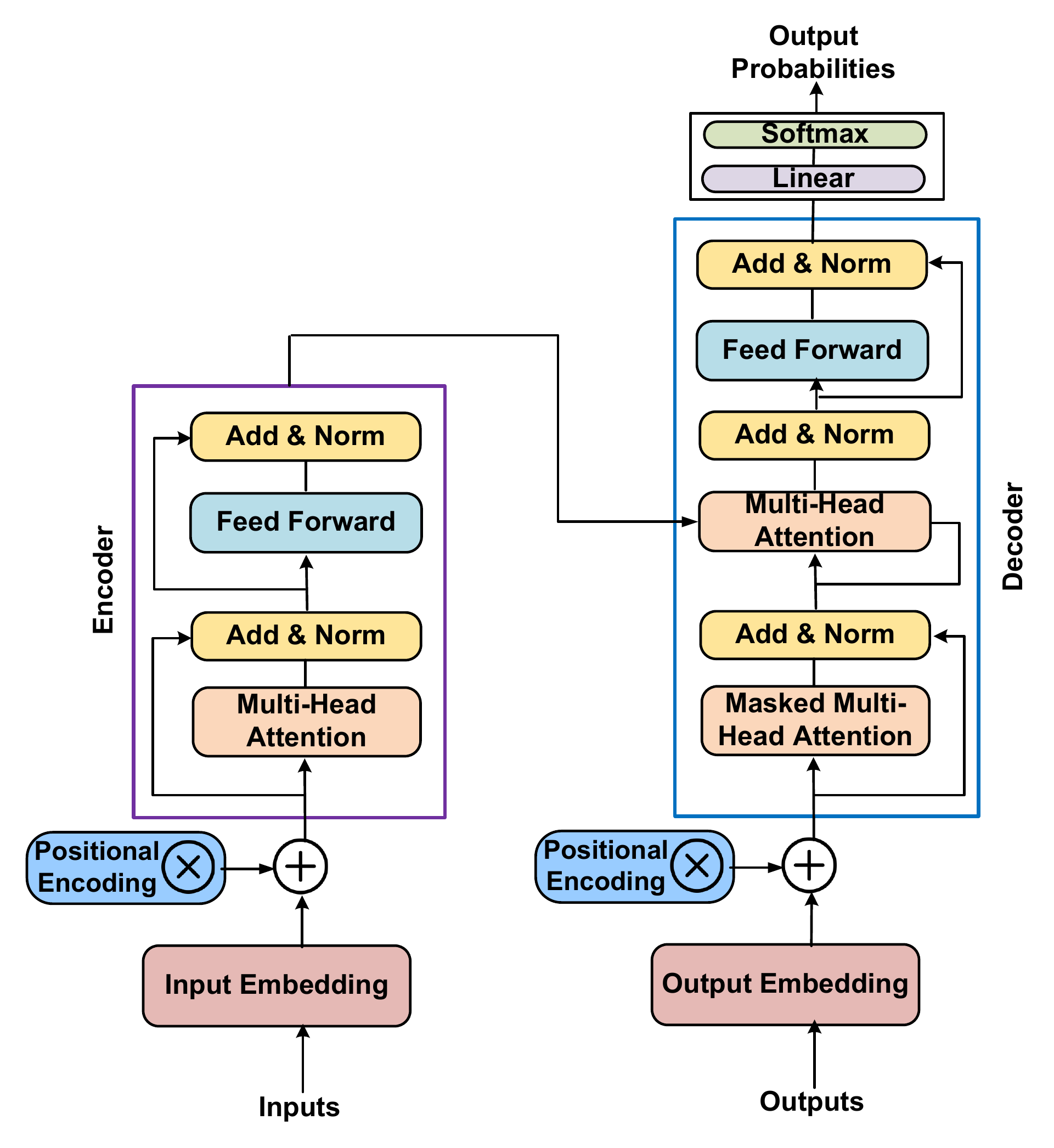}
    \caption{Transformer architecture \cite{10433480}.}
    \label{transformer}
\end{figure}
     \item \textbf{Encoder and decoder modules:} The encoder and decoder modules comprise multiple identical self-attention, feed-forward neural network layers, and normalization layers. The encoder module processes the input sequence to generate context-aware representations, capturing relationships among tokens within the input sequence. The decoder module is structured similarly to the encoder module, with self-attention, feed-forward neural network, and normalization layers, but includes an additional encoder-decoder attention mechanism. This additional mechanism allows the decoder to attend to relevant parts of the encoder's output, facilitating coherent output generation conditioned on the input sequence.

     \item \textbf{Self-attention mechanisms:} The self-attention mechanism at the encoder enables the encoder to weigh the importance of each token in relation to every other token in the sequence. It calculates the weighted representation based on the relevance of other tokens to the target token, allowing the model to capture dependencies regardless of distance. The self-attention mechanism at the decoder is slightly modified to prevent tokens from attending to future tokens, ensuring that the output generation remains auto-regressive and respects the order in which tokens are generated. The encoder-decoder attention mechanism, also known as the cross-attention mechanism, aligns encoder outputs with decoder tokens, ensuring the generated token is contextually appropriate for the input. It allows the decoder to focus on specific parts of the encoder's output.

     \item \textbf{Output layer:} The output layer generates the final prediction from the decoder's processed representations. Typically, the output layer consists of a linear layer and a softmax function, which converts the decoder's output into probabilities over the target vocabulary.
     
\end{itemize}
The operational workflow of a typical transformer is as follows: The transformer processes input through multiple layers of encoders and decoders. Initially, each input token is embedded in a high-dimensional space to capture its semantic meaning, and these embeddings are then combined with positional encodings that help to retain sequence information. Within the encoder module, a self-attention mechanism calculates how each token relates to every other token in the input sequence, producing a weighted matrix of attention scores. The matrix enables the transformer to contextualize each token in relation to the entire sequence, allowing it to process complex semantics efficiently. Then, a feed-forward neural network is applied to further refine the information. A normalization layer then stabilizes learning and improves performance across the stacked layers. The final encoded representations are then passed to the decoder. The decoder layer utilizes both self-attention and cross-attention mechanisms to enable the generation of tokens that align with the input context. Similar to the encoder, the decoder stack uses normalization to ensure model stability and efficiency. The decoder’s output representations are transformed back into probabilities over the target vocabulary by the output embedding layer, with the most likely token selected to generate the final sequence. During training, a loss function calculates the difference between predictions and actual outputs, guiding adjustments through backpropagation to improve the model’s accuracy iteratively.

\begin{table*}[ht]
    \centering
    \caption{Comparative Summary of General-Purpose and Domain-Specific LLMs}
    \label{tab:comparative_summary_llms}
    \begin{threeparttable}
        \resizebox{\textwidth}{!}{
            \begin{tabular}{|p{2.2cm}|p{2.2cm}|p{2.5cm}|p{3cm}|p{2cm}|p{3cm}|p{3cm}|}
                \hline
                \textbf{Domain} & \textbf{Model} & \textbf{Application Scenario} & \textbf{Datasets} & \textbf{Parameters} & \textbf{Tasks} & \textbf{Knowledge Base} \\
                \hline
                \multirow{5}{*}{\textbf{General-purpose}} 
                & GPT-4 \cite{achiam2023gpt} & General NLP tasks & Broad internet data, books, articles, code repositories & $\sim$1.8 trillion & Text generation, summarization, Q\&A, code completion & Cross-disciplinary knowledge, language structure, scientific and general concepts \\
                \cline{2-7}
                & LLaMA-3 \cite{dubey2024llama} & Multilingual tasks & Public internet text, academic resources & 8B - 70B & Language generation, sentiment analysis, summarization, translation & General-purpose, multilingual knowledge base \\
                \cline{2-7}
                & Mistral 7B \cite{jiang2023mistral} & General NLP tasks & Curated datasets for general NLP & 7 billion & Text generation, classification, summarization & General-purpose, cross-functional NLP knowledge \\
                \cline{2-7}
                & Turing-NLG \cite{smith2022usingdeepspeedmegatrontrain} & General NLP tasks & Microsoft's large text corpus & 17 billion & Text generation, summarization, Q\&A & Strong general-purpose knowledge with structured and conversational data \\
                \cline{2-7}
                & Gemini \cite{team2023gemini} & Multimodal processing & Multimodal data (text, image, audio) & $\sim$280 billion & Complex Q\&A, dialogue, multimodal tasks & Multimodal knowledge across NLP and visual contexts \\
                \hline
                \multirow{9}{*}{\textbf{Domain-specific}} 
                & ChatDoctor \cite{li2023chatdoctor} & Healthcare & Medical dialogue datasets, clinical guidelines, patient records & 100K & Patient support, diagnosis, medical Q\&A & In-depth medical knowledge, clinical terminology, patient-doctor interaction specifics \\ 
                \cline{2-7}
                & BloombergGPT \cite{wu2023bloomberggpt} & Finance & Financial news, economic reports, stock data & 569-770B & Market forecasting, sentiment analysis, financial reporting, trend forecasting & Financial markets, economic indicators, stock trends \\ 
                \cline{2-7}
                & DeliLaw \cite{xie2024delilaw} & Law, legal research & Legal documents, case law, statutes, contracts (Chinese legal data) & 980K & Legal counseling, case retrieval, document analysis & Comprehensive legal knowledge \\
                \cline{2-7}
                & DriveGPT4 \cite{xu2024drivegpt4} & Autonomous driving & Autonomous driving datasets (KITTI, NuScenes) & 112K & Route planning, driving decisions, obstacle detection & Driving scenarios, vehicle dynamics, traffic regulations \\ 
                \cline{2-7}
                & Mobile-LLaMA \cite{10583947} & Network analysis in 5G & Publicly available 5G network datasets (e.g., BGP routing tables, packet capture files, and performance metrics) & 13 billion & IP routing analysis, packet analysis, performance analysis & 5G network management, NWDAF (Network Data Analytics Function), network traffic patterns, anomaly detection \\
                \cline{2-7}
                & NetLLM \cite{wu2024netllm} & Networking optimization & Multimodal networking data (e.g., viewport prediction, ABR, cluster job scheduling datasets) & Varied (up to large-scale LLM) & Viewport prediction, ABR optimization, job scheduling & Network resource management, video streaming, cloud computing \\
                \cline{2-7}
                & WirelessLLM \cite{shao2024wirelessllmempoweringlargelanguage}  & Wireless communication & Multi-modal data from wireless systems (protocol data, sensor readings) & Variable, fine-tuned on wireless datasets & Spectrum sensing, power allocation, protocol understanding & Knowledge of wireless protocols, electromagnetic principles, signal processing \\
                \cline{2-7}
                & ConnectGPT \cite{10588835} & Connected vehicles & Vehicular and traffic data, C-ITS message data & GPT-4 backbone & CAV routing, hazard detection, traffic management & Traffic safety protocols, C-ITS standards, vehicle communication \\
                \cline{2-7}
                & NetGPT \cite{chen2023netgpt} & Edge and cloud synergy & Cloud-edge LLM data with fine-tuning for network environments & LLaMA-7B (cloud) and GPT-2-base (edge) & Network orchestration, trend prediction, user intent inference & AI-native network management, prompt completion, cloud-edge collaboration \\
                \hline
            \end{tabular}
        }
    \end{threeparttable}
\end{table*}
\subsection{General-purpose vs. Domain-specific LLMs}
LLMs have shown enormous potential in their application in various language-related tasks, from open-ended dialogue generation to high-level specialized technical tasks. Specifically, LLMs can be characterized by the application scenarios they are designed for, the tasks they are able to execute, the kind of datasets required to train them, and the knowledge base they can access. Based on these factors, LLMs can be broadly divided into two categories, namely, general-purpose LLMs and domain-specific LLMs. General-purpose LLMs are designed to handle a wide array of language tasks across various domains. Examples of common general-purpose LLMs are GPT-4 \cite{achiam2023gpt}, developed by OpenAI; Mistral 7B \cite{jiang2023mistral}, developed by Mistral AI; LLaMA-3 \cite{dubey2024llama}, developed by Meta; Turing-NLG \cite{smith2022usingdeepspeedmegatrontrain}, developed by Microsoft; and Gemini \cite{team2023gemini}, developed by Google DeepMind. Domain-specific LLMs are tailored to excel in particular fields or domains by integrating specialized knowledge from the domain. Examples are TelecomGPT \cite{zou2024telecomgptframeworkbuildtelecomspecfic} for telecom-specific tasks, DriveGPT4 \cite{xu2024drivegpt4} for autonomous driving and transportation-related tasks, and ChatDoctor 
\cite{li2023chatdoctor} for medical-oriented tasks. In the following, we discuss general-purpose and domain-specific LLMs considering application scenarios, datasets, tasks, and knowledge base:

\subsubsection{General-purpose LLMs}
General-purpose LLMs are versatile models trained on dense datasets spanning multiple domains. This allows them to perform a broad spectrum of language-related tasks. For instance, GPT-4 \cite{achiam2023gpt} is trained on broad internet data, books, articles, and code repositories with approximately 1.8 trillion parameters to generalize well across different topics, making them applicable to a variety of scenarios. Turing-NLG \cite{smith2022usingdeepspeedmegatrontrain} is trained on a massive text corpus collected by Microsoft with 17 billion parameters for text generation, question answering, and summarization. Their adaptability comes from exposure to vast, heterogeneous datasets, enabling them to understand, generate, and adapt language across contexts. The main purpose of general-purpose LLMs is to provide a reliable, adaptable tool that can respond efficiently to different prompts without being constrained by any specific domain.

\begin{itemize}
    \item \textbf{Application scenarios:} General-purpose LLMs are deployed in diverse applications, including conversational AI, customer service chatbots, summarization, translation, and Q\&A systems. With their flexible knowledge base, general-purpose LLMs are suitable for open-ended scenarios where the input can vary significantly. Jiang \textit{et al.} \cite{jiang2023mistral} introduced Mistral 7B, a state-of-the-art language model comprising 7 billion parameters designed to achieve high performance in general NLP tasks in research and enterprise applications. The Gemini Team at Google \cite{team2023gemini} presented Gemini, a family of multimodal models that demonstrate advanced capabilities in understanding and processing image, audio, video, and text data for application scenarios such as complex question-answering and dialogue systems.

    \item \textbf{Datasets:} The training data of general-purpose LLMs is typically extensive and heterogeneous, encompassing sources such as books, websites, news articles, and code repositories. This diverse corpus ensures that the model is exposed to varied topics and contextual knowledge. The data is generally unfiltered for specific domains, allowing these models to learn broadly applicable language patterns. For example, GPT-4 trains on broad internet data, books, articles, and code repositories, while LLaMA trains on public internet text and academic resources.

    \item \textbf{Tasks:} General-purpose LLMs are optimized to handle a wide range of tasks, including text completion, summarization, text classification, and information retrieval. They are commonly used for more advanced functions like reasoning, natural language inference, and zero-shot learning, where the model makes inferences on tasks it has not been explicitly trained on. These capabilities make them powerful for research and daily language tasks.

    \textbf{Knowledge base:} Due to broad data exposure, general-purpose LLMs exhibit a balanced understanding of many fields. The knowledge is often cross-disciplinary, covering general knowledge base in language structure, scientific concepts, and even technology knowledge. 
\end{itemize}

General-purpose LLMs have the advantage of performing a wide range of tasks due to their heterogeneous knowledge base. However, their broad knowledge scope often sacrifices specialty. That is, they may lack the intricate details needed for specialized technical or scientific inquiries. For instance, Turing-NLG \cite{smith2022usingdeepspeedmegatrontrain}, which is trained on Microsoft's large text corpus and performs well for text generation, summarization, and Q\&A tasks, may not suffice in vehicular network environments that primarily utilize video and image data for more environmental contexts and semantics.

\subsubsection{Domain-specific LLMs}
Contrasting general-purpose LLMs, domain-specific LLMs are tailored for specialized areas such as medicine \cite{li2023chatdoctor}, finance \cite{wu2023bloomberggpt}, engineering \cite{shao2024wirelessllmempoweringlargelanguage}, and law \cite{xie2024delilaw}. By focusing on a specific domain, these LLMs can achieve high accuracy and relevance in domain-specific tasks, understanding complex terminology, concepts, and regulations that general-purpose LLMs might overlook. The main goal of domain-specific LLMs is to provide robust support with specialized areas, enhancing the model's performance and reliability for experts and practitioners.

\begin{itemize}
      \item \textbf{Application scenarios:} Domain-specific LLMs are used in specialized environments where knowledge accuracy and relevance are paramount. Xie \textit{et al.} \cite{xie2024delilaw} developed DeliLaw, a Chinese legal counseling system that leverages LLM specifically trained for the legal domain, integrating both legal and case retrieval modules to mitigate issues of model hallucination. Some authors \cite{huang2024large} proposed the development of a domain-adapted LLM tailored for networking applications, highlighting the importance of mapping natural language to network-specific language.

      \item \textbf{Datasets:} Domain-specific LLMs are trained on carefully curated datasets specific for their domain. The datasets are often filtered and preprocessed to ensure that the model learns only the relevant vocabulary and procedures specific for the domain, resulting in high accuracy and domain-relevant knowledge. For example, ChatDoctor was trained on a large dataset of one hundred thousand (100,000) patient-doctor dialogues sourced from a widely used online medical consultation platform \cite{li2023chatdoctor}. The work in \cite{10588373} utilized the open-source KITTI \cite{geiger2013vision} and NuScenes \cite{caesar2020nuscenes} datasets to train their Large Vision Language Model (LVLM) for traffic scene understanding.

      \item \textbf{Tasks:} Domain-specific LLMs excel in specialized tasks unique to their domain. In healthcare, medical LLMs can understand patient inquiries and provide accurate advice. In Telecom, telecom-specific LLMs can be used for Q\&A, 3GPP technical documents classification, and telecom code summary and generation.

      \item \textbf{Knowledge base:} Domain-specific LLMs have a deep understanding of their field, allowing them to grasp complex terminologies, procedures, and knowledge structures. This specialized knowledge base allows them to generate precise, accurate, and contextually relevant responses.
\end{itemize}

Although domain-specific LLMs are well-versed in their specific domains, their capabilities may not extend well beyond the trained field due to limited cross-domain knowledge. 

Table \ref{tab:comparative_summary_llms} gives a comparative summary of general-purpose and domain-specific LLMs, considering their application scenarios, datasets, parameters, tasks, and knowledge base.

\subsection{Model Pre-Training and Fine-Tuning}
Model pre-training and fine-tuning are foundational processes in the development of LLMs \cite{sun2024amuro}. Pre-training and fine-tuning enable LLMs to learn from extensive datasets from different sources and adapt to specific tasks. If done appropriately, LLMs can yield highly versatile and powerful outcomes after being pre-trained and fine-tuned. Pre-training provides LLMs with a generalized understanding of language and patterns, equipping them with a broad representation of learning knowledge, reasoning, and comprehension capabilities. On the other hand, fine-tuning refines this knowledge, adapting the model to specialized tasks or application domains, which achieves significant performance improvements across a range of applications.

\subsubsection{Model pre-training}
Model pre-training is the initial phase of LLM development, during which the model is exposed to extensive amounts of unlabeled data to learn general language patterns, contextual relationships, and semantic understanding. Pre-training a model on extensive datasets allows it to capture complex language patterns, semantic relationships, and general world knowledge, making it more effective when applied to specific tasks. This approach significantly reduces the data and computational resources required during the fine-tuning phase, as the model already has a basic understanding of the data. 

LLM pre-training involves feeding the model with an enormous amount of data, which the model processes through supervised or self-supervised learning methods. The model predicts missing words (masked language modeling) or next words (causal language modeling) in sentences to encourage it to learn syntactic structures, contextual dependencies, and broader semantic relationships. Over time, it builds a comprehensive understanding of language, enhancing its ability to handle various tasks with minimal domain-specific training. The data for model pre-training may come from various sources such as common crawl \cite{patel2020introduction}, Wikipedia dumps, web text \cite{liu2006web}, video databases (CCTV camera feed) \cite{caesar2020nuscenes}, image files, and specific data for application-specific tasks. Common crawl is a widely used corpus that provides an extensive snapshot of the web, covering a vast range of topics and language patterns. Wikipedia dumps are highly structured and factual Wikipedia data that allow LLMs to gain reliable knowledge across diverse subjects. Web text is a curated dataset of high-quality web pages, often used to enhance the model's understanding of conversational language. The choice of dataset for model pre-training depends on whether the LLM is general-purpose or domain-specific. 
For example, using masked language modeling, BERT \cite{devlin2018bert} was pre-trained on the BooksCorpus and English Wikipedia for tasks requiring sentence-level understanding, such as Q\&A and sentence classification. For domain-specific LLMs, it is desirable to pre-train the model with datasets collected from the specific domain. For instance, Mobile-LlaMA LLM \cite{10583947} was pre-trained on publicly available 5G network datasets (e.g., BGP routing tables, packet capture files, and performance metrics). 

Pre-training LLMs are computationally intensive and often require distributed computing on GPU clusters or TPUs to process the massive datasets involved. Distributed computing offers several benefits and techniques to enhance the efficiency and stability of LLM pre-training. Techniques such as data parallelism, model parallelism \cite{shoeybi2019megatron}, pipeline parallelism \cite{li2021terapipe}, tensor parallelism \cite{wang2022tesseract}, Zero Redundancy Optimizer (ZeRO) \cite{rajbhandari2020zero}, gradient accumulation, and gradient sharding can be used to enhance LLM pre-training stability, speed, and cost-effectiveness. In data parallelism, the model is duplicated across multiple devices, such as GPUs and TPUs, and each device processes a different mini-batch of data. Gradients are calculated independently on each device and then averaged across all devices before updating the model parameters. Model parallelism divides the model across multiple devices, which is especially useful for very large models that cannot fit in the memory of a single device. Pipeline parallelism divides the model into segments (or stages), with each segment allocated to different devices. As each device completes its segment, it passes the activations to the next device in the pipeline. Tensor parallelism is a fine-grained form of model parallelism where tensors are split across multiple devices. It enables efficient computation of large matrix operations, particularly in dense layers of transformer models, and can lead to substantial performance improvements. ZeRO is an efficient optimizer that shards optimizer states across devices to reduce memory footprint. 

Efficient model pre-training stability can be achieved by carefully selecting key hyperparameters, such as learning rate, batch size, sequence length, attention heads, and dropout rate. For the learning rate, warm-up training steps, adaptive learning rate adjustments, and cosine learning rates \cite{cai2024medusa} can be employed. The batch size can be incremented step by step to enhance the stability of pre-training. A longer sequence length can help capture context but requires more computation. Multiple attention heads in the attention mechanism allow the model to focus on different parts of the input sequence, enhancing its contextual understanding. Dropout rate is a regularization technique to prevent overfitting, where certain neuron connections are randomly ignored using pre-training.

\renewcommand{\arraystretch}{0.9} 
\begin{table*}[ht]
\centering
\caption{Summary of LLMs Considering Pre-Training and Fine-Tuning}
\label{tab:pretraining_finetuning}
\tiny 
\setlength{\tabcolsep}{1pt} 
\begin{adjustbox}{width=\textwidth}
\begin{tabular}{|p{1.4cm}|p{1.1cm}|p{1.7cm}|p{1.0cm}|p{1.5cm}|p{1.7cm}|p{1.0cm}|p{1.5cm}|p{1.5cm}|}
\hline
\textbf{Model} & \textbf{Base Model} & \textbf{Pre-training Dataset} & \textbf{Tokens (Pre-training)} & \textbf{Pre-training Technique} & \textbf{Fine-tuning Dataset} & \textbf{Tokens (Fine-tuning)} & \textbf{Fine-tuning Technique} & \textbf{Hardware} \\
\hline
BERT \cite{devlin2018bert} & BERT & BooksCorpus, Wikipedia & 3.3B & Data parallelism & Q\&A, Sentiment & 500M & Task-specific fine-tuning & V100 GPUs \\
\hline
GPT-4 \cite{achiam2023gpt} & GPT-4 & Web data, etc. & 1.8T & Pipeline parallelism & Instruction-following & 100B & RLHF & A100 GPUs \\
\hline
LLaMA 3 \cite{dubey2024llama} & LLaMA 3 & Multilingual web corpus & 15T & ZeRO & Human-feedback & 1B & RL-based fine-tuning & H100 GPUs, 128K context \\
\hline
ChatDoctor \cite{li2023chatdoctor} & LLaMA & 52K Alpaca instructions & 10M & Data parallelism & Patient dialogues & 50M & Domain-specific Fine-tuning & A100 GPUs \\
\hline
DeliLaw \cite{xie2024delilaw} & ChatGLM & Chinese legal texts & 2B & Data parallelism & Legal Q\&A & Medium & Domain-specific Fine-tuning & A100 GPUs \\
\hline
Mobile-LLaMA \cite{10583947} & LLaMA 2 & 5G network data & 15K samples & ZeRO & IP routing, analysis & Small & Instruction Fine-tuning & A100 GPUs \\
\hline
Tele-LLMs \cite{maatouk2024tele} & TinyLLaMA, etc. & Tele-Data & 1B & ZeRO & Telecom Q\&A & 750K & Full Fine-tuning & A100 GPUs, 1B-8B \\
\hline
ConnectGPT \cite{10588835} & GPT-4 & V2X, road data & 1B & Pipeline parallelism & C-ITS messages & Small & Task-specific Fine-tuning & A100 GPUs, Edge hardware \\
\hline
GenFollower \cite{chen2024genfollower} & GPT-based & Waymo dataset & 10B* & Model parallelism* & Car-following & 1M & Task-specific Fine-tuning & V100 GPUs \\
\hline
MistralBSM \cite{hamhoum2024mistralbsm} & Mistral-7B & VeReMi vehicular data & 500M & Data parallelism & Safety Messages & 100M & Task-specific Fine-tuning & A100 GPUs, Edge deployment \\
\hline
DriveGPT4 \cite{xu2024drivegpt4} & GPT-4 & BDD-X data & 500M* & Pipeline parallelism* & Visual instructions & Small & Mix-multimodal Fine-tuning & A100 GPUs \\
\hline
DriveLLM \cite{cui2023drivellm} & LLaMA & Autonomous driving data & 1B* & Data parallelism* & Real cases & Medium & Domain-specific Fine-tuning & V100 GPUs \\
\hline
DynamicRouteGPT \cite{zhou2024dynamicroutegpt} & LLaMA 3 & Traffic data, dynamics & 500M & ZeRO & Path selection & Medium & RL + Fine-tuning & A100 GPUs \\
\hline
Traj-LLM \cite{lan2024traj} & LLaMA & Trajectory data & 500M* & Data parallelism* & nuScenes & Medium & Parameter-Efficient Fine-tuning & V100 GPUs \\
\hline
IoV-BERT-IDS \cite{fu2024iov} & BERT-based & IoV traffic data & 200M* & Data parallelism* & Intrusion detection & 50M & Task-specific Fine-tuning & A100 GPUs \\
\hline
NetGPT \cite{10466747} & GPT-2, LLaMA & Network, computing data & 1B & ZeRO & Edge-cloud tasks & 10M & LoRA & Edge 7.8GB, Cloud 112GB \\
\hline
NetLLM \cite{wu2024netllm} & LLaMA 2 & Multimodal networking & 1B & ZeRO* & Viewport prediction, bitrate streaming & Medium & Data-Driven Low-Rank Adaptation (DD-LRNA) & A100 GPUs \\
\hline
CoLLM \cite{li2024collm} & GPT-3, LLaMA & Collaborative filtering data & 1B & Data parallelism* & User-item recommendation & Medium & LoRA-based fine-tuning & RTX 3090, Multi-GPU setup \\
\hline
\end{tabular}
\end{adjustbox}
\begin{threeparttable}
		\begin{tablenotes}
			\small 
			\item M: Million, B: Billion, T: Trillion; [$\ast$] Indicates estimated values
		\end{tablenotes}
\end{threeparttable}
\end{table*}

\subsubsection{Model fine-tuning}
Model fine-tuning takes a pre-trained LLM and adapts it to perform specific tasks or operate within a specific domain by further training it on task-specific or domain-specific datasets. Fine-tuning leverages the base knowledge from pre-training, refines it, and optimizes the performance of the model for particular domain applications or use cases. Fine-tuning also enhances the model's ability to generate coherent, accurate responses, adapt to instructions, and minimize errors, making it suitable for practical deployment in diverse fields such as healthcare, legal analysis, and network management.

LLM fine-tuning can be done in several ways and may vary based on the type of data involved, the level of supervision, and the specific objectives. Examples of LLM fine-tuning methods are supervised fine-tuning \cite{jiang2024supervised}, unsupervised fine-tuning \cite{tanwisuth2023pouf}, zero-shot and few-shot fine-tuning \cite{wortsman2022robust}, parameter-efficient fine-tuning \cite{han2024parameter}, and RL-based fine-tuning \cite{chan2024dense}. 

\textbf{i) Supervised fine-tuning:} Supervised fine-tuning uses labeled data, where input-output pairs are explicitly provided. This method helps the model learn to generate desired outputs for specific inputs, refining its ability to perform defined tasks. Examples of supervised fine-tuning are instruction fine-tuning, task-specific fine-tuning, and domain-specific fine-tuning. Instruction fine-tuning adapts the LLM to follow human-like instructions by training it on large datasets containing human-written prompts and responses. For instance, models fine-tuned on instruction-based data can answer questions, generate summaries, or perform data-specific analysis based on human-like commands. 
Some authors \cite{10583947} utilized an instruction-fine-tuned variant of the LLaMA 2 13B model for network analysis in 5G networks. The data for fine-tuning were BGP routing tables, packet capture files, and performance metrics. In task-specific fine-tuning, the model is trained on input-output examples directly related to the task. For instance, a sentiment analysis model might be fine-tuned on datasets with labeled positive and negative statements, enhancing its accuracy in classifying sentiment \cite{ding2024dynamic}. In domain-specific fine-tuning,  the model is fine-tuned on data from a specific field, like medicine, finance, or law. For instance, a medical LLM like ChatDoctor \cite{li2023chatdoctor} is fine-tuned on patient-doctor conversations and medical literature to enhance its responses to medical queries. To enhance the acquisition and understanding
of intents prompted by users while adapting to their
profile and way of expression, domain-specific fine-tuning can be employed for domain-specific knowledge, injected
by suitable fine-tuning procedures \cite{fontana2024exploring}. Domain-specific fine-tuning can be used for specialized applications where high accuracy is essential.

\textbf{ii) Unsupervised fine-tuning:}  Unsupervised fine-tuning trains the model without explicit labels, typically using large corpora related to the desired domain or task. The model is allowed to continue learning language patterns and context from new domain-specific corpora without explicit input-output labels. Examples of unsupervised fine-tuning methods are continual learning \cite{jie2022alleviating} and self-supervised fine-tuning \cite{schaldenbrand2024cofrida}. Continual learning involves exploding the model to new unlabled data within the domain to help it adapt to evolving language and trends. For instance, an LLM could be fine-tuned on recent news articles to keep its knowledge up to date without labeled input-output pairs. Although unsupervised, this helps the model remain relevant for applications that require a current understanding. Self-supervised fine-tuning leverages patterns within the data itself. For example, the model might predict masked tokens in a domain-specific corpus, enhancing its contextual knowledge. This method is valuable for continual improvement without requiring task-specific labels.

\textbf{iii) Zero-shot and few-shot fine-tuning:} These fine-tuning methods aim to make models effective in scenarios where little to no data is available. Zero-shot fine-tuning, also known as zero-shot adaptation, is technically not fine-tuning. It involves training the model during pre-training to handle various tasks so that it can perform new tasks without any fine-tuning on that specific task. Zero-shot capabilities are often achieved through extensive pre-training on diverse tasks and prompts. In few-shot fine-tuning, the model is given only a small number of examples to learn from. This is valuable when labeled data is scarce or costly, as in highly specialized domains like legal and medical fields. The model learns to generalize based on limited examples, which helps achieve decent performance with minimal data. The work in \cite{mekrache2024llm}  leveraged an open-source LLM, trained using few-shot examples from a Knowledge Base (KB), which include
user intents and their corresponding network slice descriptors for intent-driven service configuration
in next-generation networks. In summary, zero-shot leverages broad pre-training and few-shot refines performance with minimal labeled data, making them ideal for specialized and resource-limited scenarios.

\textbf{iv) Parameter-efficient fine-tuning:} Parameter-efficient fine-tuning techniques allow fine-tuning with fewer resources, which is beneficial for large models that would otherwise be resource-intensive \cite{ding2023parameter}. Examples are Low-Rank Adaptation (LoRA) \cite{hu2021lora} and prompt tuning \cite{peng2024model}. LoRA adds a small number of trainable parameters to the existing model weights, allowing for effective fine-tuning without updating all model parameters. This reduces the memory and computation needed for fine-tuning and is particularly useful in distributed setups. NetGPT, an AI-native network architecture for provisioning beyond personalized generative services, leveraged LoRA to achieve parameter-efficient fine-tuning on consumer-level hardware \cite{10466747}. Rather than updating model parameters, specialized prompts are optimized to guide the model's responses in a specific direction in prompt-tuning. These prompts act as instructions within the model's input, allowing it to generate outputs relevant to the desired domain or task without adjusting its core parameters.  

\textbf{v) RL-based fine-tuning:} This approach uses feedback mechanisms to refine the model based on the quality of its outputs. Examples of RL-based fine-tuning are RL from Human Feedback (RLHF) and reward modeling. RLHF fine-tunes the model based on human-provided feedback on its responses. It is widely used in applications where human-like interactions and ethical considerations are important, such as chatbots and support systems \cite{bill2023fine}. The model parameters are adjusted to optimize for reward signals derived from human feedback, resulting in higher alignment with user expectations. In reward modeling, a model is fine-tuned to maximize certain desired outcomes, defined by a reward function. For example, in content moderation, the reward function may penalize inappropriate content. This allows the model to generate responses that align with predefined standards of appropriateness or accuracy. 

Table \ref{tab:pretraining_finetuning} summarizes various LLMs, reporting their base models, pre-training and fine-tuning datasets, and hardware. Note that [$\ast$] indicates the values are estimated considering their base models.

\subsection{Summary}
In summary, LLM pre-training and fine-tuning are fundamental processes that endow LLMs with robust, generalizable language understanding and then adapt them to specialized applications. However, it is paramount to consider the potential challenges that may arise in pre-training and fine-tuning LLMs, especially domain-specific ones, customizing them for communication networks. Communication networks require data from various sources, such as network logs, packet data, and specific domain data (Waymo open dataset for car-following prediction in vehicular networks \cite{chen2024genfollower}, WikiTableQnA and TabFac for prompting in generative IoT \cite{xiao2024efficient}, Andromeda dataset for automated repairing of Ansible script in cloud-edge infrastructures \cite{kwon2023exploring}, ImageNet1K for characterizing disparity between edge models and high-accuracy based models for vision tasks \cite{wang2024characterizing}, and resource configuration intents for zero-touch network configuration management in mobile networks \cite{lira2024largelanguagemodelszero}). These data may differ greatly in structure, content, and quality, and can be sparse, noisy, and heavily domain-specific, limiting the LLM's generalizable learning. Implementing data pre-processing techniques to clean and augment network data can reduce noise and improve the model's output quality \cite{gu2021domain}.

Pre-training LLMs on communication network data requires substantial computational resources, particularly when aiming to capture the unique dynamics of the specific network. Considering the need for real-time or near real-time processing in these networks, the scalability of training LLMs poses practical deployment and financial constraints. Methods like model parallelism, tensor parallelism, and ZeRO can be employed to enhance efficiency, which is highly essential for resource-limited edge and fog networks. For instance, tensor parallelism can serve as a basis for collaborative LLM inference in resource-constrained devices \cite{li2024collm}.

Communication networks are constantly changing with varying traffic loads, device mobility, and network configurations. Fine-tuning communication network-specific LLMs to respond to these variations requires frequent updates, which can be resource-intensive and disrupt model performance. Utilizing continual learning techniques to fine-tune the model periodically as new data arrives may help the model adapt to dynamic network environments without requiring full retraining. Besides, many network applications require real-time responses, especially in mobile and vehicular networks. Fine-tuning LLMs to meet these latency requirements is challenging, as complex models may be computationally intensive and slow to respond. Parameter-efficient techniques like LoRA and prompt tuning allow models to be fine-tuned without updating all parameters. These techniques are valuable for edge networks with limited processing power, where LoRA reduces the model’s memory and computational demands, enabling efficient deployment.

\section{Proposed Taxonomy of LLMs for NSM}\label{sec:taxonomy}
The communication network ecosystem has evolved over the years, with new and improved applications and services emerging at every phase of the network evolution process. It is envisioned that communication networks will realize tremendous revolution and mature technological advancements every ten years. A notable instance is the maturity of 5G mobile network technology that supports autonomous driving in vehicular networks, which was impossible in the predecessor Fourth Generation (4G) mobile network technology era \cite{HAKAK2023100551}. To complement critical latency issues in cloud computing technology, edge computing and fog computing technologies have been birthed in the past years. All these developments are a testament to the fact that communication networks have evolved and will continue to improve in the future. With this revolution comes cutting-edge techniques for managing such powerful networks and their services. NSM has improved proportionally to the advancements in the rapid growth of communication networks. From traditional optimization techniques such as Particle Swarm Optimization (PSO) \cite{6264109} to AI/ML methods such as DRL \cite{MEKRACHE2022100398} and deep learning-based \cite{9165550} algorithms, NSM has been successful in managing complex, dynamic infrastructures and applications. 

\begin{figure*}
\centerline{\includegraphics[width=6.5in]{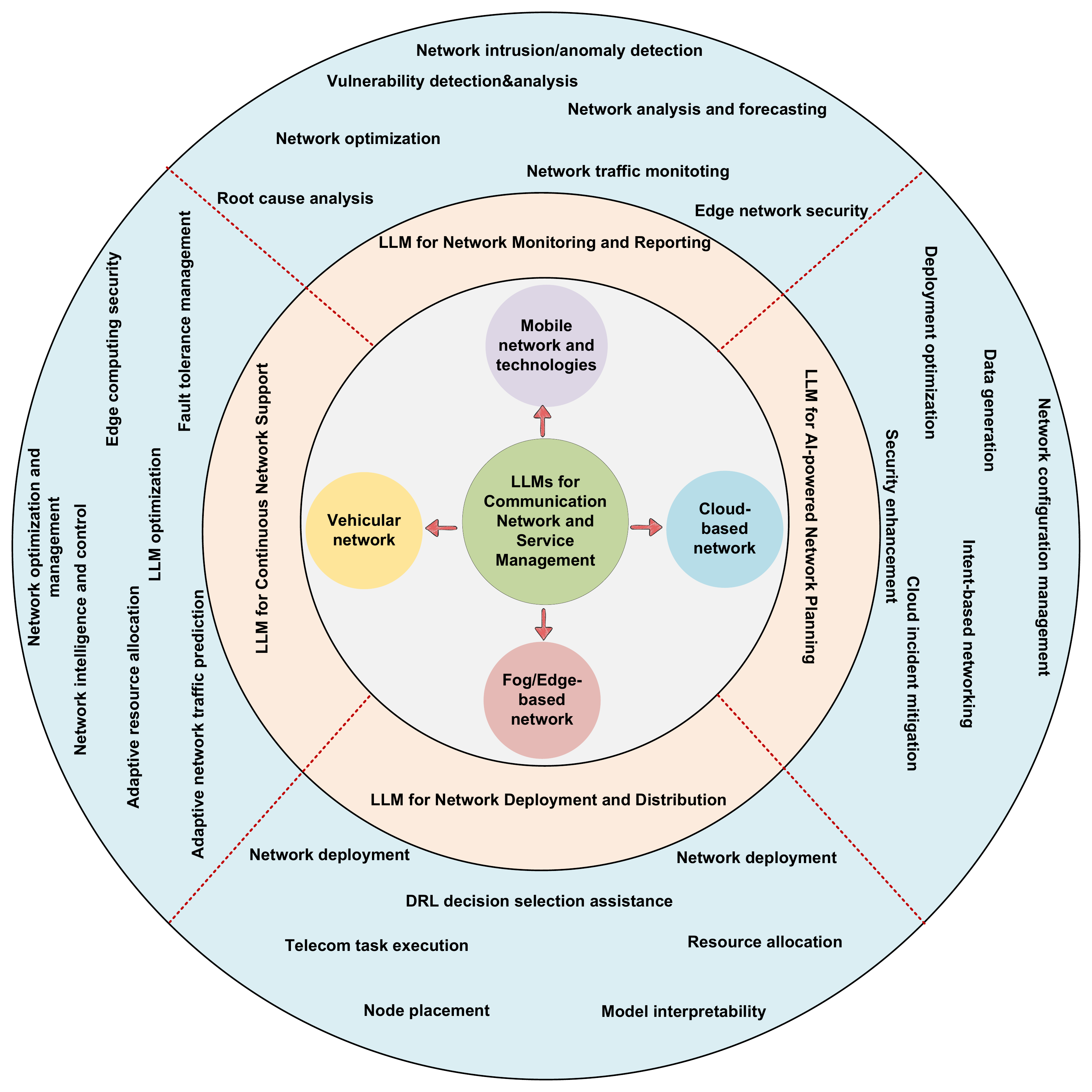}}
\caption{Proposed taxonomy of LLMs for communication NSM.}
\label{tax}
\end{figure*}
Recently, LLMs have gained root in their application for NSM in various communication networks. LLMs present a unique transformative approach to NSM, offering significant capabilities in automation, adaptability, and scalability. For instance, Huang \textit{et al.} \cite{10614634} posited that LLMs can be effectively adapted for networking applications, facilitating the translation of natural language into network-specific language through techniques like parameter-efficient fine-tuning and prompt engineering. The proposed ChatNet framework exemplifies this approach by integrating LLMs with external network tools, thereby enhancing efficiency in network planning and addressing the challenges of generalization and integration in existing AI models. Communication network engineering in enterprise settings is characterized by its complexity and susceptibility to errors due to the manual nature of tasks such as designing network topologies and configuring devices. In \cite{ifland2024genetmultimodalllmbasedcopilot}, a multimodal co-pilot utilizing LLM was presented to enhance network design workflows by integrating visual and textual data to interpret and update network configurations according to user intents.

While expanding the capabilities of LLMs in networking, their full-scale introduction into communication networks will transcend complex task features and designs. This section formulates a novel taxonomy that is used to categorize the application of LLMs for communication NSM. We first fragment the categorization considering the diverse networks in the broad scope of communication networks as mobile networks and IoT technologies, vehicular networks, cloud-based networks, and fog/edge-based networks. Under this main umbrella, the proposed taxonomy discusses LLM-based NSM and related works for each of the different networks in the following aspects: (i) \textit{network monitoring and reporting}, (ii) \textit{AI-powered network planning}, (iii) \textit{network deployment and distribution}, and (iv) \textit{continuous network support}, as shown in Fig. \ref{tax}.

\subsection{LLM for Network Monitoring and Reporting}
LLMs can play a critical role in network monitoring and reporting by automating and enhancing the detection, analysis, and reporting of network conditions. In communication networks, LLMs can support a variety of monitoring and reporting tasks, such as network intrusion detection, network forecasting, traffic monitoring, and root cause analysis. Network intrusion detection tasks include detecting threats and anomalies and ensuring network security and reliability. For instance, pre-trained general-purpose LLMs can be leveraged to create robust vector embeddings, which capture the semantic content of log messages while adapting to changes in log formats due to software evolutions and mitigating the cold-start problem in cloud environments \cite{ott2021robust}. LLMs can enhance the capacity for real-time and predictive network analysis that will offer network operators valuable insights into usage trends and potential future network demands. LLMs can be used to effectively predict VNF resource requirements by realizing real-world consumption data \cite{10588943}.

Scene analysis and traffic monitoring examine complex network conditions to identify meaningful patterns or configurations. LLMs can be utilized to accurately detect hazards during the transition to automated traffic by designing digital maps for road conditions for improved routing \cite{10588835}. When network issues occur, identifying the root cause can be challenging, especially in complex network environments. LLMs can enhance root cause analysis by identifying the underlying issues that caused the network disruptions \cite{goel2024x}. For instance, in fog and edge-based networks, LLM could identify if a bottleneck originates at a specific edge node rather than at the cloud, and this could reduce the time spent on troubleshooting.


In summary, traditional optimization and ML techniques face limitations in handling unstructured data, such as log messages, which are critical for detecting anomalies and understanding evolving network conditions. These methods often require extensive task-specific feature engineering, struggle with adapting to dynamic changes, and lack the ability to generalize across diverse scenarios. LLMs address these challenges with their ability to process unstructured data seamlessly, leveraging pre-trained knowledge and context awareness to deliver real-time insights, predictive analysis, and more efficient troubleshooting, making them indispensable for critical network monitoring and reporting tasks.

\subsection{LLM for AI-powered Network Planning}
AI-powered network planning can leverage machine intelligence to automate and optimize various aspects of network configuration, resource allocation, and incident management. By integrating LLMs into the network planning process, network operators can address service demands more accurately and ensure optimal network management and orchestration. LLM-enabled network planning tasks include Intent-Based Networking (IBN), network configuration management, agent route planning, LLM deployment optimization, and data generation. IBN enables network administrators to specify desired outcomes through high-level intents and language translations. LLMs can be utilized to create a pipeline that progressively breaks down intents into a sequence of policies, facilitating automated execution through a closed control loop that integrates monitoring, analysis, planning, and execution \cite{dzeparoska2023llm}. Automated network configuration management analyzes existing network setups, applies necessary adjustments, and generates optimized configurations. LLMs can automatically generate, verify, and implement network configurations based on high-level intents articulated in natural language \cite{lira2024largelanguagemodelszero}. Agent route planning, especially in vehicular networks, involves determining optimal data paths to ensure efficient communication across the network. LLMs can be utilized to enhance situational awareness, decision-making, and trajectory planning through multimodal data synthesis from various sources, including vehicle cameras and infrastructure sensors \cite{you2024v2xvlmendtoendv2xcooperative}. 

Cloud incident mitigation allows for the detection of patterns that typically precede incidents, such as unexpected spikes in resource usage or application errors. This assists network operators to take corrective action before the service is disrupted. LLMs can be used to assist engineers with incident resolution through a comprehensive analysis of incidents utilizing bot semantics and human evaluations \cite{ahmed2023recommending}. For deployment optimization, the most suitable locations and resources can be earmarked for deployment across network nodes. LLMs can identify which edge nodes are best suited to host computationally intensive tasks, minimizing latency by processing data closer to the source. LLM inference services can be shifted to edge devices to ensure real-time responses and enhanced privacy, while addressing the critical challenges of insufficient memory and computational resources \cite{dhar2024empirical}.

In summary, traditional and ML techniques for AI-powered network planning tasks often struggle with interpreting high-level intents, managing dynamic configurations, and synthesizing multimodal data, limiting their adaptability and effectiveness. These approaches typically require extensive manual effort for task-specific modeling and lack the ability to seamlessly integrate real-time insights across complex scenarios. LLMs address these shortfalls by leveraging natural language processing to translate high-level intents into actionable policies, automating network configuration, optimizing deployment strategies, and enhancing situational awareness through multimodal data synthesis, making them essential for efficient network planning.


\subsection{LLM for Network Deployment and Distribution}
Network deployment and distribution involves setting up and distributing network resources, services, and applications across the infrastructure to achieve optimal performance, accessibility, and scalability. With the increased complexity of modern communication networks, effective deployment and distribution are critical for maintaining robust, scalable, and responsive networks. LLMs can facilitate and optimize these tasks by leveraging their enormous data processing and predictive capabilities to streamline deployment, enhance service placement, and manage resources dynamically. 
For network deployment, LLMs can be combined with RL to optimize wireless network deployment in urban environments \cite{sevim2024largelanguagemodelsllms}. Telecom networks require frequent and precise task execution, including updates, patches, and routine maintenance activities. LLMs can be integrated with telecom network components for data retrieval, planning, and evaluation tasks \cite{10638533}.  

Service placement involves determining the optimal locations within the network for deploying services or applications to minimize latency, enhance performance, and maximize resource efficiency. 
Effective resource management is crucial to maintaining balanced network loads, particularly in environments like cloud, fog, and edge-based networks. To avoid SLA violations and inefficient resource utilization, LLMs can intelligently allocate queries and scale computing resources based on predicted requirements and latency models \cite{nie2024aladdinjointplacementscaling}.

In summary, traditional and ML techniques often face limitations in addressing the complexity and dynamic nature of modern network deployment and distribution tasks, such as suboptimal resource allocation and static service placement. These methods struggle with adapting to real-time changes in network conditions and require extensive manual intervention for precise configuration and scaling. LLMs overcome these challenges by leveraging their predictive and data processing capabilities to optimize resource management, enhance service placement, and dynamically configure network components, ensuring efficient, scalable, and responsive deployments in complex communication networks.


\subsection{LLM for Continuous Network Support}
Continuous network support refers to the ongoing maintenance and improvement of network infrastructure to ensure stable, optimized, and uninterrupted service delivery. It involves activities aimed at adapting to real-time network conditions, identifying and resolving faults, and optimizing the overall network over time. LLMs can enhance continuous network support by automating real-time adjustments and proactively managing faults, thus creating more resilient and adaptive networks. Continuous network support tasks that can be performed by LLMs include network optimization, network intelligence and control, real-time monitoring and adjustments, IoT honeypot, code repair, and fault tolerance management. Network optimization involves adjusting network parameters to achieve optimal performance, often based on varying conditions and demand. LLMs can enable adaptive network optimization by analyzing data in real-time and suggesting adjustments to bandwidth allocation and resource distribution. For instance, LLMs can be integrated into current management and orchestration frameworks to enable the translation of user intents into actionable technical requirements, efficient mapping of network functions to infrastructure, and comprehensive lifecycle management of network slices \cite{dandoush2024largelanguagemodelsmeet}. Real-time monitoring and adjustments are essential for detecting and responding to fluctuations in network conditions. LLMs can facilitate real-time adjustments by analyzing network metrics continuously and executing changes to prevent potential issues. LLMs can be integrated as decision-making components to enable nuanced reasoning that mimics human commonsense understanding in complex driving situations \cite{Sha2023LanguageMPCLL}.

An IoT honeypot is a specialized security mechanism designed to attract and analyze malicious activity targeting IoT devices. LLMs can enhance IoT honeypots by identifying evolving patterns in attacks and adapting the honeypot environment to capture richer data on malicious behaviors. For example, fine-tuned LLMs can be utilized on dataset of attacker-generated commands to create interactive honeypots capable of sophisticated engagement \cite{Otal_2024}. Fault tolerance management ensures that networks can withstand and recover from unexpected failures without significant downtime. LLMs can contribute to fault tolerance by identifying fault patterns, predicting potential failures, and implementing preemptive strategies to minimize service interruptions.

In summary, traditional and ML techniques often fall short in continuous network support tasks due to their limited adaptability to real-time conditions, reliance on predefined rules, and inability to process complex patterns in evolving network scenarios. These approaches struggle with dynamic fault detection, nuanced reasoning, and proactive optimization. LLMs address these limitations by automating real-time monitoring, enabling adaptive fault tolerance, and enhancing IoT security through intelligent honeypots, making networks more resilient, responsive, and capable of delivering uninterrupted service.



\section{LLMs for Mobile Networks and Technologies-based NSM}\label{sec:mobile}
The groundbreaking advancements in mobile networks and technologies, including 5G and beyond mobile networks and IoT, have placed unprecedented demands on network performance and scalability \cite{8879484}. LLMs have increasingly been explored as a tool to enhance the performance of these mobile networks and technologies. The ability of LLMs to process vast amounts of mobile network data and make remarkable real-time inferences renders them valuable in network analysis \cite{10583947}, network configuration \cite{10575237}, anomaly detection \cite{zhang2024largelanguagemodelswireless}, network intelligence \cite{shao2024wirelessllmempoweringlargelanguage}, and overall NSM. Specifically, LLMs can be integrated as intelligent agents to analyze network conditions, allocate resources, and enable continuous network support to ensure overall network and service efficiency. 

In mobile networks, where latency, bandwidth management, and seamless connectivity are critical, LLMs offer significant advantages by learning from historical data, predicting future network behavior, and automating resource allocation based on contextual analysis and outcomes \cite{lee2024llmempoweredresourceallocationwireless}. For instance, in 5G and beyond mobile networks, LLMs can be deployed to orchestrate Network Functions (NFs), dynamically adjusting network configuration based on real-time demand, and supporting IBN, which allows network operators to define their desired outcomes and have the network intelligently configure itself to meet those requirements. Additionally, LLMs can augment traditional ML methods in network optimization and deployment decision-making, complementing their ability to improve network management decisions \cite{10588921,10592370}. Moreover, in the IoT ecosystem, LLMs play a crucial role in managing the massive influx of data generated by billions of connected devices. By providing data analytics, LLMs can facilitate well-informed decision-making in terms of network deployment and resource distribution. For example, LLMs can help optimize resource usage of IoT devices, ensuring seamless resource requests and responses.

This section explores the evolving role of LLMs in mobile networks and related technologies NSM, especially in 5G and beyond networks and IoT technologies. The focus is on the role of LLMs in \textit{network monitoring and reporting}, \textit{AI-powered network planning}, \textit{network deployment and distribution}, and \textit{continuous network support}. Through a review of existing literature and use cases, this section will provide insights into how LLMs are reshaping the mobile network and IoT landscape with respect to NSM. Fig. \ref{mobile} depicts a typical application of LLM for mobile networks and technologies. Network data such as system logs, network configuration intents, bandwidth consumption, and candidate DRL decisions are generated from the mobile network environment. Based on this data, the LLM input and its corresponding expected output for each task are designed and serve as the input to the LLM. Through data preprocessing, embedding, encoding, decoding, and post-processing, the LLM performs model pre-training and fine-tuning and provides context-aware inferences for each input task. Based on these model inferences, well-informed optimization decisions are made and applied accordingly in the mobile network environment. 
\begin{figure*}
\centerline{\includegraphics[width=7.2in]{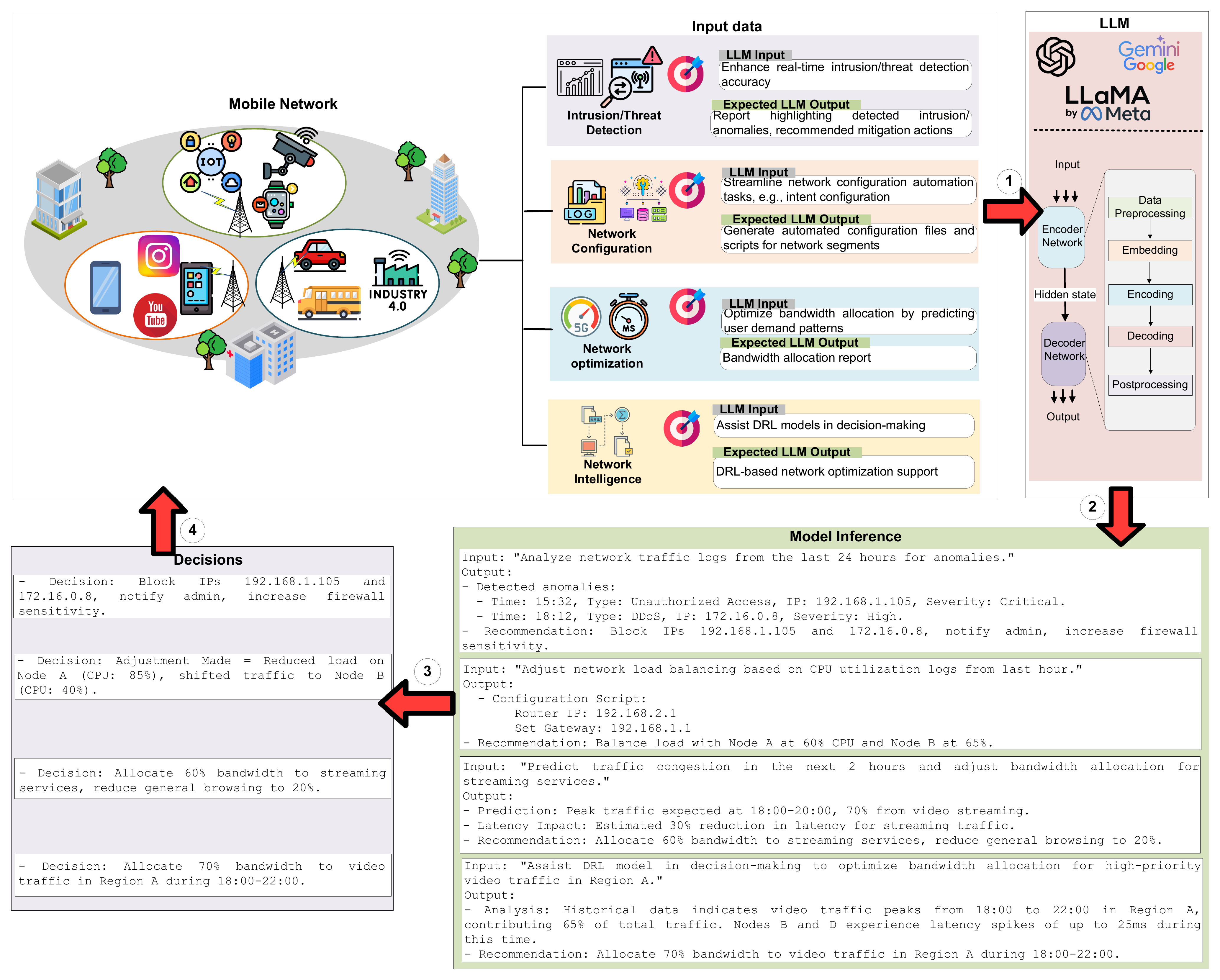}}
\caption{LLM for mobile networks and technologies-based NSM.}
\label{mobile}
\end{figure*}
\subsection{LLM for Network Monitoring and Reporting}
Network monitoring and reporting in mobile networks, as well as associated technologies, are instrumental for optimal network performance, scalability, and security. In mobile networks, monitoring involves continuous observation of network components and traffic, collecting data on key performance metrics such as packet loss, bandwidth usage, latency, and signal strength. These metrics are vital for accurate and efficient network analysis, which has the potential to raise awareness of anomalies or malicious activities in the network. The main objective of network monitoring is to detect anomalies, forecast potential performance degradation, and ensure that the network satisfies the SLAs of subscribers. On the other hand, network reporting involves generating insights from the monitoring data and conveying actionable decisions to network orchestrators, enabling them to make informed decisions on resource allocation, network re-configurations, and predictive maintenance. 

With the advent of 5G and IoT, the complexity and scale of network environments have expanded significantly. Massive connectivity of IoT devices, involving billions of sensors generate huge amounts of network traffic at any point in time. This introduces challenges in monitoring a much larger and more heterogeneous set of services (eMBB, uRLLC, and mMTC for 5G) and devices (for IoT). For example, the introduction of network slicing in 5G, where virtualized networks are created to offer specific services, requires sophisticated monitoring mechanisms to ensure each slice meets its required performance standards. Similarly, IoT devices continuously generate vast amounts of data, and managing this influx in real time becomes crucial to ensure smooth operations.

Traditional network monitoring systems rely on pre-configured rules and statistical models to monitor network performance and detect anomalies. However, these methods may struggle to cope with the complexity and scale of emerging mobile networks, especially when dealing with vast amounts of multimodal data and previously unseen network traffic. LLMs have the potential to revolutionize network monitoring and reporting by introducing more intelligent, scalable, and adaptive capabilities. Notably, LLMs can enhance network analysis by analyzing large volumes of network data in real-time and interpreting various performance metrics collected from different sources. Leveraging the underlying NLP capabilities, LLMs can extract semantic insights from logs, system reports, and periodic analytics data. This enables more accurate and timely identification of performance bottlenecks, suggesting practical ways of solving such problems.
Additionally, LLMs excel in recognizing patterns and detecting anomalies by analyzing complex datasets. For network monitoring, they can analyze historical data and learn normal patterns of network behavior. This allows them to detect deviations in real-time, such as unexpected latency, which may indicate potential security breaches.

\subsubsection{Network intrusion/threat detection}
As mobile networks and IoT systems grow more complex, they become increasingly susceptible to cyber threats. Traditional Intrusion Detection Systems (IDS) often rely on signature-based methods, making them ineffective against new, evolving threats. LLMs offer a novel approach to enhancing network security via advanced pattern recognition and inference capabilities to detect anomalies and identify potential threats in real time. The authors in \cite{zhang2024largelanguagemodelswireless} explored the use of LLMs for network intrusion detection in wireless communication networks. Fig. \ref{llmids} illustrates a pre-trained LLM-based IDS framework with four steps: feature selection, data collection, prompt building, and decision extraction. These steps streamline detection by identifying key indicators, transforming data into LLM-compatible formats, configuring task-specific prompts, and processing outputs into actionable decisions for automated, context-aware intrusion detection. Three distinct in-context learning methods that enhance LLM performance without necessitating additional training or fine-tuning were introduced. Utilizing GPT-3.5, GPT-4, and LlaMA base models, the authors evaluated the feasibility on LLMs for network intrusion detection based on a real network intrusion detection dataset. Experimental results showcased over 90\% improvement in testing accuracy and F1-score with GPT-4, achieving above 95\% on various attack types using only 10 in-context learning examples. Furthermore, the proposed framework enables fully automated network intrusion detection by effectively leveraging the capabilities of pre-trained LLMs to address security challenges within wireless communications. 

Satellite communication networks are a potential target for cyber-attacks due to the heterogeneity of their components and the multiple domain characteristics. Hassanin \textit{et al.} 
\cite{HASSANIN2025103645} proposed PLLM-CS, a pre-trained LLM to enhance cyber threat detection in Smart Satellite Networks (SSNs). They incorporated a specialized module that transforms network data into contextually suitable SSN environment inputs, allowing the model to effectively encode relevant contextual information from cyber data. Due to the unavailability of real-world satellite data, IoT-based traffic datasets such as UNSW\_NB 15 and TON\_IoT were used to train the model. The proposed PLLM-CS method achieved significant benchmark performance by attaining 100\% accuracy on the UNSW\_NB 15 dataset while outperforming conventional ML techniques like BiLSTM, GRU, and CNN. 
LLMs have also been extended to detect wireless symbols. In \cite{abbas2024leveraginglargelanguagemodels}, LLM was employed for wireless symbol demodulation tasks via in-context learning (prompting) without any training or fine-tuning. Results indicated that LLMs not only surpass traditional Deep Neural Network (DNN) approaches but also produce highly confident predictions when combined with advanced calibration techniques. 
\begin{figure}[!t]
    \centering
    \includegraphics[width=3.5in]{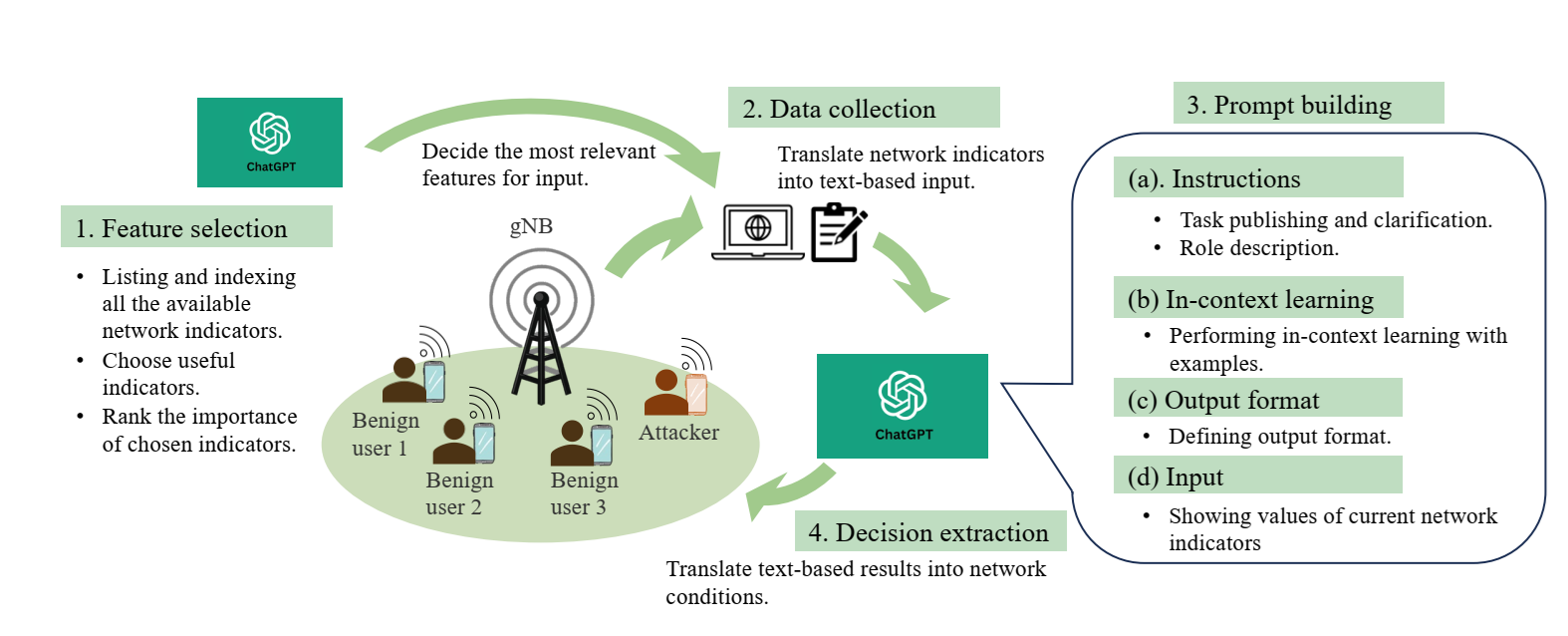}
    \caption{Illustration of LLM-empowered IDS \cite{zhang2024largelanguagemodelswireless}.}
    \label{llmids}
\end{figure}
\subsubsection{Network analysis and forecasting}
By leveraging the advanced capabilities of LLMs, network analysis and forecasting can become more intelligent and proactive. This will result in adaptable and more resilient mobile networks that can support the fluctuating demands of modern services and applications. Network analysis entails examining mobile network performance to understand the behaviors and efficiency of network infrastructure and services. Kan \textit{et al.} \cite{10583947} introduced \textit{Mobile-LLaMA}, an instruction-fine-tuned open-source LLM based on LLaMA 2 13B, specifically designed for network analysis via the Network Data Analytics Function (NWDAF) in 5G networks. Using real-world 5G datasets, \textit{Mobile-LLaMA} excelled in tasks such as packet analysis, IP routing analysis, and performance analysis, confirming its contribution to the automation and AI-driven management of 5G networks.

Forecasting extends network analysis by predicting future network behavior based on trends in historical data. It enables network operators and service providers to anticipate network traffic fluctuations and resource demand, streamlining potential issues before they occur, and allowing proactive and reactive NSM. Some authors \cite{10621076} proposed TimesLM, a novel framework that leveraged lightweight LLMs for automated time series analytics, addressing the need for general-purpose solutions that can adapt to diverse applications in future AI-native networks like 6G. By enabling zero-shot and few-shot inference, TimesLM enhances decision-making, demonstrating improved accuracy and reduced latency in real-world applications. Virtual Network Function (VNF) resource forecasting is a cornerstone for optimizing network resource allocation and ensuring service continuity. Capitalizing on the advanced pattern recognition and next-token prediction capabilities of LLM, Su \textit{et al.} \cite{10588943} presented a robust alternative to conventional probability-based forecasting models for analyzing real-world VNF resource consumption data and enhancing overall network resource management.


\subsection{LLM for AI-powered Network Planning}
As mobile networks and IoT technologies continue to evolve beyond the 5G paradigm, managing and configuring these networks efficiently becomes increasingly challenging. Network planning tends to be an arduous choice to address these challenges by configuring network operations via intent generation and translation. However, traditional network planning methods struggle with adapting to dynamic environments, managing complexity, scaling effectively, processing unstructured data, and relying heavily on manual, time-consuming processes. Thanks to AI in general and ML in particular, network planning tasks can transition from partial automation to full automation, limiting the involvement of human operators and annotators. This will enhance network configuration tasks for planning, creating more efficient networks devoid of errors. At the forefront of AI-empowered network planning for mobile networks and IoT technologies is IBN, a paradigm that allows network operators to define high-level business intents, such as minimizing latency and maximizing bandwidth for specific applications without relying entirely on human network engineers. The network then autonomously configures and reconfigures itself to achieve these outcomes, dynamically adapting to changing conditions.

LLMs can be instrumental in enhancing AI-powered network planning in mobile networks and IoT technologies, especially in implementing and managing IBN and overall network planning and configurations. With their advanced intent understanding and contextual inference abilities, LLMs can interpret and translate high-level intent statements into precise network configurations and actionable policies. Therefore, LLMs can act as a conduit between human operators and the network, allowing for intuitive and efficient network management. In this subsection, we elaborate on attempts to utilize LLMs for AI-powered network planning for mobile networks and technologies NSM, with emphasis on IBN and network configuration management.

\begin{figure}[!t]
    \centering
    \includegraphics[width=3.5in]{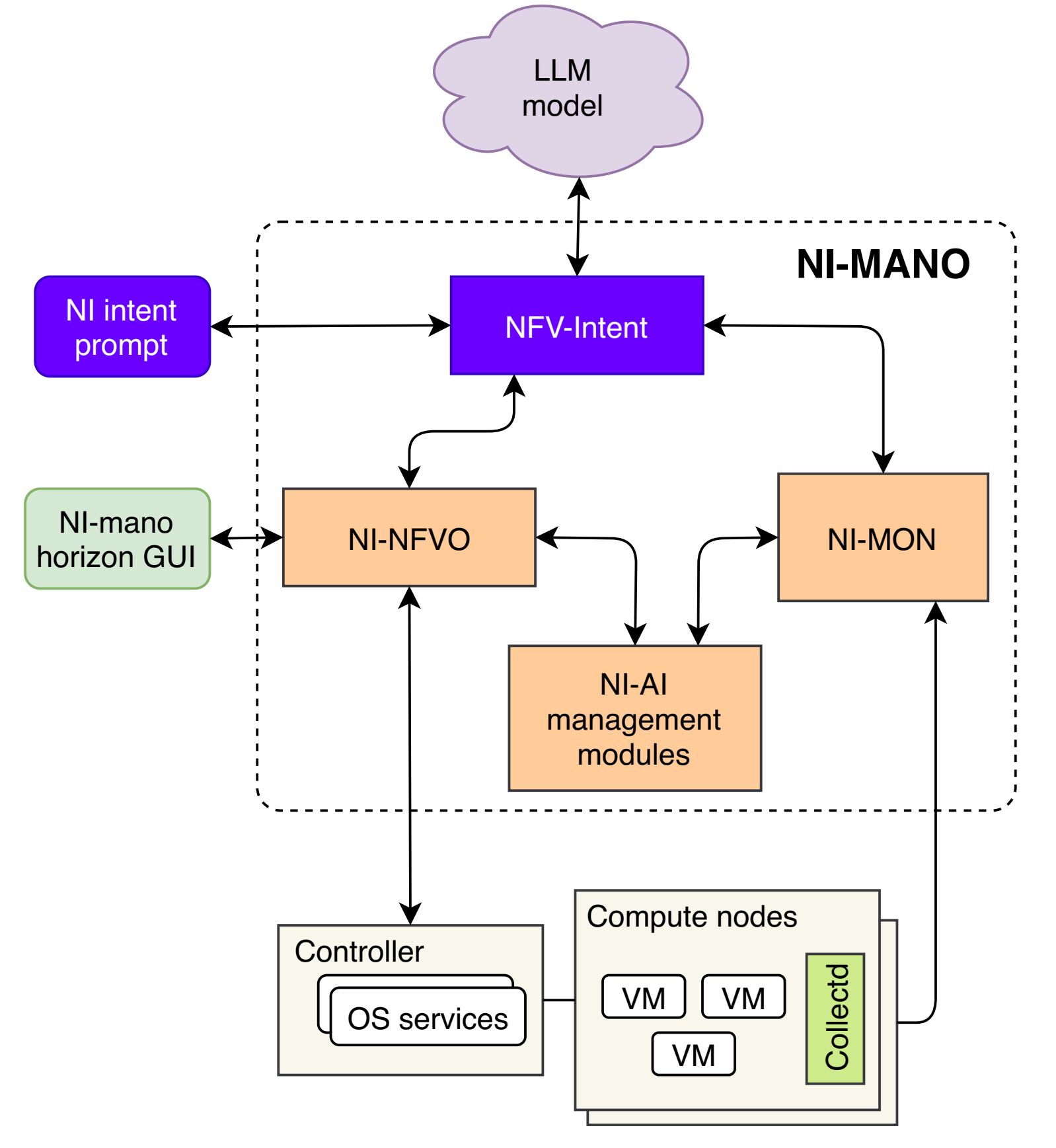}
    \caption{NI-testbed architecture with NFV-intent integrated \cite{10575237}.}
    \label{llmibn}
\end{figure}
\subsubsection{Intent-based networking}
IBN automates network management by translating high-level business objectives (intents) into precise, actionable network configurations \cite{10574890,10588879}. LLMs can enhance IBN by converting natural language intents in text to executable network policies. This approach allows the network to autonomously adjust operations in real-time, aligning with dynamic performance requirements and minimizing manual or human intervention. For instance, Fontana \textit{et al.} \cite{10588924} integrated LLMs to enhance user interaction and improve the understanding of user intents through intelligent chatbots. They emphasized a user-centric design approach, ensuring that solutions are intuitive and accessible to end-users. In \cite{10539172}, the authors discussed the creation of a custom LLM tailored for intent-based networking in 5G and beyond, emphasizing its role in automating the extraction and interpretation of user intents to streamline network operations and enhance overall network intelligence. Fuad \textit{et al.} \cite{10588879} proposed a modular framework that integrates LLMs to facilitate automated configurations, highlighting their potential to enhance IBN while maintaining data privacy and the integrity of configuration outcomes.
Additionally, significant challenges and unresolved issues that persist in LLM-based IBN were discussed. Multimodal LLMs (MLLMs) have been developed to manage 6G networks via IBN. For instance, some authors \cite{10597022} presented an intent-based management framework that translates high-level user requirements and supplementary deployment descriptors into actionable configuration files for network systems, addressing the complexities of resource orchestration across various network domains. By integrating multimodal Generative AI, particularly LLMs, the framework enhances the dynamic interpretation of user intent while ensuring compatibility with existing orchestration platforms and next-generation operating support systems.

\subsubsection{Network configuration management}
LLMs can enable continuous monitoring and adaptation of network configurations, ensuring that the network remains aligned with the defined intents. For example, LLMs can analyze real-time data to automatically configure resources and allocate bandwidth to meet performance goals. Lira \textit{et al.} \cite{lira2024largelanguagemodelszero} proposed the Network Configuration Generator (LLM-NetCFG), which leverages LLMs and Zero Touch Network \& Service Management (ZSM) agents to autonomously generate, verify, and implement network configurations based on high-level intents articulated in natural language. By architecting ZSM configuration agents through LLMs, LLM-NetCFG not only streamlines the complexity of communication networks but also enhances automation and accuracy in network management processes. In \cite{10575237}, a Network Function Virtualization (NFV)-intent system was proposed that utilizes in-context learning with LLMs to facilitate high-level natural language intent translation without the need for extensive retraining, employing a JSON template for the desired output. To demonstrate the feasibility of NFV-intent, it was implemented and integrated into the NI-testbed, which is a developed system for AI-based NFV lifecycle management, as shown in Fig. \ref{llmibn}.

\subsection{LLM for Network Deployment and Distribution}
Efficient network deployment and distribution are crucial in managing modern mobile networks, especially as 5G and IoT technologies introduce more complex architecture and heterogeneous service types. Network deployment in mobile networks implies the initial setup and configuration of network infrastructure and resources for service provisioning. It determines the optimal deployment strategy while considering infrastructure and resource limitations. On the other hand, network distribution involves the continuous reconfiguration of network infrastructure and resources to automate the distribution of services where they are needed. Conventional methods for network deployment and distribution often require significant manual inputs, which struggle to keep up with the unpredictable nature of mobile networks.

Leveraging their capabilities in pattern analysis, audit, and decision-making, LLMs present a potential avenue for tackling the above-mentioned challenge. LLMs can automate the network deployment process, using real-time data to determine the optimal configuration of network elements such as base stations and antennas. Utilizing historical planning and deployment data, as well as perceived user traffic patterns, LLMs can effectively determine the most suitable deployment strategies. Moreover, LLMs excel in distributing and reallocating network resources following fluctuating traffic load, device connections, and service demands. After deployment, LLMs can continuously monitor the fluctuations and make intelligent decisions about where and when to scale up or scale down resources and infrastructure. For instance, in a typical 5G network, LLMs can deploy and distribute slices by distributing resources to the different slices based on their service requirements. This subsection focuses on studying existing literature on the application of LLMs for network deployment and distribution in mobile networks and technologies NSM, with emphasis on mobile network deployment, telecom tasks, and assisting DRL decision selection.

\subsubsection{Network deployment/development}
LLMs provide a scalable and automated solution for network deployment and development by leveraging real-time data to optimize network configuration, resource allocation, and telecom task execution. They can predict optimal deployment strategies, such as network slicing and dynamically allocating resources, enhancing the efficiency and feasibility of network and service management. Sevim \textit{et al.} \cite{sevim2024largelanguagemodelsllms} presented a novel framework that integrates LLMs with RL to optimize network deployment in urban environments, focusing on maximizing coverage by training an RL agent to navigate urban complexities and determine optimal network parameters. The study demonstrated that LLM-assisted models can achieve superior or comparable performance to CNN-based models, highlighting the potential of LLMs in enhancing automated decision-making in wireless systems. Deploying base stations on demand is a feasible way to ensure service continuity in mobile networks. Traditional base station siting methods rely heavily on drive testing and user feedback, which require extensive expertise in communication, networking, and optimization. To solve this challenge, Wang \textit{et al.} \cite{wang2024largelanguagemodelsbase} introduced a novel LLM-empowered base station siting optimization framework, demonstrating through empirical evaluation that these AI-driven methods can enhance the efficiency, cost-effectiveness, and reliability of network deployments while minimizing the need for manual intervention. Developing hardware prototypes for mobile wireless networking requires Hardware Description Language (HDL) code, which requires more complex computation tasks. LLMs can be used for code refactoring and validation in terms of wireless communication system development. The work in \cite{du2024powerlargelanguagemodels} investigated the role of LLMs in FPGA-based hardware development for advanced signal processing algorithms in wireless communication networks. The study explored LLM-assisted code refactoring, reuse, and validation using an open-source software-defined radio project, demonstrating significant productivity gains for researchers and developers.

\begin{figure}[!t]
    \centering
    \includegraphics[width=3.5in]{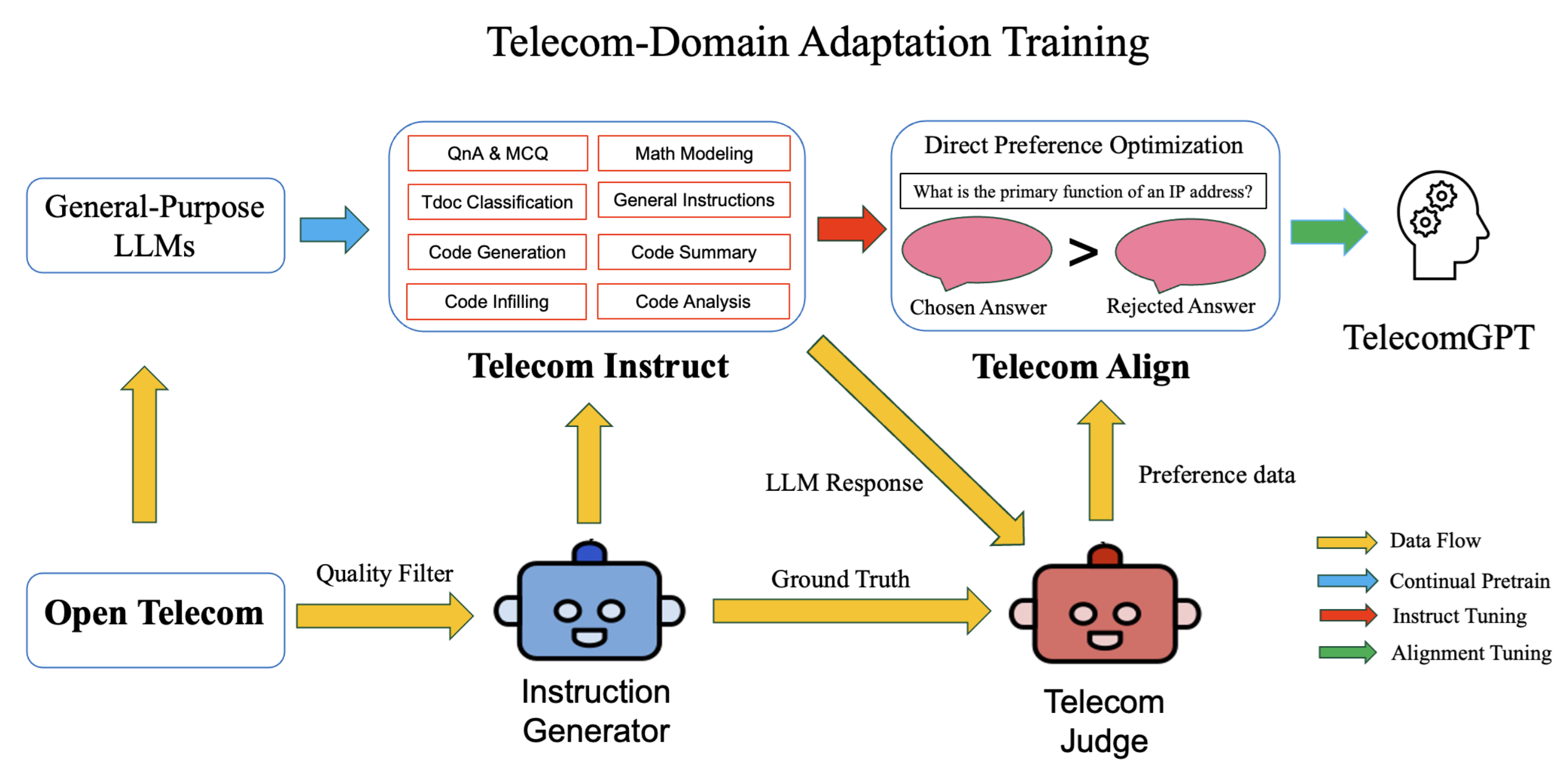}
    \caption{The training pipeline of TelecomGPT framework \cite{zou2024telecomgptframeworkbuildtelecomspecfic}.}
    \label{telecomgpt}
\end{figure}
\subsubsection{Telecom task execution}
Telecom task execution involves managing and automating various operational processes within telecommunication networks. LLMs offer a potential solution for executing telecom tasks by automation through intelligent analysis and decision-making. By leveraging their natural language understanding capabilities, LLMs facilitate more intuitive operator instructions, automating complex workflows and enhancing network responsiveness. Zou \textit{et al.} \cite{zou2024telecomgptframeworkbuildtelecomspecfic} introduced TelecomGPT, a novel pipeline for adapting general-purpose LLMs into telecom-specific models by creating tailor-made datasets for continual pre-training, instruction tuning, and alignment tuning, as illustrated in Fig. \ref{telecomgpt}. Based on new evaluation benchmarks, TelecomGPT demonstrated superior performance compared to state-of-the-art models like GPT-4 and LlaMA-3 in relevant telecom tasks. Some authors \cite{maatouk2024telellmsseriesspecializedlarge} addressed the limitations of general-purpose LLMs that struggle with domain-specific terminologies and mathematical representations by introducing Tele-LLMs. Tele-LLMs are a series of open-source models ranging from 1B to 8B parameters. A comprehensive dataset called Tele-Eval was curated from relevant telecommunications sources for specialized question-answering in the telecommunications field. Evaluation results proved that the Tele-Eval benchmark achieved better performance on domain-specific tasks while mitigating the issue of catastrophic forgetting. LLMs have been employed to provide faster access to documents on telecom standards. The work in \cite{10624786} explored the application of LLMs as Question-Answering assistants to enhance access to the Third Generation Partnership Project (3GPP) standards. They introduced a new model named TeleRoBERTa, which achieved competitive results with significantly fewer parameters in terms of troubleshooting and network operations. In \cite{10638533}, CommLLM, a multi-agent system that integrates LLMs with retrieval, planning, and evaluation functionalities, was proposed to enable more effective communication-related task resolution and expand the knowledge base in the context of 6G communications.

\subsubsection{DRL decision selection assistance}
DRL methods have been proposed in the literature to optimize decision-making in mobile network environments, including network slicing and resource optimization. However, DRL models often require significant computational resources and extensive training to adapt to changing scenarios efficiently. LLMs can assist in the DRL decision-selection process by providing additional contextual insights and guiding policy updates based on real-time analysis. By integrating LLMs, DRL models gain a broader understanding of the mobile network environment, enabling them to make more informed and adaptive decisions. Aneur \textit{et al.} \cite{10588921} proposed a Composable eXplainable RL (XRL) framework that utilizes LLMs and prompt engineering to clarify the decision-making process of DRL algorithms in 6G network slicing optimization, making them more interpretable and accessible to a broader audience, including stakeholders and engineers. Experimental results confirmed the efficacy of the proposed framework compared to traditional methods, highlighting its practical effectiveness for real-world 6G applications. Some authors \cite{10592370} introduced a Mixture of Experts (MoE) framework augmented with LLMs to effectively analyze user objectives and constraints, select specialized DRL experts, and weigh each decision from the participating experts for intelligent networks. The advanced reasoning capability of LLM was utilized to manage expert model selection for joint decision-making. The effectiveness of this LLM-enabled MoE approach was validated through applications in both general maze navigation tasks and specific utility maximization scenarios within network service providers, showcasing its practical applicability for complex optimization problems in networking systems.

\begin{table*}[!t]
\scriptsize 
\centering
\caption{Summary of Related Works on LLM for Mobile Networks and Technologies-based NSM}
\begin{tabular}{|p{1.3cm}|p{1.8cm}|p{1.5cm}|p{3.2cm}|p{2.3cm}|p{1.4cm}|p{1.4cm}|p{1.4cm}|}
\hline
\textbf{Taxonomy} & \textbf{NSM Task} & \textbf{Authors} & \textbf{Main Contributions} & \textbf{Problem Addressed} & \textbf{LLM Solution} & \textbf{Dataset Used} & \textbf{Evaluation Metrics} \\
\hline
\multirow{5}{*}{\parbox{1.5cm}{Network Monitoring and Reporting}} & \multirow{3}{*}{\parbox{1.8cm}{Network intrusion/ threat detection}} & Zhang \textit{et al.} (2024) \cite{zhang2024largelanguagemodelswireless} & Pipeline for automated LLM-based network security & Intrusion detection with LLM & GPT-3.5, GPT-4, LLaMA & Real-world intrusion detection dataset & Acc., F1-score \\ \cline{3-8}
& & Hassanin \textit{et al.}. (2025) \cite{HASSANIN2025103645} & Attention-based IDS with PLLM-CS for SSNs & Cyber threat detection in SSNs & - & UNSW-NB15, TON\_IoT & Acc., Prec., Recall, AUC \\ \cline{3-8}
& & Abbas \textit{et al.} (2024) \cite{abbas2024leveraginglargelanguagemodels} & LLM-based symbol detection in wireless systems & Wireless symbol demodulation & GPT-J, LLaMA-2 (7B, 13B) & - & Acc. \\ 
\cline{2-8}
& \multirow{2}{*}{\parbox{1.8cm}{Network analysis and forecasting}} & Kan \textit{et al.} (2024) \cite{10583947} & Mobile-aware LLM for NWDAF analysis & Network analysis (NWDAF) & Fine-tuned LLaMA-2 (13B) & Real-world 5G datasets & Score (5G QoS, UE traffic volume) \\ \cline{3-8}
& & Samaras \textit{et al.} (2024) \cite{10621076} & AutoML for intelligence and prediction & Forecasting and optimization & Fine-tuned Gemma & Dolly, CDN, 5G traffic data & MAPE, Acc., Latency \\ 
\hline
\multirow{5}{*}{\parbox{1.5cm}{AI-powered Network Planning}} & \multirow{3}{*}{\parbox{1.8cm}{Intent-Based Networking (IBN)}} & Fontana \textit{et al.} (2024) \cite{10588924} & Uses LLMs for intent recognition and translation, adapting to user profiles and expressions through fine-tuning & LLM for intent acquisition and translation & - & - & Acc., prec., recall \\ \cline{3-8}
& & Fuad \textit{et al.} (2024) \cite{10588879} & Translates network intents into configurations while addressing privacy and hallucinations & LLM for IBN & GPT-4, GPT-3.5, LLaMA2, Mistral & - & Prompt engineering, few-shot learning \\ \cline{3-8}
& & Brodimas \textit{et al.} (2024) \cite{10597022} & Framework using AI and ML tools with standards-based interfaces for inclusive NaaS & LLM for IBN & GPT-3.5-turbo & - & - \\ 
\cline{2-8}
& \multirow{2}{*}{\parbox{1.8cm}{Network configuration management}} & Lira \textit{et al.} (2024) \cite{lira2024largelanguagemodelszero} & Uses local LLM for zero-touch network configuration & LLM for zero-touch config. mgmt. & - & 90 config. intents (CP, RP, ACL, TN) & Acc., time \\ \cline{3-8}
& & Tu \textit{et al.} (2024) \cite{10575237} & NFV-Intent prototype for configuring VNFs and SFCs via natural language intents & LLM for NFV configuration & GPT-3.5-turbo & - & - \\ 
\hline
\multirow{9}{*}{\parbox{1.5cm}{Network Deployment and Distribution}} & \multirow{3}{*}{\parbox{1.8cm}{Network deployment}} & Sevim \textit{et al.} (2024) \cite{sevim2024largelanguagemodelsllms} & Integrates LLMs and RL for wireless network deployment & LLM-assisted RL for network deployment & DistilBERT & Textual & Reward \\ \cline{3-8}
& & Wang \textit{et al.} (2024) \cite{wang2024largelanguagemodelsbase} & Proposes LLM-based base station siting framework with four implementations & LLM for base station deployment & ChatGPT 4o & Coordinates, traffic data & Success rate, base station planning \\ \cline{3-8}
& & Du \textit{et al.} (2023) \cite{du2024powerlargelanguagemodels} & Explores LLMs in generating HDL code for FPGA-based wireless development & LLMs for FPGA-based development & GPT-3 & - & Time, quality score \\ 
\cline{2-8}
& \multirow{4}{*}{\parbox{1.8cm}{Telecom task execution}} & Zou \textit{et al.} (2024) \cite{zou2024telecomgptframeworkbuildtelecomspecfic} & Adapts general LLMs to telecom-specific tasks through specialized tuning & LLM for Telecom-specific tasks & Llama2-7B, Mistral-7B, Llama3-8B & TeleQnA, non-contextual QA & Loss, CDF \\ \cline{3-8}
& & Maatouk \textit{et al.} (2024) \cite{maatouk2024telellmsseriesspecializedlarge} & Tailors LLMs for telecoms domain with Gemma-2B, LLaMA-3-8B & LLM for Telecoms & Gemma-2B, LLaMA-3-8B & Tele-Data & Perplexity, SemScore \\ \cline{3-8}
& & Karapantelakis \textit{et al.} (2024) \cite{10624786} & Evaluates LLMs for QA on 3GPP standards & LLMs for Telecom standards & TeleRoBERTa & TeleQuAD & BERTScore, GPT-4 Ref (Acc.) \\ \cline{3-8}
& & Jiang \textit{et al.} (2024) \cite{10638533} & Develops CommLLM for communication tasks using domain knowledge & LLM for communication tasks & GPT-3.5 & 6G-related research papers & Score, cosine similarity \\ 
\cline{2-8}
& \multirow{2}{*}{\parbox{1.8cm}{DRL decision selection assistance}} & Ameur \textit{et al.} (2024) \cite{10588921} & Uses LLMs and Prompt Engineering to offer personalized DRL decision explanations for diverse users & LLM-assisted DRL decision explanation for network slicing & GPT-3.5 Turbo & Runtime dataset from DRL model & Sensitivity, stability, comprehensibility \\ \cline{3-8}
& & Du \textit{et al.} (2024) \cite{10592370} & Combines LLMs with MoE to guide multiple DRL models in network optimization & LLM-assisted DRL for expert selection in network optimization & GPT-3.5 Turbo-1106 & - & Success rate, reward \\  
\hline
\multirow{5}{*}{\parbox{1.5cm}{Continuous Network Support}} & \multirow{3}{*}{\parbox{1.8cm}{Network optimization and management}} & Dandoush \textit{et al.} (2024) \cite{dandoush2024largelanguagemodelsmeet} & Automates resource allocation, network configurations, and real-time analysis & LLM for network slicing MANO & - & - & - \\ \cline{3-8}
& & Tong \textit{et al.} (2024) \cite{tong2024wirelessagentlargelanguagemodel} & WirelessAgent: an LLM framework with modules for perception, memory, planning, and action & LLM for network slicing management & GPT-4o-128K & - & - \\ \cline{3-8}
& & Zhou \textit{et al.} (2024) \cite{zhou2024largelanguagemodelllmenabled} & Uses LLMs for in-context learning in wireless network optimization & LLM for power control & Llama3-8B, Llama3-70B, GPT-3.5 turbo & - & Avg. power, service quality \\ 
\cline{2-8}
& \multirow{2}{*}{\parbox{1.8cm}{Network intelligence and control}} & Shao \textit{et al.} (2024) \cite{shao2024wirelessllmempoweringlargelanguage} & WirelessLLM: adapts LLMs for expertise in wireless systems & LLM for wireless intelligence & WirelessLLM & TeleQnA dataset & - \\ \cline{3-8}
& & Rong \textit{et al.} (2024) \cite{10487933} & Explores LLMs for intelligent control in 6G TN-NTN networks & LLM for intelligent control & - & - & - \\ 
\hline
\end{tabular}
\label{tab:nsmtaxonomy}
\end{table*}

\subsection{LLM for Continuous Network Support}
Maintaining continuous support and optimizing network performance in mobile networks and IoT technologies is crucial, as it ensures that networks operate efficiently, dynamically adapting to varying demands, and delivering consistent QoS. LLMs provide a transformative approach to achieving these goals by offering network optimization and intelligent control practices. 
Network optimization requires real-time analysis to adjust network resource allocations and load balancing. For example, 5G and beyond networks can utilize network slicing to optimize multiple virtual networks, each tailored for specific use cases like eMBB, uRLLC, and mMTC. LLM plays a critical role in managing these slices, dynamically monitoring and adjusting resource allocations based on real-time service demands and performance metrics. This capability ensures that each slice meets its specific requirements, enhancing the flexibility and scalability of the network. LLMs can also adjust network parameters and implement reconfigurations to maintain service quality and prevent potential disruptions. By leveraging their predictive capabilities, LLMs enable proactive management, anticipating network traffic imbalance and making the desired adjustments before they impact user experience. Integrating LLMs into mobile and IoT networks not only improves service continuity and efficiency but also enhances the ability to scale and adapt to new demands, making these networks more robust, flexible, and responsive.

\begin{figure}[!t]
    \centering
    \includegraphics[width=4.5in]{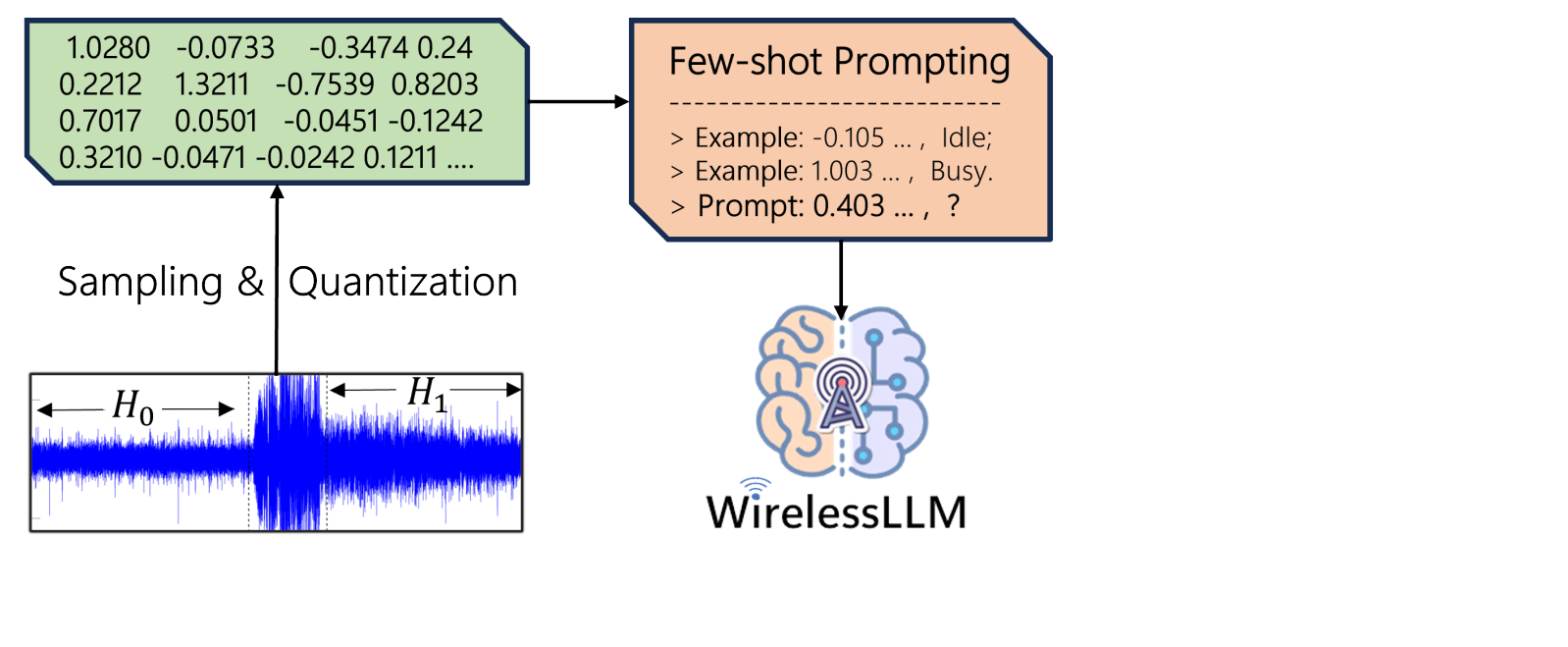}
    \caption{WirelessLLM framework for spectrum sensing with prompting \cite{shao2024wirelessllmempoweringlargelanguage}.}
    \label{wirelessllm}
\end{figure}
\subsubsection{Network optimization and management}
Network optimization and management are critical for ensuring efficiency and reliability for mobile networks, such as network slicing and power control. Network slicing enables the creation of multiple virtualized network segments, each tailored for specific applications, requiring precise management to maintain QoS. LLMs can optimize and manage these network slices by analyzing real-time performance data and dynamically adjusting allocated resources based on demand. Current network slicing orchestration and management approaches face limitations in handling the complexity of new service demands within multi-administrative domain environments. Dandoush \textit{et al.} \cite{dandoush2024largelanguagemodelsmeet} proposed a novel framework that integrates LLMs and multi-agent systems into current network slicing management and orchestration frameworks to enable the translation of user intent into actionable technical requirements, efficient mapping of network functions, and comprehensive lifecycle management of network slices, while addressing the challenges of implementation and collaboration across domains. In \cite{tong2024wirelessagentlargelanguagemodel}, a novel approach named WirelessAgent that leverages LLMs to develop AI agents capable of network slicing management was proposed. WirelessAgent utilizes LLMs to manage intricate tasks within these networks, enhancing performance through sophisticated reasoning and autonomous decision-making. Experimental results showed that WirelessAgent is capable of accurately understanding user intent, effectively allocating slice resources, and consistently maintaining optimal performance.

LLMs can also intelligently control power levels across the network, ensuring optimal energy use while maintaining connectivity and performance. Zhou \textit{et al.} \cite{zhou2024largelanguagemodelllmenabled} proposed an in-context learning algorithm that leverages the inference capabilities of LLMs for base station power control. By formulating tasks in natural language and employing tailored examples of both discrete and continuous-state problems, the proposed approach achieved performance on par with conventional DRL techniques, highlighting its efficiency and potential for enhancing future wireless network management.

\subsubsection{Network intelligence and control}
Network intelligence involves monitoring, analyzing, and optimizing network operations to ensure seamless and continuous connectivity and QoS. LLMs provide an avenue for enhancing network intelligence by processing vast amounts of network data, detecting patterns, and identifying anomalies in real time. To effectively address the unique challenges emanating from current telecom language understanding deficiencies, Shao \textit{et al.} \cite{shao2024wirelessllmempoweringlargelanguage} introduced a framework called WirelessLLM. By establishing foundational principles such as knowledge alignment, fusion, and evolution and exploring enabling technologies like prompt engineering and multimodal pre-training, WirelessLLM demonstrated its practical applicability through case studies while also identifying key challenges and future research directions. Fig. \ref{wirelessllm} depicts the WirelessLLM framework for spectrum sensing using prompt engineering. The work in \cite{10487933} explored the potential of LLMs in network control for 6G integrated Terrestrial Networks and Non-Terrestrial Networks (TN-NTN).

Table \ref{tab:nsmtaxonomy} provides a summary of related works on LLM for mobile networks and technologies-based NSM.

\begin{figure*}[!t]
    \centering
    \includegraphics[width=\linewidth]{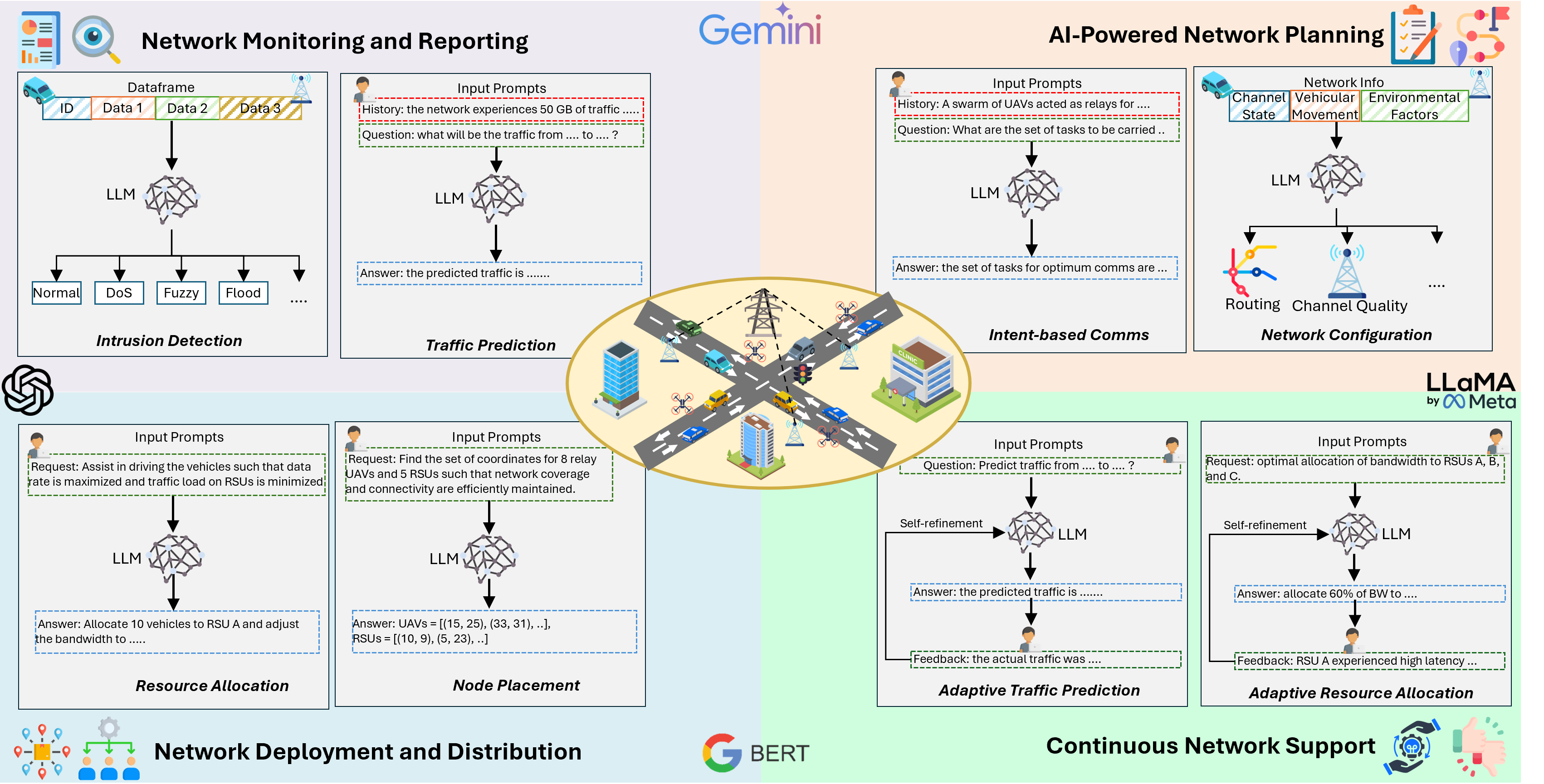}
    \caption{Overview of existing domains of using LLM for vehicular networks.}
    \label{fig:overview_vehicular}
\end{figure*}
\section{LLM for Vehicular Networks-based NSM}\label{sec:vehicular}

Vehicular networks have evolved significantly over the past few years, going from Vehicular Ad-hoc Networks (VANETs) to the more advanced IoV. Modern vehicular networks consist of various components, such as On-Board Units (OBUs) that are installed in vehicles, RSUs, cellular base stations, cloud infrastructure, and intelligent agents, including vehicles, UAVs, and even pedestrians \cite{hussein2022comprehensive}. In such networks, real-time interactions among the different components and agents are facilitated in various forms of communications, such as in-vehicle Controller Area Networks (CANs), Vehicle-to-Vehicle (V2V), Vehicle-to-Infrastructure (V2I), Vehicle-to-Pedestrian (V2P), and more, encapsulated under the umbrella of V2X \cite{tang2021comprehensive, noor2020survey}. Space-Air-Ground Integrated Networks (SAGINs) further enhance vehicular networks by integrating satellite, aerial, and terrestrial systems to improve connectivity and support communication in diverse and remote environments. These networks are important in several critical applications, including traffic management, autonomous driving, road safety, and environmental monitoring \cite{talpur2021machine}. Despite their advancements and critical applications, such dynamic networks come with several challenges, such as latency constraints, communication scalability, resource allocation and scheduling, and security issues \cite{hussein2022comprehensive}.

Lately, LLMs have emerged as powerful tools that support the deployment and maintenance of vehicular networks. Most existing LLM-based solutions in vehicular systems operate on the application level. Such LLM-empowered applications include traffic monitoring \cite{tong2024connectgpt}, scene analysis and understanding \cite{10588373, cao2024maplm}, trajectory/path planning \cite{zhong2024safer, lan2024traj, you2024v2x}, and vehicle dispatching \cite{chen2024llm}. However, only few recent works explored the utilization of LLM on the network level in vehicular networks, with the aim of facilitating intelligent network monitoring, planning, deployment, and continuous support. This section explores the utilization of LLMs within vehicular networks for network monitoring and reporting, AI-powered network planning, network deployment and distribution, and continuous network support. It provides insights into the potential of LLMs to transform vehicular network management by offering predictive and adaptive solutions to emerging challenges, ensuring seamless connectivity, and enhancing overall operational efficiency. Fig. \ref{fig:overview_vehicular} depicts an overview of the existing domains of using LLM for vehicular networks, which are discussed in the next subsections.

\subsection{LLM for Network Monitoring and Reporting}
In vehicular networks, monitoring and reporting are crucial to maintain network health and ensure seamless communication between vehicles and infrastructure. LLMs can play a key role by processing large volumes of real-time data to detect anomalies and potential intrusions and generate detailed performance reports. They also help identify patterns that may lead to congestion or disruptions and facilitate real-time alerts for operators through natural language summaries, reducing the cognitive load on human administrators.
intrusion/anomaly detection}

In the context of vehicular networks, an IDS is tasked with monitoring and analyzing network traffic and communications to identify unauthorized access attempts, malicious behaviors, and potential security breaches that could disrupt the functionality of the network. With vast amounts of data exchanged across various entities, vehicular networks become more susceptible to cyber threats. In this context, LLMs offer advanced capabilities to analyze patterns in network traffic and identify anomalies and irregular behaviors. To this end, this section surveys existing LLM-based intrusion detection solutions for vehicular networks.
\begin{figure*}[!t]
    \centering
    \includegraphics[width=\textwidth]{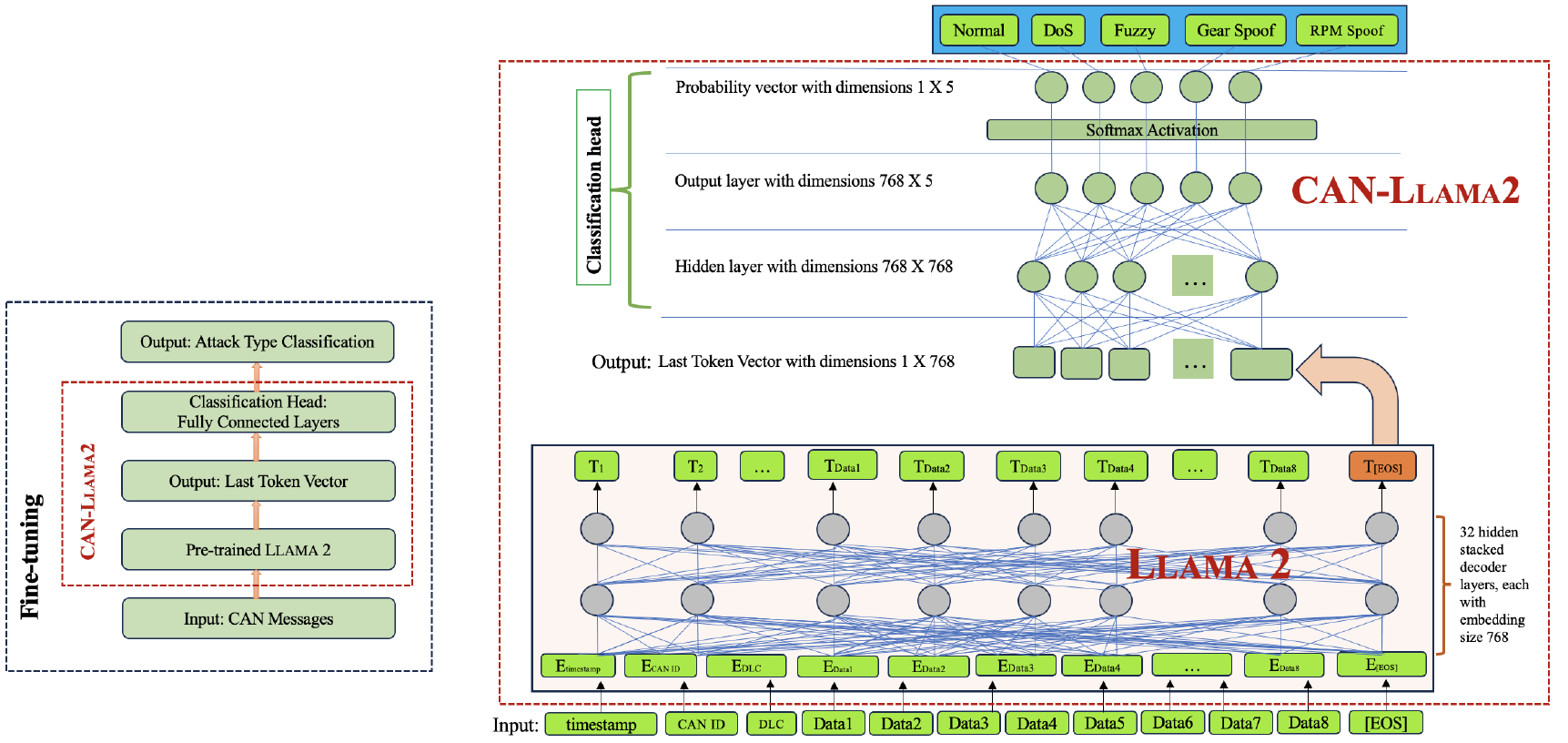}
    \caption{CAN-LLAMA2 finetuning model  \cite{li2023securebert}.}
    \label{CAN-LLAMA2}
\end{figure*}

In the context of CANs, LLMs can be utilized for intrusion detection by analyzing CAN messages exchanged among Electronic Control Unit (ECU) components in a vehicle. The authors in \cite{nwafor2022canbert} proposed CANBERT, a language model that extends BERT into the realm of intrusion detection for CANs. A CAN is a vehicle-bus protocol that allows vehicular components (i.e., ECUs) to communicate on the network. The main aim of this work is to develop a language model that is capable of understanding CAN messages and classifying them as malicious or normal. The model is trained on an open-source dataset \cite{lee2017otids} consisting of logs of CAN traffic from a real vehicle while various message injection attacks were performed. The dataset contains scenarios such as attack-free, denial of service (DoS) attacks, fuzzy attacks, and impersonation attacks. Following a two-stage process, the authors initially pre-trained a variation of BERT, i.e., RoBERTa \cite{liu2021robustly}, on the dataset to learn CAN syntax, then fine-tuned an additional classification layer for intrusion classification, resulting in the final CANBERT model. The model was benchmarked against several commonly used ML techniques, such as XGBoost, SVMs, and GANs, showing outperformance. Various metrics were used to evaluate the performance, including accuracy, precision, recall, and F1 scores, to detect different attacks. A similar work in \cite{alkhatib2022can} proposed CAN-BERT for intrusion detection in CANs, where ID sequences on the CAN bus are analyzed and classified as normal or anomalies. Unlike previous methods, this work uses a masked language model in an unsupervised learning setting. During training, the CAN-BERT model randomly masks some of the CAN IDs in the input, with the aim of anticipating the original IDs that have been masked. Here, the model is only trained on normal sequences of CAN messages to learn regular CAN ID patterns in the network. During testing, the model is given a sequence of CAN IDs, some of which have been masked out. If the model is able to retrieve these IDs, the sequence is considered normal; otherwise, it is flagged as an anomaly. The model is tested on an in-vehicle intrusion detection dataset \cite{kang2021car}, which considers several attacks including flooding, fuzzy, and malfunctions. The performance was analyzed using F1 score, showing outperformance when compared to existing methods such as AutoEncoders and Principal Component Analysis (PCA). 

In \cite{li2023securebert}, the authors proposed three LLMs, namely CAN-C-BERT, CAN-SecureBERT, and CAN-LLAMA2, to detect and classify CAN attacks. These models adapt existing pre-trained models such as BERT, SecureBERT, and LLAMA2 for intrusion detection in CANs. The CAN-C-BERT model uses BERT's bidirectional text comprehension, which is pre-trained on a large text corpus. CAN-SecureBERT relies on the SecureBERT architecture \cite{aghaei2022securebert}, which is a version of RoBERTa tailored for cybersecurity, and which has been pre-trained on a large corpus of cybersecurity-related data. Lastly, CAN-LLAMA2 is based on Meta's LLAMA2 model that is pre-trained on an extensive dataset covering various domains. The three models were fine-tuned with a classification head on the car hacking dataset \cite{song2020vehicle}, which includes attacks such as DoS, fuzzy, and spoofing. Fig. \ref{CAN-LLAMA2} illustrates the fine-tuning process for the CAN-LLAMA2 model. The authors used methods such as LoRA to reduce computational overhead when dealing with large models like LLAMA2. Using metrics such as detection rate, false alarm rate, F1 score, and accuracy, the proposed models were analyzed and compared with state-of-the-art models, showing outperformance. Specifically, CAN-LLAMA2 was found to outperform all existing benchmarks, even while trained on only 10\% of the data. The authors in \cite{nam2021intrusion} treated CAN IDs as words in a sequence and proposed a combination of two GPT networks in a bi-directional manner for intrusion detection in CANs. This method allows past and future CAN IDs to be used in the estimation of a given CAN ID. The model was trained on normal data to minimize the Negative Log-Likelihood (NLL), which is used to determine whether a sequence is an attack or not by comparing the NLL value to a predetermined threshold. A private dataset collected using CAN bus signals from the 2020 Hyundai Avante CN7 was used to test the performance of the proposed model and compared to existing benchmarks, such as uni-directional GPT, GANs, and bi-directional LSTM, considering several attacks such as flooding, fuzzy, spoofing, and replay. The proposed method showed better performance in terms of true positive rate and F-measure.

\begin{figure*}[!t]
    \centering
    \includegraphics[width=\linewidth]{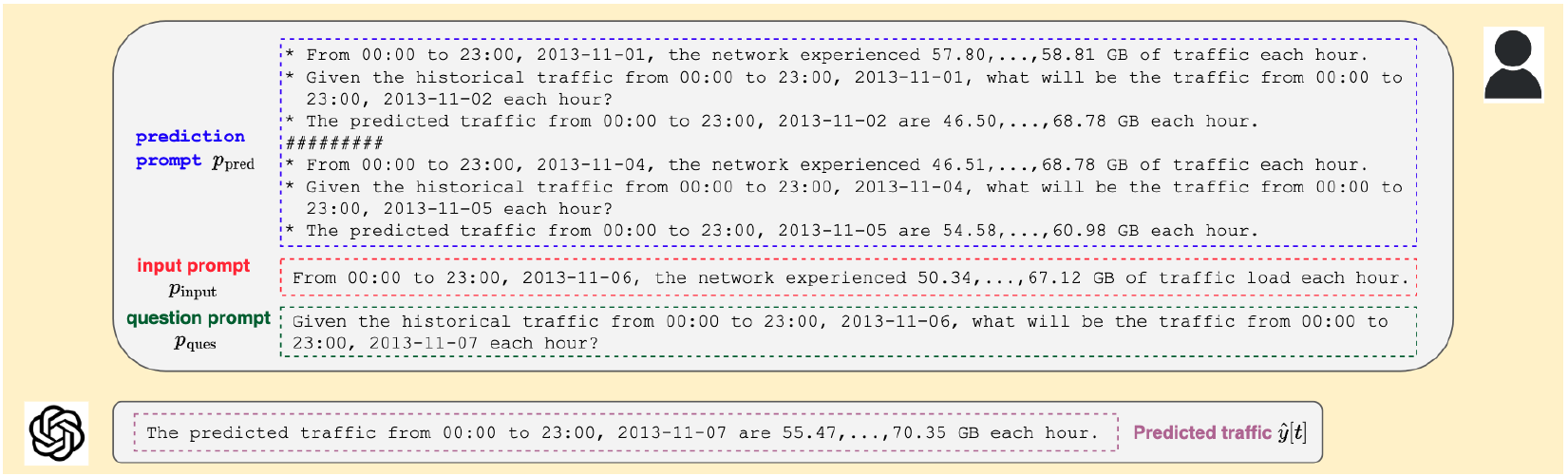}
    \caption{Sample scenario using the TrafficLLM model \cite{hu2024self}.}
    \label{TrafficLLM}
\end{figure*}

\begin{figure*}[!t]
    \centering
    \includegraphics[width=\linewidth]{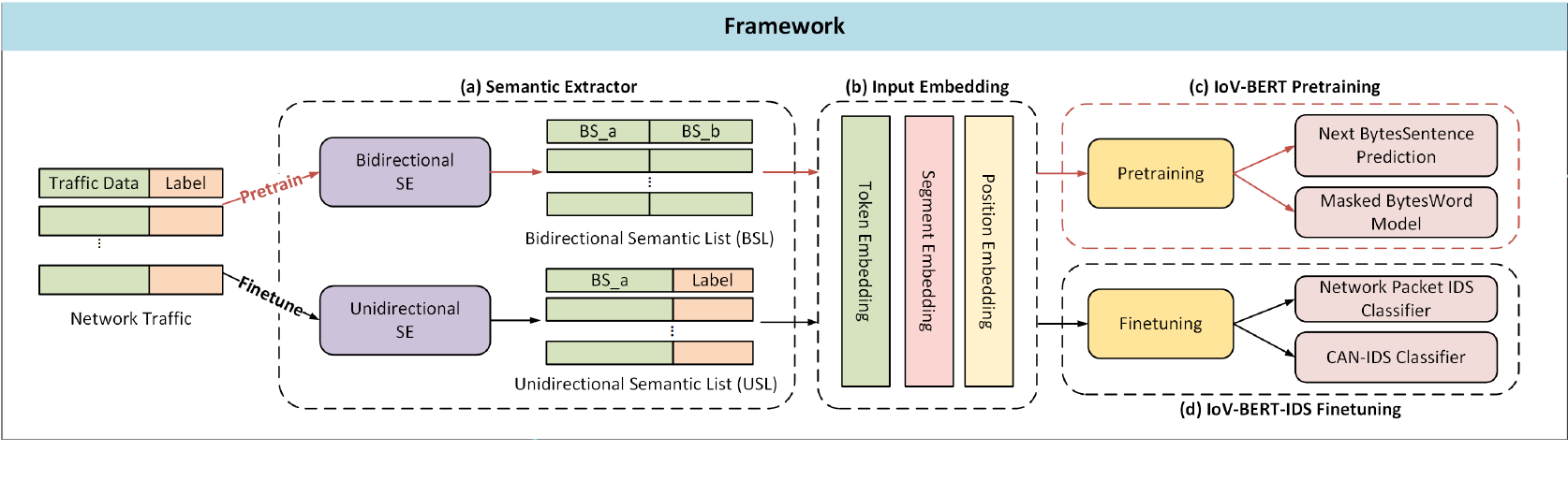}
    \caption{The overall framework of IoV-BERT-IDS  \cite{fu2024iov}.}
    \label{IoV-BERT-IDS}
\end{figure*}
In the field of IoVs, LLMs offer powerful capabilities for intrusion detection by processing and understanding complex communication patterns across vehicles. In \cite{fu2024iov}, the authors proposed IoV-BERT-IDS, a hybrid network traffic IDS designed for in-vehicle and extra-vehicle networks based on the BERT model, as illustrated in Fig. \ref{IoV-BERT-IDS}. Specifically, the proposed method converts raw network traffic data into structured byte sequences (hexadecimal strings), upon which the model can learn representations of the traffic and be fine-tuned to address intrusion detection tasks. The first component of the proposed framework is a semantic extractor, which is a module that transforms network packets into byte sentences using unidirectional and bidirectional processing. This is done to ensure compatibility with BERT's input structure and capture contextual and sequential relationships in the data. The proposed framework then creates embeddings of the input data by combining token, segment, and position embeddings. In the final stage, the authors first use unlabeled data to pre-train the model to create general representations of traffic using methods such as Masked Byte Word Model (MBWM) and Next Byte Sentence Prediction (NBSP). Then, using labeled data, the model is adapted to downstream tasks for intrusion detection. The authors used a combination of private and public datasets for network traffic and intrusion, such as CICIDS \cite{sharafaldin2018toward}, BoT-IoT \cite{koroniotis2019towards}, and Car-Hacking \cite{song2020vehicle}, outperforming traditional IDS approaches in metrics such as precision, recall, F1 score, and accuracy. In \cite{he2024collaborative}, a collaborative framework was proposed with the aim of enriching the fine-tuning process of LLMs in IoVs with more high-quality vehicular threat data (VTD). The framework leverages vehicular honeypots to collect VTD while using methods like Local Differential Privacy (LDP) to add noise to the VTD before sharing for the purpose of protecting user privacy. Additionally, the authors proposed a contract-based incentive mechanism using self-disclosure, which ensures that vehicles select contracts that align with their privacy preferences. The proposed framework was tested on the Qwen2-1.5B-Instruct language model \cite{yang2024qwen2}, which is fine-tuned for thread detection on the CSE-CIC-IDS-2018 dataset \cite{cic_ids_2018} under several types of attacks, such as brute force, DoS, botnet, and DDoS. The authors used metrics such as training loss, accuracy, recall, and F1 score to assess the effect of the proposed framework on the performance of the LLM. In another work \cite{hamhoum2024mistralbsm}, the authors proposed MistralBSM, a real-time Misbehavior Detection System (MDS) empowered by an edge-based LLM. Specifically, a fine-tuned Mistral-7B LLM is deployed at the edge (RSUs) to analyze vehicles' Basic Safety Messages (BSMs) and classify their behaviors into normal or misbehaving. The model is quantized to reduce memory requirements, allowing its deployment on edge RSUs. The collected BSMs are preprocessed and transformed into textual prompts that include the feature names and their corresponding values before being tokenized and fed to the LLM model. The model was fine-tuned and tested on the VeReMi dataset \cite{van2018veremi, kamel2020veremi} covering attack types such as DoS, delayed messages attack, and traffic Sybil attacks. Using metrics such as recall, precision, F1, and accuracy, the proposed method outperformed other LLM models, including LLAMA2-7B and RoBERTa.

\subsubsection{Network traffic prediction}
\label{Network Traffic Prediction}
In vehicular networks, real-time analysis of traffic flow is significant to optimize network communication. Through the analysis of historical traffic data and real-time vehicular inputs, future traffic patterns can be predicted to help identify potential bottlenecks and optimize network routes. To this end, the authors in \cite{hu2024self} proposed TrafficLLM, which is a model designed for wireless traffic prediction in dynamic 6G networks. The proposed model leverages in-context learning without parameter fine-tuning, using a pre-trained LLM (GPT-4) for wireless traffic prediction. Using a template-based description \cite{xue2022translating}, the model takes an input prompt describing historical traffic and a question prompt inquiring about future traffic and returns traffic predictions, as illustrated in Fig. \ref{TrafficLLM}. The proposed model also incorporates self-refinement through feedback to improve its performance (to be discussed later in Section \ref{Section: Adaptive Network Traffic Prediction}). The model was tested in scenarios of vehicular networks using a V2I radio channel measurement dataset \cite{skocaj2023vehicle}. The performance of the method was analyzed using Mean Squared Error (MSE)/Mean Average Error (MAE) and compared against existing methods based on LSTM and base GPT-3.5, showing better performance.

\subsection{LLM for AI-powered Network Planning}
In the context of vehicular networks, LLMs can prove efficient in managing and configuring these networks. Their ability to process high-level and abstract information can help translate user objectives into communication protocols and network actions. Furthermore, LLMs can help optimize the network's configuration to adapt to the fast-changing demands of dynamic vehicular networks, which helps ensure efficient connectivity. This section covers a few works in the literature utilizing LLMs for AI-powered network planning.

\subsubsection{Intent-based communications}

\begin{figure}[!t]
    \centering
    \includegraphics[width=\columnwidth]{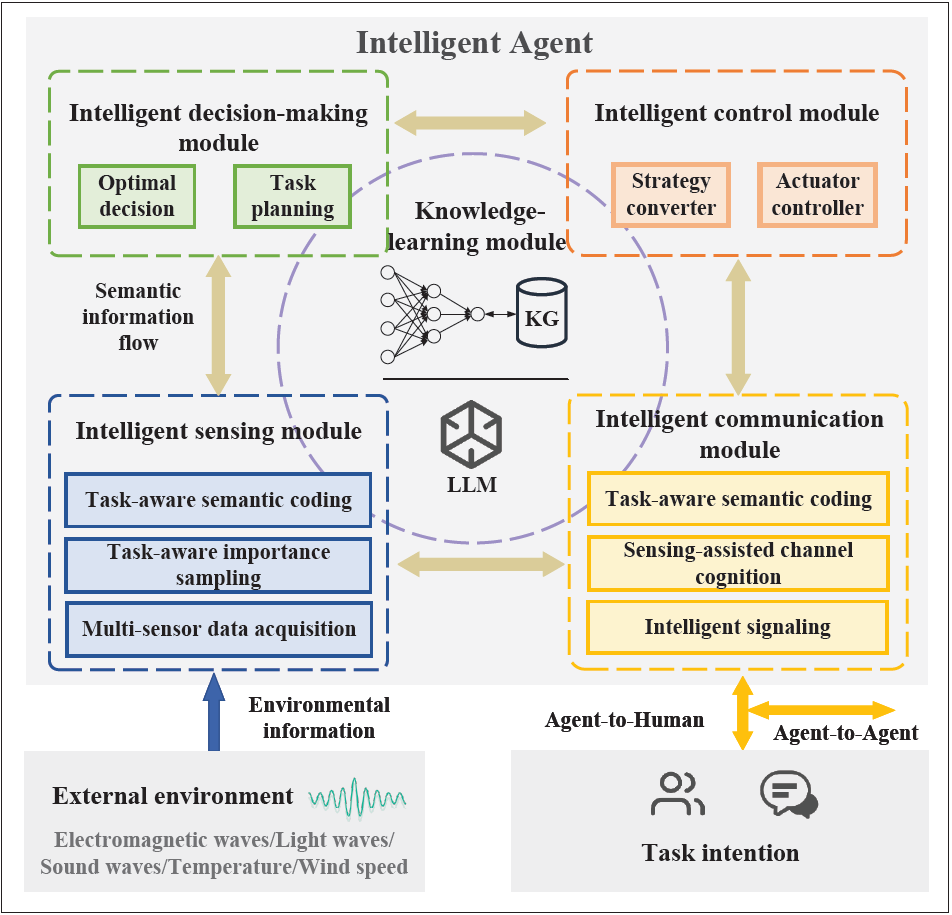}
    \caption{The MAIN architecture \cite{gao2024space}.}
    \label{MAIN}
\end{figure}

Objectives in vehicular networks can be better demanded in a high-level and semantic manner by users and systems. LLMs can excel in interpreting users' natural language intents and converting them into detailed tasks, which helps bridge the gap between abstract objectives and practical implementations. While no works specifically addressed intent-based communications within standalone vehicular networks, the authors in \cite{gao2024space} proposed a Multi-Agent Intelligent Networking (MAIN) architecture in SAGIN, which incorporates LLMs to allow for communication between humans and agents (i.e., UAVs, ground vehicles, etc) and between different agents. The MAIN architecture, shown in Fig. \ref{MAIN}, consists of interconnected intelligent agents across space, air, and ground, which are equipped with intelligent sensing, decision-making, control, and communication modules. All of these modules are coordinated through an LLM-powered knowledge-learning module, which provides semantic understanding and intent translation. In the intent-driven network management and control layer of the architecture, LLMs are used to interpret high-level mission objectives or user intents and translate them into actionable tasks. The authors analyzed the proposed architecture in a use case of task-driven satellite-UAV-ship networking, where agents feed task knowledge to LLMs that coordinate various physical and functional entities. The proposed architecture is conceptual, as the article does not provide details about actual implementations.

\subsubsection{Network nonfiguration}
Efficient network configuration is essential for adapting to the fast-changing demands of vehicular networks. LLMs can analyze large volumes of network data and automate the configuration of the network by optimizing parameters such as bandwidth, frequency, and device connections, with the aim of reducing latency and ensuring efficient communications. In this context, the researchers in \cite{liu2024llm} aimed to enhance Reconfigurable Intelligent Surfaces (RIS) using LLMs, aiming to achieve energy efficiency and reliable communication in 6G IoV networks. LLMs were used to construct a wireless transmission system model based on data such as channel status, vehicle status, QoS requirements, and environmental surroundings, which were then used to establish a resource configuration strategy. Specifically, an LLM takes as input channel state information, vehicular movement data, and surrounding environmental factors and deduces optimized strategies for Resource Block (RB) allocation and signal decoding order that significantly impact the performance and efficiency of communication systems. While the specifics of the LLM model and training are not discussed in the article, the proposed method shows promising results when analyzing the system sum rate (total data rate achieved by all users) under different vehicular environments.  

\subsection{LLM for Network Deployment and Distribution}
Efficient network deployment and distribution are crucial for ensuring reliable vehicular communications. This involves tasks like the optimized allocation of network resources and the optimal placement of nodes (base stations, RSUs, etc). In this context, LLMs can be used as intelligent and adaptive methods for analyzing real-time data and optimizing network coverage, minimizing latency, and ensuring proper resource utilization.

\begin{figure*}[!t]
    \centering
    \includegraphics[width=\linewidth]{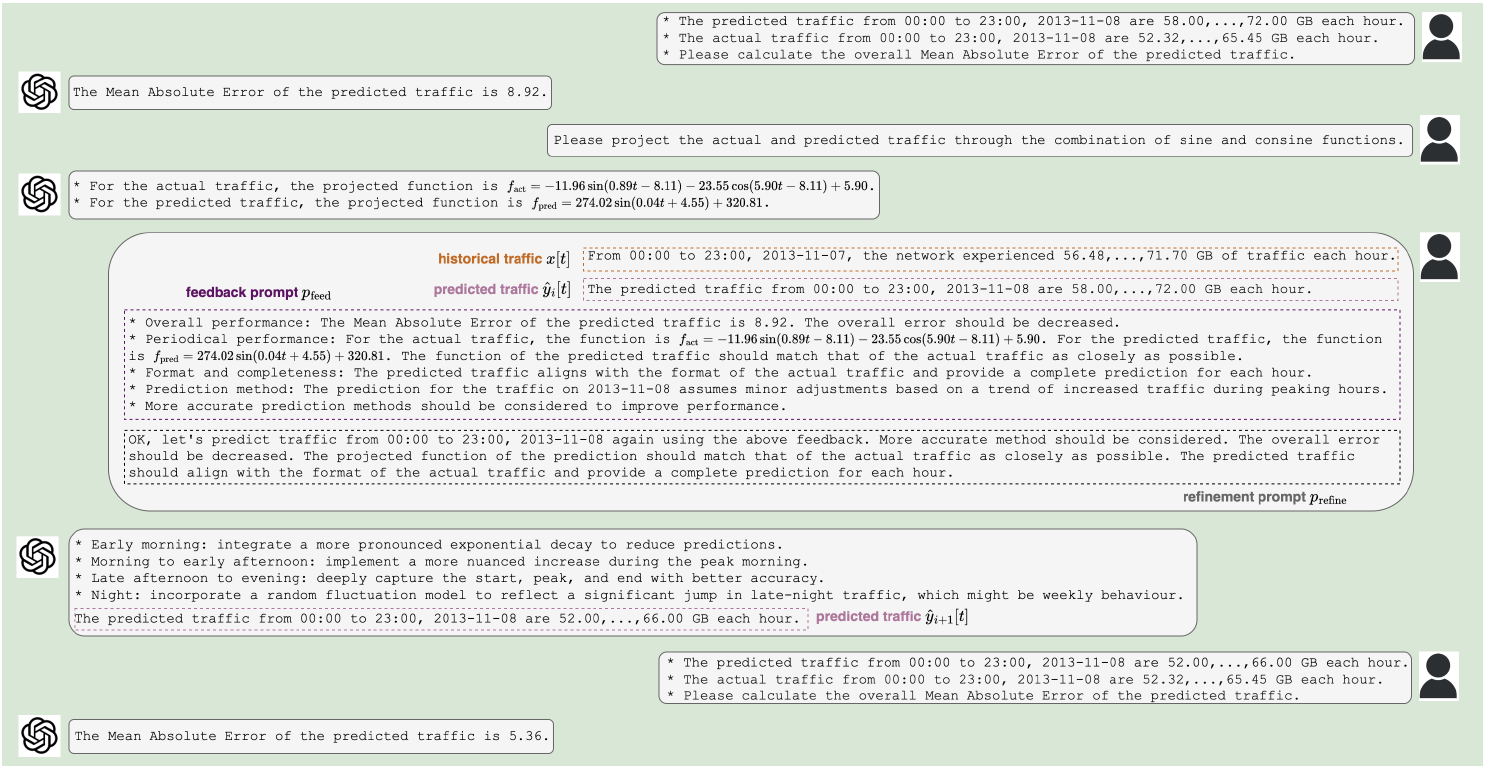}
    \caption{Feedback and refinement in the TrafficLLM method \cite{hu2024self}.}
    \label{TrafficLLM2}
\end{figure*}
\subsubsection{Resource allocation}
\label{Resource Allocation}
This task focuses on the efficient distribution of network resources, such as bandwidth, power, and computational capacity across the network, with the aim of meeting the demands of vehicles, users, and the infrastructure. LLMs can be fine-tuned on diverse datasets of traffic patterns, channel conditions, and user requirements to predict resource needs and optimize their allocation. 

In the domain of vehicular metaverses, Vehicle Twins (VTs) serve as the digital counterparts of the physical vehicles, by which vehicles can access vehicular metaverses to obtain services such as pedestrian detection and 3D entertainment. Since VTs are resource-intensive, vehicles can offload VT tasks to nearby edge computing servers (in RSUs). Due to the dynamicity of the environment and the resource limitations of such RSUs, there is a need for a constant migration of a vehicle's VT from one RSU to another. To this end, the authors in \cite{tong2024multi} proposed a GPT-based DRL algorithm for resource allocation to optimize VT migration across Metaverse Service Providers (MSPs), i.e., RSUs. A DRL algorithm was proposed to solve the sequence problem in the Multi-Attribute Double Dutch Auction (MADDA)-based mechanism, in which vehicular users (bidders) and MSPs (auctioneers) are matched with the aim of optimizing the users' experiences. DRL governs the sequential process of the auction, where a GPT-based model takes in auction attributes and produces auction decisions. The performance of the method was validated using a simulated scenario and compared to existing methods in terms of social welfare, showing outperformance. In another work \cite{yan2024hybrid}, an LLM-powered DRL method was proposed for the joint optimization of V2I communications and AD policies. In the first layer of the proposed method, an LLM takes in textual descriptions of the task (e.g. assistance in driving on a 3-lane highway) and a set of examples of previous experiences (good or bad), and produces an AD decision (acceleration, lane change, etc). The produced decisions by the LLM are then fed as part of the state to the DRL method, based on Double Deep Q-Network (DDQN), which aims to continuously select the most suitable base station for optimized data rate and reduced handovers. While the LLM is used mostly on the application level, its decisions highly impact those of the DDQN with regard to V2I communications. The method was tested using a simulated scenario of autonomous vehicles, where several LLMs (Llama3.1-8B and Llama3.1-70B) are combined with DDQN. The performance was measured using collision rate and handover probability across different scenarios of varying numbers of autonomous vehicles. In the domain of UAVs, the researchers in \cite{sun2024large} explored the integration of LLMs with graph-based dynamic networking for enhanced decision-making. Graph structures were used to analyze complex relationships and interactions between network entities. With LLMs, nodes and edges in a graph are transformed into vector representations, which are then integrated into word embeddings in LLMs that develop an enhanced understanding of entities and relationships. For the problem of UAV networking, the authors proposed an LLM-enabled graphs framework that optimizes UAV trajectory and communication resource allocation. The proposed framework takes as input requests representing information related to a graph in a dynamic network (starting position of the UAV, distances between monitoring points, amount of data to be extracted from the graph). A graph-to-text layer processes the graph information and translates it into embeddings capturing semantic features. The decision layer uses an LLM, such as GPT-4 or Bard, to process the generate a response regarding the recommended trajectory of the UAV and the allocation of resources. The article, however, does not conduct an analysis of the proposed LLM method and only focuses on analyzing the effectiveness of the encoder (graph-to-text) in comparison to existing methods.

\begin{table*}[!t]
\scriptsize 
\centering
\caption{Summary of Related Works on LLM for Vehicular Network-based NSM}
\begin{tabular}{|p{1.3cm}|p{2.3cm}|p{1.5cm}|p{3.2cm}|p{2cm}|p{1.4cm}|p{1.6cm}|}
\hline
\textbf{Taxonomy} & \textbf{NSM Task} & \textbf{Authors} & \textbf{Main Contributions} & \textbf{LLM Solution} & \textbf{Dataset Used} & \textbf{Evaluation Metrics} \\
\hline

\multirow{8}{*}{\parbox{1.5cm}{Network Monitoring and Reporting}} & \multirow{7}{*}{\parbox{1.8cm}{Network intrusion/ anomaly detection}} & Nwafor \textit{et al.} (2022) \cite{nwafor2022canbert} & LLM-based intrusion detection on CAN & CANBERT & Open-source can traffic dataset with attacks & Acc., Prec., Recall, F1-score \\ \cline{3-7}

& & Alkhatib \textit{et al.}. (2022) \cite{alkhatib2022can} & LLM-based intrusion detection on CAN in unsupervised learning setting & CAN-BERT & In-vehicle intrusion detection dataset & F1-score \\ \cline{3-7}

& & Li \textit{et al.}. (2023) \cite{li2023securebert} & Intrusion detection on CAN using several LLMs with reduced computational overhead & CAN-C-BERT, CAN-SecureBER,  CAN-LLAMA2 & Car-hacking dataset & Detection rate, false alarm rate, F1-score, Acc. \\ \cline{3-7}

& & Nam \textit{et al.}. (2021) \cite{nam2021intrusion} & Combine two GPT networks in a bi-directional manner for intrusion detection in CANs & GPT & Private dataset from 2020 Hyundai AVante CN7 & True positive rate, F1-Score \\ \cline{3-7}

& & Fu \textit{et al.}. (2024) \cite{fu2024iov} & LLM-based hybrid network traffic IDS for in- and extra- vehicle networks & IoV-BERT-IDS & CICIDS, BoT-IoT, Car-Hacking & Acc., Prec., Recall, F1-score \\ \cline{3-7}

& & He \textit{et al.}. (2024) \cite{he2024collaborative} & LLM fine-tuning for vehicular threat with honeypots for data enrichment & Qwen2-1.5B-Instruct & CSE-CIC-IDS-2018 & Training loss, Acc., Recall, F1-Score \\ \cline{3-7}

& & Hamhoum  \textit{et al.} (2024) \cite{hamhoum2024mistralbsm} & Edge-based LLM for real-time misbehavior detection & MistralBSM & VeReMi & Acc., Prec., Recall, F1-score \\ 

\cline{2-7}
& \multirow{1}{*}{\parbox{1.8cm}{Network traffic monitoring}} & Hu \textit{et al.} (2024) \cite{hu2024self} & LLM-based wireless traffic prediction in dynamic 6G vehicular networks & TrafficLLM & V2I radio channel measurement dataset & MSE, MAE \\ 
\hline

\multirow{2}{*}{\parbox{1.5cm}{AI-powered Network Planning}} & \multirow{1}{*}{\parbox{1.8cm}{Intent-based communications}} & Gao \textit{et al.} (2024) \cite{10588924} & LLM-empowered networking architecture in SAGIN allowing communication between humans and agents and between different agents & - & - & - \\ 

\cline{2-7}

& \multirow{1}{*}{\parbox{1.8cm}{Network configuration}} & Liu \textit{et al.} (2024) \cite{lira2024largelanguagemodelszero} & LLM-enhanced RIS for reliable communication in 6G IoV networks & - & - & - \\ 
\hline

\multirow{5}{*}{\parbox{1.5cm}{Network Deployment and Distribution}} & \multirow{3}{*}{\parbox{1.8cm}{Resource allocation}} & Tong \textit{et al.} (2024) \cite{tong2024multi} & LLM-based DRL algorithm for resource allocation to optimize VT migration across vehicular metaverses & GPT & Simulated & Social welfare \\ \cline{3-7}

& & Yan \textit{et al.} (2024) \cite{yan2024hybrid} & LLM-powered DRL method for the joint optimization of V2I communications and autonomous driving policies & Llama3.1-8B, Llama3.1-70B & Simulated scenarios & collision rate, handover probability\\ \cline{3-7}

& & Sun \textit{et al.} (2024) \cite{sun2024large} & Integration of LLMs with graph-based dynamic UAV networking for enhanced UAV trajectory and communication resource allocation & GPT-4, BARD & - & - \\ 
\cline{2-7}

& \multirow{2}{*}{\parbox{1.8cm}{Node placement}} & Wang \textit{et al.} (2024) \cite{wang2024multi} & LLM-driven optimization framework for multi-UAV placement in IAB networks & GPT-4-turbo & - & data rate score \\ \cline{3-7}

& & Li \textit{et al.} (2024) \cite{li2024large} & Investigate using an LLM to assist in solving a multi-objective optimization problem for the deployment and power control in multi-UAV networks & GPT-3.5-Turbo & - & HV \\ 
\hline

\multirow{2}{*}{\parbox{1.5cm}{Continuous Network Support}} & \multirow{1}{*}{\parbox{2.3cm}{Adaptive Network traffic prediction}} & Hu \textit{et al.} (2024) \cite{hu2024self} & Self-refining LLM-based wireless traffic prediction in dynamic 6G vehicular networks & TrafficLLM & V2I radio channel measurement dataset & MSE, MAE \\ 
\cline{2-7}

& \multirow{1}{*}{\parbox{2.3cm}{Adaptive resource allocation}} & Sun \textit{et al.} (2024) \cite{sun2024large} & Adaptive integration of LLMs with graph-based dynamic UAV networking for enhanced UAV trajectory and communication resource allocation &  - & - & - \\ 
\hline
\end{tabular}
\label{tab:nsmtaxonomy}
\end{table*}

\subsubsection{Node placement}
This task is critical in vehicular networks, as it ensures optimal placement of communication infrastructure such as RSUs, base stations, and UAVs. Effective node placement significantly impacts network performance in terms of coverage, signal quality, and resource availability, which are challenging to maintain due to the dynamicity of the vehicular environment. LLMs can prove efficient in providing adaptive node placement solutions by analyzing mobility patterns, network conditions, and geographic constraints.

In \cite{wang2024multi}, the authors proposed an LLM-driven optimization framework for multi-UAV placement in Integrated Access and Backhaul (IAB) networks. Following an optimization by prompting approach, the LLM takes as input a prompt describing the objective in the form of a request (e.g., to find the set of UAV coordinates that maximize a certain score) and previous examples (previous coordinates and their scores), to produce a response regarding the optimal placement. The score is given as a function of the data rates observed by each UAV. The authors used GPT-4-turbo as the LLM but gave no details on the data pre-processing and model fine-tuning steps. The performance of the proposed method was measured in terms of data rate score across varying number of UAVs. In another work \cite{li2024large}, the researchers investigated using an LLM to solve a Multi-objective Optimization Problem (MOP) to optimize the deployment and power control in multi-UAV networks. An LLM-Enabled Decomposition-based Multi-objective evolutionary Algorithm (LEDMA) was proposed to decompose the problem into a series of sub-problems. Here, UAVs act as aerial base stations, and their deployment aims to maximize the total network utility of all users. LLM was used as a search (crossover and mutation) operator that generates new points. It takes the problem description (objectives and variables of the MOP), a few sample solutions, and instructions to generate new solution points as input with the aim of enhancing the evolutionary algorithm. The proposed solution was built using GPT-3.5-Turbo as the search operator and compared against existing optimization methods using the Hypervolume (HV) metric that measures the algorithm's convergence and diversity.

\subsection{LLM for Continuous Network Support}
The sustainability of vehicular networks demands continuous solutions that adapt to the dynamic nature of the environment and improve over time. Due to their fine-tuning capabilities, LLMs can dynamically leverage feedback from the users in real-world vehicular networks to iteratively refine their models and improve performance. Few of the aforementioned surveyed articles have also incorporated refinement strategies to enhance the performance of LLMs in vehicular networks, which are discussed in this subsection.

\subsubsection{Adaptive network traffic prediction}
\label{Section: Adaptive Network Traffic Prediction}

Due to the dynamicity of network traffic patterns, it is important for a deployed LLM to continuously refine itself using real-time feedback and historical data. In this context, the TrafficLLM model proposed in \cite{hu2024self}, which was previously discussed in Section \ref{Network Traffic Prediction}, has self-refinement capabilities through feedback to improve its performance. The model is designed for wireless traffic prediction in dynamic 6G networks, which translates historical traffic and request prompts into information about future traffic. Following the initial traffic prediction, the model receives a feedback prompt encompassing information such as the prediction performance (e.g., using MSE), prediction format and completeness, and prediction method, as shown in Fig. \ref{TrafficLLM2}. The model is then continuously refined through more and more experiences. This method was implemented using a GPT-4 model for traffic prediction, and compared against a base GPT-4 model that has no feedback and refinement, showing significant outperformance in terms of MSE/MAE.

\subsubsection{Adaptive resource allocation}
The dynamic and fluctuating demands in vehicular networks require real-time adjustments to the model parameters to ensure quick recovery from abnormal situations such as network failures or cyber-attacks. To this end, the authors in \cite{sun2024large} explored using LLMs in graph-based dynamic UAV networking for enhanced decision-making with regards to UAV trajectory and communication resource allocation, as discussed in Section \ref{Resource Allocation}. Additionally, through iterative graph-to-text transformations, the LLM continuously refines its predictions and strategies based on updated network data. This self-tuning ability allows UAVs to autonomously adjust their flight paths and resource allocations to respond to changing conditions, optimizing both network performance and resource efficiency.

Table \ref{tab:nsmtaxonomy} provides a summary of related works on LLM for
vehicular network-based NSM.

\begin{figure*}[!t]
    \centering
    \includegraphics[width=\linewidth]{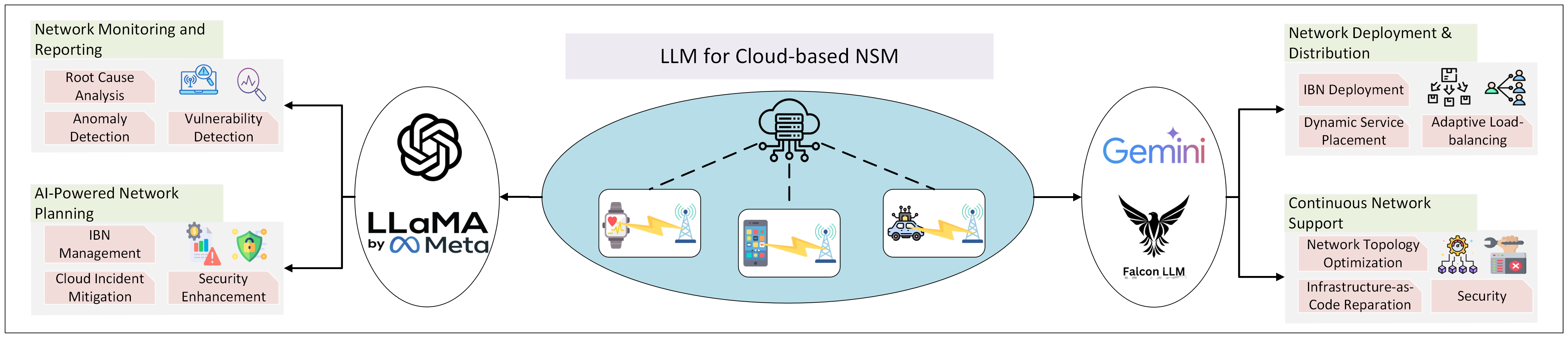}
    \caption{Overview of existing domains of using LLM for cloud-based NSM}
    \label{fig:overview_cloud}
\end{figure*}

\section{LLM for Cloud-based NSM}\label{sec:cloud}
In this section, we discuss the recent literature efforts contributing to the utilization of LLMs to support the cloud network and its service management. A cloud network is typically characterized by powerful computing machines, scalable networks, and management services to provide scalable, flexible, and on-demand network resources. It allows users to manage, deploy, and monitor network components, such as routers, switches, and firewalls, over the cloud, often using virtualized resources. Cloud networks are crucial for modern businesses, especially with the growth of distributed applications, remote work, and IoT devices. LLMs can play a transformative role in enhancing cloud networks by providing AI-driven automation, real-time decision-making, and predictive analytics. Specifically, we divide the LLM support into four distinctive categories: (1) network monitoring and reporting, (2) AI-powered network planning, (3) Network deployment and distribution, and (4) Continuous network support. In this sequel, we review the literature efforts to support the cloud using LLMs for each of these four categories. Fig. \ref{fig:overview_cloud} presents an overview of the different literature domains where LLMs are implemented as a solution to support cloud-based NSM.

\subsection{LLM for Network Monitoring and Reporting}
In this context, LLMs perform network monitoring by providing advanced tools for root cause analysis, anomaly detection, and network load analysis. By processing vast amounts of real-time network data, LLMs can automatically detect performance issues and unusual patterns, helping network engineers quickly identify the root cause of problems. This enhances decision-making and allows for real-time adjustments, ensuring the network operates efficiently and securely. LLMs can also generate natural language reports, simplifying complex data into actionable insights. 

\subsubsection{Root cause analysis}
Under this scenario, we study the literature contributions to using LLM for performing root cause analysis, which refers to identifying the main reason behind a certain problem. Root cause analysis requires troubleshooting, which is ideally performed by analyzing the cloud network to reach the root cause. We particularly look into the use of LLM for generating a report, studying the root cause of the incident, and evaluating the confidence of the generated report through a score.

Goel \textit{et al.} \cite{goel2024x} focused on the use of LLMs for root-cause analysis of incidents over the cloud. The main contribution of this work revolves around the importance of providing contextual information related to the Software Development Life-Cycle (SDLC) to improve the effectiveness and accuracy of the generated analysis. This contextual information includes X-lifecycle service functionality and upstream dependency information. Furthermore, the authors study the impact of this contextual information on the improvement of accuracy for the Service Level Optimization (SLO) classification. The ultimate goal was to reduce the burden on the on-call engineers who rely on the SDLC for manual root-cause analysis. The experiments were performed on a dataset of a couple of hundred incidents provided by Microsoft. In addition, GPT-4 was used as the main model for evaluating the mechanism and impact of prompt engineering. Various metrics were used for evaluating the accuracy of the root cause analysis and SLO classification, including BLEU, METEOR, ROUGE, Bert Score \cite{Zhang_2020BERTScore}, and NUBIA \cite{kane2020nubia}.

In \cite{zhang2024lm}, the authors focused on identifying the problem of hallucination of LLM models when supporting on-call engineers in the diagnosis of incidents' root causes over the cloud. To this end, the article proposed a confidence estimation score to assign a value determining the effectiveness of the analysis provided by the model. The engineers would rely on this score to determine if the analysis performed is accurate. Furthermore, an augmented calibration mechanism was performed to accommodate for the evolution of incidents and the need to upgrade the confidence rate accordingly. An LLM was proposed to generate the confidence score, identified as Large Language Models Prompted with Augmentation for Calibrated Confidence Estimation (LMPACE). Through prompting, the proposed framework is composed of a two-step procedure to ensure reliable results. The first stage is responsible for historical incident retrieval, which is sent to the LLM to identify if there is a similar incident encountered in the past and if there is enough information for the model for evaluation. A chain-of-thought (CoT) approach is applied to the LLM to divide complex large problems into manageable sub-problems, which is proven to improve the capability of LLMs in reasoning and analysis \cite{dhuliawala2024chain, nye2022show}. Additionally, multiple queries of analysis were executed in parallel to make sure several versions of analysis were available. Instead of asking for a binary confidence score to evaluate the analysis, the proposed framework divides the scoring for different parts of the analysis to ultimately compute the mean. The mechanism of parallel execution of LLMs has been extensively studied, where additional optimization can be applied towards obtaining separate results \cite{chen2024design}. \\
In their experiments, the authors in \cite{zhang2024lm} used a private dataset from a list of incidents provided by Microsoft. Furthermore, the GPT-4, GPT-3.5-Turbo, and Text-DaVinci-003-3shot models (with and without human feedback) were used and evaluated for building the LLM for confidence scoring. The evaluation metrics used include reliability diagrams and ECE scores, in addition to human evaluation. The reliability diagram is a widely used mechanism to rapidly assess a machine learning model performance by comparing the probability of the output proposed against the frequency of observations. The obtained results showed superior performance in terms of reliability and accuracy over the various metrics for root cause analysis evaluation using LLM-based models.

Similarly, the work in \cite{roy2024exploring} leveraged LLM, named ReAct, to perform root cause analysis of cloud incidents. The LLM is based on the ReAct framework \cite{yao2023react}. Similar to \cite{zhang2024lm}, the CoT mechanism is applied while studying the importance of contextual information about the services when passed to the LLM. The evaluation and testing were performed in a real-world production environment with a set of incidents collected internally. Furthermore, the OpenAI GPT4-8k was used as the main LLM during the experiments. The evaluation metrics used are C-BLEU, S-BLEU, ROUGEL, METEOR, and BertS. Different benchmarks were used for comparison with the base LLM, including a Retrieval Baseline based on the RAG framework, CoT, and Interleaving Retrieval-CoT. The results showed the effectiveness of the proposed LLM using ReAct framework when considering information retrieval and the CoT mechanism.

Additionally, researchers in \cite{chen2024automatic} proposed Root Cause Analysis Copilot (RCACopilot). In this work, on-call engineers collect the necessary logs and information for diagnostics before sending them to an LLM that can identify the root cause and provide detailed reporting. The LLM models used for evaluation include GPT-4, GPT-3.5 fine-tuned, XGBoost, and FastText. In terms of evaluation, F1-score and the inference time were used. Similar to all other works, the dataset used was privately collected from the Microsoft transport system.

As can be seen from our thorough literature review on root-cause analysis of cloud network incidents, we can identify the main limitation lying around the dataset availability. Furthermore, there is no clear study on the computational overhead caused by the LLM inference time and its impact on the decision and mitigation strategies. Furthermore, these papers mainly focused on the root-cause analysis with no mention of potentially using LLMs to offer resolutions. 

\subsubsection{Anomaly detection}
In the cloud context, anomaly detection refers to the different solutions that study network behaviors to identify suspicious traffic or behavior that falls outside the range of normal ones. The cloud-based problem that falls under that category of monitoring and reporting is anomaly detection. Existing cloud systems lack an effective anomaly detection method that utilizes the servers and network metric data. Traditional DL methods usually suffer from generalization problems and require further fine-tuning, especially when faced with real-life production environments. To this end, the role of LLM for anomaly detection is to assist engineers in detecting potential risks that can adversely affect the cloud network. In large cloud systems, identifying anomalies through a clear definition is critical to mitigate the large number of false alarms that exist using traditional DL approaches. Specifically, anomaly detection refers to the statistical data point divergence from regular occurring events. Furthermore, the correlation of other incidents is important for anomaly identification. The power of LLM in this context and its potential success relies on the capability of handling long-term time dependencies between various adversary incidents.

According to \cite{nunes2024leveraging}, LLMs can significantly enhance cloud network monitoring and reporting by leveraging their advanced NLP capabilities to interpret and analyze large volumes of network data in real-time. They can detect anomalies and predict potential failures, thus enabling proactive maintenance and reducing downtime. Additionally, LLMs can strengthen security by identifying unusual patterns indicative of cyber threats and automating incident response actions. Their integration ultimately improves the reliability, scalability, and efficiency of cloud network operations.

The work in \cite{yu2024monitorassistant} proposed MonitorAssistant, an end-to-end anomaly detection system that automates the process of model configuration and automation for effective knowledge inheritance towards real-time detection using the cloud metric data. This system is primarily based on an LLM model, in particular GPT-4 to process natural language and integrate domain knowledge from historical incident-metric data. It automates the anomaly detection process across three phases: Configuration Recommendation, Anomaly Alert, and Feedback Loop. The system uses the Monitor Configuration Infusion technique to recommend configurations based on historical experiences and a combination of time-series data and descriptive information similarity. Practical Alert Generation integrates expert knowledge for detailed anomaly reports, while LLM-Engineer-In-The-Loop interaction facilitates efficient feedback incorporation and rule-based adjustments, enhancing the model's adaptability in real-time deployment. The system was deployed within Microsoft's cloud services, enabling practical and automated anomaly detection for industrial applications. The dataset and source code used for the experiments are not provided in this work, limiting its impact on the community.

The authors in \cite{patil2024leveraging} leveraged LLMs for cloud network monitoring and reporting by fine-tuning pre-trained models on domain-specific datasets, including network traffic logs, security alerts, and incident reports. This fine-tuning process involves training the LLMs to recognize patterns in network behavior, enabling them to interpret and analyze complex log data effectively. Additionally, the models are optimized to respond to user queries about network health, security incidents, and performance metrics, generating detailed reports that provide actionable insights for network administrators. There is no comprehensive evaluation following the proposed methodology.

The work of \cite{ott2021robust} presents a novel approach to cloud network monitoring and reporting by utilizing LLMs for anomaly detection in log data. By employing sentence-level embeddings from models like BERT, GPT-2, and XL-Transformers, the framework enables the identification of both ground truth and synthetic anomalies generated through intentional alterations in log messages. The researchers conducted experiments using the CloudLab OpenStack log dataset, which includes both normal and anomalous log data. In the experiments, the normal dataset is manipulated to create additional test sets that simulate various degrees of log alterations, allowing for an assessment of model transferability during software updates. Evaluation metrics, including precision, recall, and F1 scores, indicate that BERT consistently outperformed other models in both regression and classification tasks, showcasing the efficacy of LLMs in enhancing the robustness of anomaly detection in dynamic cloud environments.

The first main limitation of these works is related to the limited or complete lack of information about the experimental setup, dataset used, and implementation. In terms of technical limitations, these works include the dependency on manual feedback for refining the anomaly detection model, which can be time-consuming and may not always generalize well across different metrics. Additionally, the unified similarity calculation relies on LLM-generated similarity scores, making it computationally expensive, and real-time processing may be challenging. Another limitation is that the model configuration recommendations may not always adapt well to new metrics with limited historical data, potentially affecting detection accuracy in novel scenarios.

\subsubsection{Vulnerability detection \& analysis}
The authors in \cite{cao2024llm} introduced LLM-CloudSec, a novel approach leveraging LLMs with Retrieval Augmented Generation (RAG) and the Common Weakness Enumeration (CWE) for fine-grained vulnerability detection, classification, and analysis in cloud applications. This framework employs a dual agent, comprising a detection agent and an analysis agent, to automate vulnerability detection and analysis. The detection agent identifies potential vulnerabilities using CWE-based categorization through few-shot learning, while the analysis agent verifies and performs fine-grained analysis using detailed CWE descriptions and examples, ensuring reliable results through context-enhanced retrieval and encoding. The experiments, conducted using the Juliet and D2A datasets, showed that LLM-CloudSec achieves over 70\% accuracy in detecting 60 CWE vulnerability types and provides detailed analysis with precise localization of vulnerabilities. GPT-4 API was used as the main LLM for evaluations. The results highlighted its effectiveness in automating high-quality detection and analysis, even in real-world scenarios like buffer overflow vulnerabilities in the D2A dataset.
While this work is one of the pioneering efforts in building vulnerability analysis and detection, it still focuses on a single dataset scope and applicability of code-based LLMs and could face potential challenges in handling incomplete or ambiguous code snippets. These aspects require further investigation. 



\subsection{LLM for AI-powered Network Planning}
When applied to network planning, LLMs assist with code completion, network configuration, predictive maintenance, and security enhancement. LLMs enable the automation of network design tasks, generating configuration scripts and predicting network needs based on historical data and traffic patterns. They improve the resilience of the network by forecasting maintenance needs, dynamically scaling resources to match demand, and enhancing security by anticipating and mitigating risks before they escalate. 

\subsubsection{Intent-based network management}

This initial category of literature attempts to perform cloud-based network planning through LLMs and is referred to as intent-based network \cite{leivadeas2022survey}. These management solutions utilize the cloud to control and manage network services. In this section, the focus is explicitly on controlling network configuration, excluding network security over the cloud, which is discussed in Section \ref{sec:cloud_sec_plan}.

Assuming that a user requests network and computing resources to run a Metaverse-based application, which has its own application requirements that are typically intensive \cite{10507201}. The user intent includes specific demands for a network that can handle three Extended Reality applications: an augmented reality server, a mixed reality collaboration platform, and an engine for virtual reality. The exact GPU, CPU, memory, and disk requirements are specified for each of the applications. For instance, mixed reality requires five vCPU and six gigabytes of memory. Furthermore, additional network specifications that are related to the user's quality of experience, such as the tolerated network delay, can be provided. For example, only a 6ms delay is tolerated for augmented reality applications. The IBN lifecycle is essentially composed of five steps: (1) intent decomposition, (2) intent translation, (3) intent negotiation, (4) intent activation, and finally (5) intent assurance \cite{leivadeas2022survey}. First, information is extracted from the intent to specify the technology domain and requirements. Second, the extracted intent is translated into infrastructure-level intents specific to each extracted domain. Third, in case requirements cannot be fully satisfied, alternative intent is proposed to the creator for acceptance or rejection. Fourth, the agreed intents are translated into actions that software can perform. Fifth is the process of ensuring that the operational state of the network matches its defined intentions. 

\begin{figure}
    \centering
    \includegraphics[width=\linewidth]{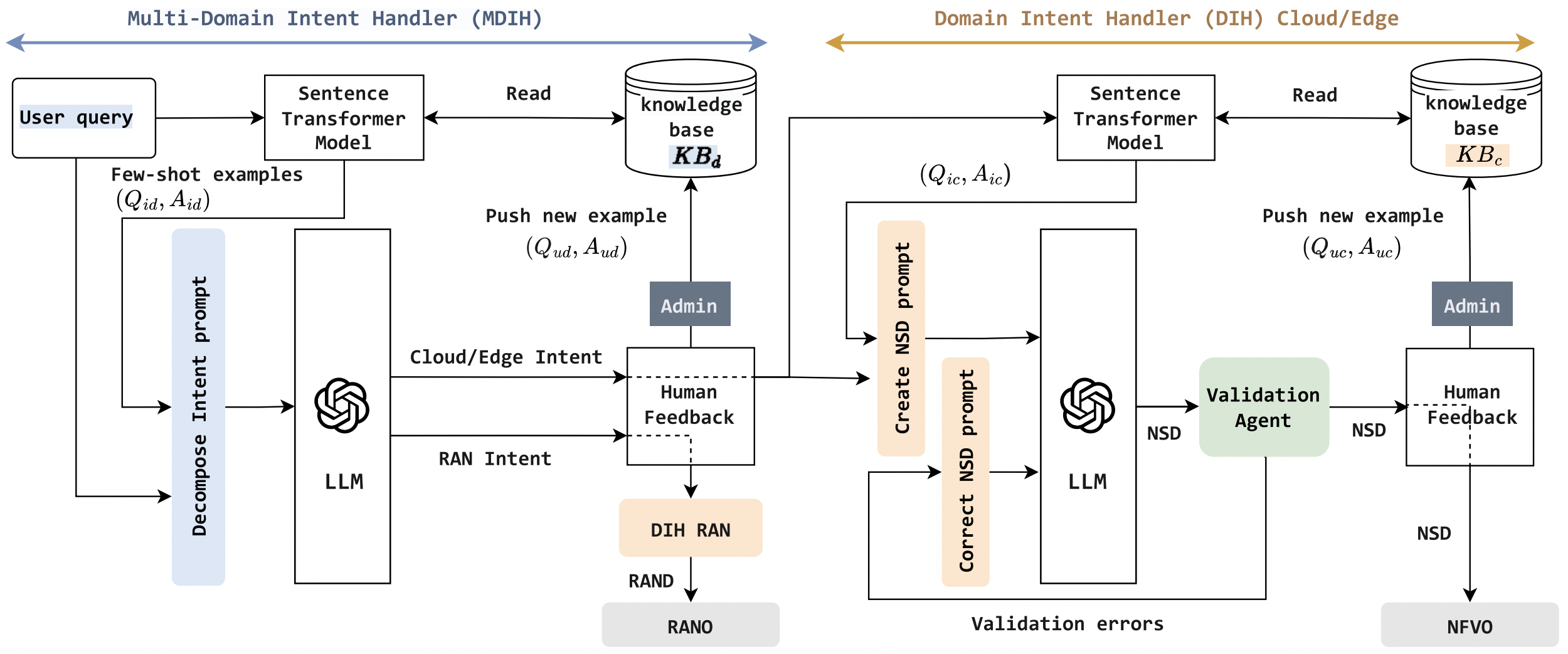}
    \caption{LLM system for decomposition and translation \cite{mekrache2024intent}.}
    \label{fig:llm_ibn}
\end{figure}

In \cite{mekrache2024intent}, researchers proposed the use of natural language understanding through LLMs to process user intent in IBN. As seen in Fig. \ref{fig:llm_ibn}, the approach prepares the LLM for use using a few-shot learning mechanism, which utilizes smaller amounts of data while still achieving a promising learning performance. Additionally, human feedback is incorporated into the proposed system, enabling the use of RL and allowing the LLM to learn from previous experiences. This work proposed a high-level architecture for handling intent lifecycle in multi-domains, leveraging natural language-based intent decomposition using LLMs. The architecture, supported by a Multi-Domain Intent Handler, abstracts underlying infrastructure complexities, decomposes intents into sub-parts for different domains (e.g., cloud, edge, RAN), and validates them for seamless cross-domain deployment.
Fig. \ref{fig:llm_ibn} details an LLM-based framework for intent decomposition and translation, consisting of three stages: retrieving few-shot examples from Knowledge Bases (KBs), using in-context learning to decompose intents or generate ILIs, and validating outputs for correctness. Human feedback is incorporated into the KB to refine future performance, enabling an iterative improvement process for intent handling.
Evaluations in real-world settings were performed using the Code LLAMA LLM on a single NVIDIA A100 GPU with 40GB of vRAM. The evaluation showed efficient intent decomposition with an average user rating of 4.5, while intent translation initially required minor adjustments but improved over time, maintaining an average rating above 3.5 and achieving accurate Network Service Descriptor generation.

The Open Network Automation Platform (ONAP) acts as an operating system for the network, which provides functionalities to manage the different end-to-end lifecycle processes for services, such as Software Defined Network (SDN) and Network Function Virtualization (NFV). This framework also covers the different types of physical, virtual, and cloud-native network entities, including the support for 5G and 6G cellular networks. A Service Design and Creation (SDC) component in ONAP is responsible for the design and creation of orchestrated services over the network. In \cite{panchal2024simplifying}, the authors extended the ONAP capabilities through an integration of LLM for automated network services orchestration. This short paper was dedicated to presenting the framework and idea of LLM integration in ONAP without a formal presentation of the framework, theoretical study, or statistical analysis of its performance. 

The authors in \cite{dandoush2024large} introduced the potential of LLMs for intelligent, collaborative management of dynamic network slices in cloud environments. LLMs enable seamless communication across different domains by interpreting user intents in natural language and abstracting network topologies for collaboration. They enhance resource allocation, network configuration, and real-time monitoring through automated processes, ensuring efficient network functioning. No experiments were conducted in this work. However, the article highlights the main limitations of the proposed mechanism, including a limitation in dataset availability and the requirement for huge computing power, in addition to the need for unified standards and interoperability across the heterogeneity of network environments.

Up to the time of writing this survey, the idea of using LLM for network control over the network is starting the grow. However, there is no proper investigation of its feasibility, a testing or simulation environment.

\subsubsection{Cloud incident mitigation}
In cloud incident mitigation, LLMs process vast amounts of system logs and network traffic data in real-time to detect anomalies and identify root causes. They use their knowledge to suggest tailored resolutions, such as reconfiguring services, restarting impacted nodes, or isolating problematic components. LLMs can also automate routine mitigation steps, like triggering failovers or scaling infrastructure dynamically. Furthermore, they enhance communication by generating incident summaries for stakeholders and providing actionable insights to improve system resilience.


\begin{table*}[h]
    \centering
    \caption{Examples of using the Davinci model for incident resolution by \cite{ahmed2023recommending}}
    \label{tab:cloud_incident}
    \begin{tabular}{@{}|c|c|p{6cm}|p{6cm}|@{}}
        \hline
        \textbf{Examples} & \textbf{Tasks} & \textbf{Original Text} & \textbf{Generated by Davinci model} \\
        \hline
        \textbf{I} & Root Cause & \makecell[l]{The customer was sending very large amounts \\ of data as their payload. \\ The size in addition to the rate at which \\ these transactions were being sent led to \\ the initial network crash.} & \makecell[l]{Large payloads that exceed the max \\ payload size limit of the ledger \\ caused the ledger to crash.} \\ 
        \cline{2-4}
         & Mitigation & \makecell[l]{A code upgrade with higher \\ payload size solved the problem.} & \makecell[l]{Republished the ledger with \\ a higher payload size.} \\
        \hline
        \textbf{II} & Root Cause & \makecell[l]{An unexpected power loss caused \\ multiple VMs to go down in East US \\ causing DB services.} & \makecell[l]{Datacenter outage impacting \\ multiple services.} \\
        \cline{2-4}
         & Mitigation & \makecell[l]{DB team mitigated the upstream issue \\ in the West US region.} & \makecell[l]{DB team mitigated the issue \\ by restarting the service.} \\
        \hline
    \end{tabular}
\end{table*}

The work in \cite{ahmed2023recommending} proposed the use of LLM for automated mitigation mechanisms against analyzed root cause incidents over the cloud. The methodology involves preparing a dataset of incident summaries, root causes, and mitigations from Microsoft's incident database covering the period between January 2018 and July 2022. Unfortunately, this dataset is not shared as it unveils sensitive information, making the methodology and evaluation results difficult to reproduce. After data preparation, including deduplication and filtering, the dataset includes 57,520 root causes and 8,300 mitigations, with severe incidents prioritized for analysis. Various transformer-based LLM models are fine-tuned and evaluated to generate root causes and mitigate them. Some of these models are RoBERTa, CodeBERT, and OpenAI's GPT-3.x family (i.e., Curie, Codex, Davnici, and Code-DaVinci-002). The work showcases the advantages of pre-trained models in handling domain-specific tasks, with a particular focus on both language and code-based models to address the unique terminologies in incident management. Comparing the various benchmarks showed that GPT-3.x models outperformed baselines in most metrics, with Code-Davinci-002 leading in correctness and readability. Table \ref{tab:cloud_incident} showcases an example of the performance of this model for mitigation techniques generation against two examples of incidents. Multi-task learning shows limited improvement, and root cause information significantly enhances mitigation quality. The human evaluation highlights the practical utility of the Davinci-002 model. Up to the time of writing this paper, this work presents the most comprehensive result analysis in the context of incident analysis and mitigation techniques through a detailed and well-structured comparison with various benchmarks.

In \cite{hamadanian2023holistic}, the authors proposed the use of LLMs for incident management over the network in the cloud context. To this end, this work establishes principles for designing automated On-Call Engineer helpers in incident response and mitigation mechanisms. Through iterative prediction, the LLM is able to hypothesize, test, and reassess decisions in feedback loops for handling complex and novel incidents, which is typically beyond the capabilities of one-shot learning as opposed to literature efforts. Furthermore, the integration of risk assessments and confidence measures to avoid costly mistakes and maintain operator trust is of paramount importance. Moreover, helpers based on LLMs need to be adaptive and account for changes in code basis or network environment, ensuring continuous support. Thus, continuous support in cloud networks combined with the automation strategies of LLMs open the door for further research investigations, which is detailed in Section \ref{sec:cloud_continuous_support} of this survey.

This work primarily focused on the proposal of the idea and opening research directions for the community without formal architecture design or evaluation results, limiting its impact. Some of the research directions essentially focus on: (1) Enhancing expertise in dealing with LLMs; (2) Developing network-focused embedding; (3) Improving the network planning and mitigation capabilities and capacities of LLMs; (4) Advancing risk assessment by integrating LLM-based qualitative reasoning with analytical risk models; and finally (5) Promoting toolbox abstraction and encouraging verifiable LLM-based tools by leveraging formal verification and consistency checks to ensure reliability in automated pipelines.

\subsubsection{Security enhancement}
\label{sec:cloud_sec_plan}
In the security context, root cause mitigation over the cloud focuses on preventing threats and minimizing vulnerabilities, while traditional mitigation focuses primarily on performance and reliability. The use of LLMs in this context brings novelty in terms of automation and policy updates over both the network and software levels, leveraging the advancements of large language understanding and solution planning. For example, LLM would be capable of generating security-based policy rules adjustments, such as a firewall or IAM policies, of blocking malicious activities \cite{dizdarevic2019survey}. At the network level, security mitigation entails identifying and isolating malicious traffic, enforcing zero-trust models, and preventing lateral movements \cite{dizdarevic2019survey}. In terms of software-level adjustments, security focuses on patching vulnerabilities, detecting backdoors, and restricting unauthorized access \cite{kumar2019cloud}. 

The work in \cite{rigaki2024hackphyr} introduced Hackphyr, a locally fine-tuned 7 billion parameter LLM designed as a red-team agent for network security. The authors created a novel cybersecurity dataset to enhance the model's capabilities, outperforming larger commercial models like GPT-4 and Q-learning agents in complex scenarios. Key contributions include Hackphyr's deployment in the NetSecGame environment, an ablation study analyzing the impact of dataset components on performance, and a detailed behavioral analysis of multiple LLM-based agents, providing insights into their decision-making and planning abilities in network security contexts. The NetSecGame environment is a network security simulation designed for both agents and humans to engage in attack-defense scenarios, with an interactive interface for human players and high-level actions for RL agents. It is highly configurable, allowing different network topologies, goals, and attack parameters. The environment’s state includes networks, hosts, services, and data, and the agent can perform actions like scanning, exploiting services, or exfiltrating data, with rewards assigned based on goal achievement or detection. The LLM-based agent in the game uses a profile, memory, and planning to generate actions, with its performance fine-tuned using a specialized dataset and supervised learning methods. In their experiments, GPT-4 consistently outperformed all other models across three data exfiltration scenarios. The Hackphyr agent, while not as strong as GPT-4, demonstrated robust performance, especially in the small and full scenarios, outperforming other LLM agents. However, all agents faced significant performance degradation in the more complex three subnets scenario, with GPT-4's win rate dropping by 20\%. The results highlight the challenges of adapting to new, unseen network configurations and the potential need for further model refinement.

There are still limited efforts in building LLM security managers for cloud networks and services. Thus, a promising avenue of research is ahead in this domain.

\subsection{LLM for Network Deployment and Distribution}
For deployment, LLMs focus on enabling IBN and resource management. They interpret high-level business objectives and translate them into network configurations that can be automatically implemented. Through intelligent resource allocation, LLMs ensure efficient distribution of network services across cloud and edge environments, reducing manual intervention and the risk of misconfigurations. By optimizing deployments, LLMs minimize downtime and ensure that network resources are allocated where they are most needed.

\subsubsection{Intent-based network deployment}
\begin{figure}
    \centering
    \includegraphics[width=\linewidth]{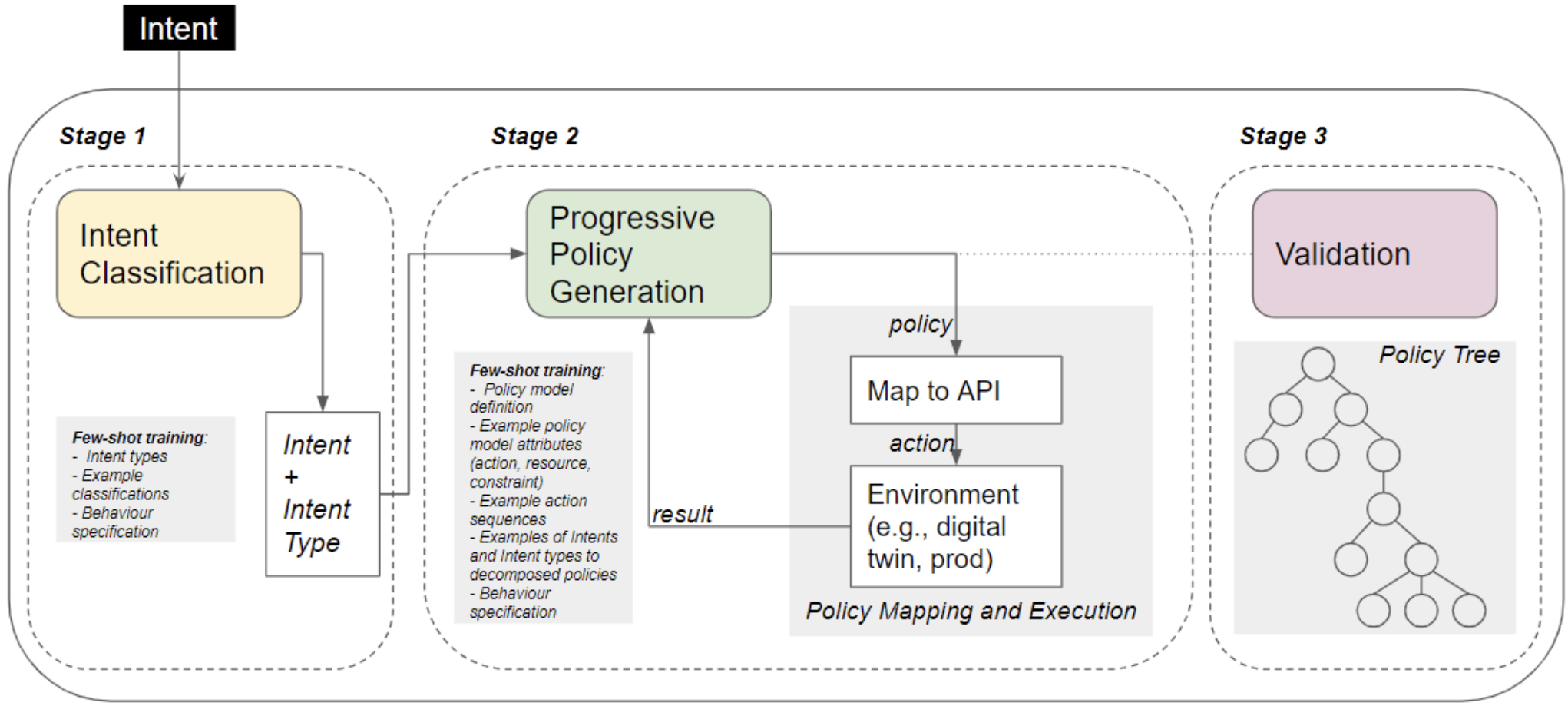}
    \caption{LLM Pipeline for IBN deployment \cite{dzeparoska2023llm}.}
    \label{fig:LLM_intent}
\end{figure}

In \cite{dzeparoska2023llm}, the authors proposed an Emergence system, which is an LLM-based pipeline for automated deployment of IBN to meet the user intent. Specifically, the pipeline is composed of three stages: intent classification, progressive policy generation, and validation, as shown in Fig. \ref{fig:LLM_intent}. 
Each stage involves training LLMs to classify intents, break them down into policy actions, and validate the policies for errors before deployment. The system uses a policy model abstraction to represent actions, constraints, and metadata, which are converted into APIs for execution.
The evaluation uses OpenAI's ChatGPT (GPT-3.5 and GPT-4) models trained with few-shot learning to classify, decompose, and validate intents into executable policies. The system was tested through intent fulfillment, where it generated 11 policies for deploying a service function chain, and intent assurance, where it successfully handles failure scenarios by generating corrective actions. Results showed that the system performs efficiently, with negligible response time from the ChatGPT API, and demonstrates strong generalization capabilities, effectively handling both familiar and unseen intent requirements.


While LLMs demonstrate effective intent decomposition, challenges remain in validating policies, ensuring proper sequence and logic, and handling context length limitations in transformer models. Future avenues of research can focus on improving validation through enhanced logic checks, such as utilizing Finite State Machines (FSMs) and overcoming context constraints with longer token limits in newer models. Additionally, further improvements are needed in policy generation to avoid unintended deviations from the original intent, potentially by embedding more detailed reasoning algorithms in the training process. Finally, a widely used benchmarking dataset is essential to enable larger research communities to expand and develop novel solutions addressing these challenges.

\subsubsection{Dynamic service placement}
The service placement problem involves optimally assigning services to network nodes to minimize latency, maximize resource utilization, and meet QoS requirements \cite{sami2020dynamic}. Existing solutions, such as heuristics and RL \cite{sami2021demand, sami2021ai}, often struggle with dynamic network conditions, high-dimensional state speed, and most importantly the increased learning time (primarily in AI-based solutions). RL models require extensive training and fail to adapt quickly, while heuristics lack scalability and flexibility. To this end, LLMs can overcome these limitations by analyzing complex, real-time network traffic, predicting service demands, and generating adaptive placement strategies. Their ability to process unstructured data, reason contextually, and integrate with other AI models (such as RL) makes them ideal for dynamic and effective service placement.

The authors in \cite{rao2024eco} introduced ECO-LLM, a middleware leveraging LLMs like ChatGPT for dynamic microservice placement in edge-cloud infrastructures. Its main contribution is enabling natural language-based customization and automatic code generation for task placement, eliminating manual effort and overcoming the solution of existing AI-based techniques, such as time-series and RL. ECO-LLM optimizes workloads for latency and cost by adapting to changes in application demand. Experiments with video analytics applications (face recognition, human attribute detection, license plate recognition) showed that ECO-LLM matches manual baselines in detecting workload thresholds while differing by only 1.45\% or less in placement accuracy across two days. 

The main technical limitations of using LLMs for service placement in cloud networks include challenges with scalability, real-time adaptability, and integration with existing cloud management systems. While LLMs can process large datasets, their performance degrades in dynamic environments with fluctuating workloads and strict resource constraints. They also lack mechanisms for real-time feedback, making them less effective for immediate decision-making. LLMs offer a promising solution but require further advancements to handle these limitations, particularly in real-time service placement, seamless integration, and resource optimization in large-scale systems.

\subsubsection{Adaptive load balancing}
Load balancing ensures even distribution of network traffic and application workloads across multiple servers to prevent overloading, optimize resource usage, and maintain service availability \cite{kumar2019issues}. It involves tasks like traffic routing, resource allocation, and real-time adjustment to handle dynamic workloads and failures. In this context, LLMs can analyze large-scale traffic data in real-time, predict traffic patterns, and identify bottlenecks. They can automate dynamic decision-making by generating efficient load distribution strategies, forecasting service demands, and adapting to anomalies like server failures. This improves scalability, reduces downtime, and optimizes resource utilization with minimal manual intervention.

McAuley \textit{et al.} \cite{mcauley2023adaptive} presented a novel integration of LLMs with adaptive load balancing in cloud networks, enabling dynamic resource allocation and efficient workload distribution. LLMs enhance real-time data processing, predictive analytics, and intelligent decision-making, addressing the limitations of traditional load-balancing methods like static configurations and limited metrics. This approach improves performance, scalability, and resource utilization, offering a transformative solution for managing modern cloud infrastructures.

Current literature on adaptive load balancing in the cloud often relies on rule-based or heuristic methods, which struggle to handle complex, dynamic workloads and unforeseen traffic patterns in real time. These approaches lack the ability to generalize across diverse environments or learn from historical data effectively. Additionally, limited integration of AI, particularly LLMs, means existing systems are often reactive rather than proactive. Future research should focus on leveraging LLMs to process real-time traffic, predict workload trends, and generate adaptive strategies dynamically. This requires developing frameworks that combine LLMs with reinforcement learning for continuous optimization while addressing challenges like model interpretability, latency, and energy efficiency.

\subsection{LLM for Continuous Network Support}
\label{sec:cloud_continuous_support}
In continuous support, LLMs facilitate code repair and ongoing network support through AI-driven automation. They identify network issues in real-time, proactively fixing code and network configurations to prevent failures. This includes automatically patching vulnerabilities and performing updates to keep the network functioning smoothly. Through continuous learning, LLMs evolve with the network, ensuring that support mechanisms remain effective as traffic patterns, hardware, and software evolve, allowing for seamless, self-sustaining network operations.

\subsubsection{Network topology design and optimization}
\begin{figure}
    \centering
    \includegraphics[width=\linewidth]{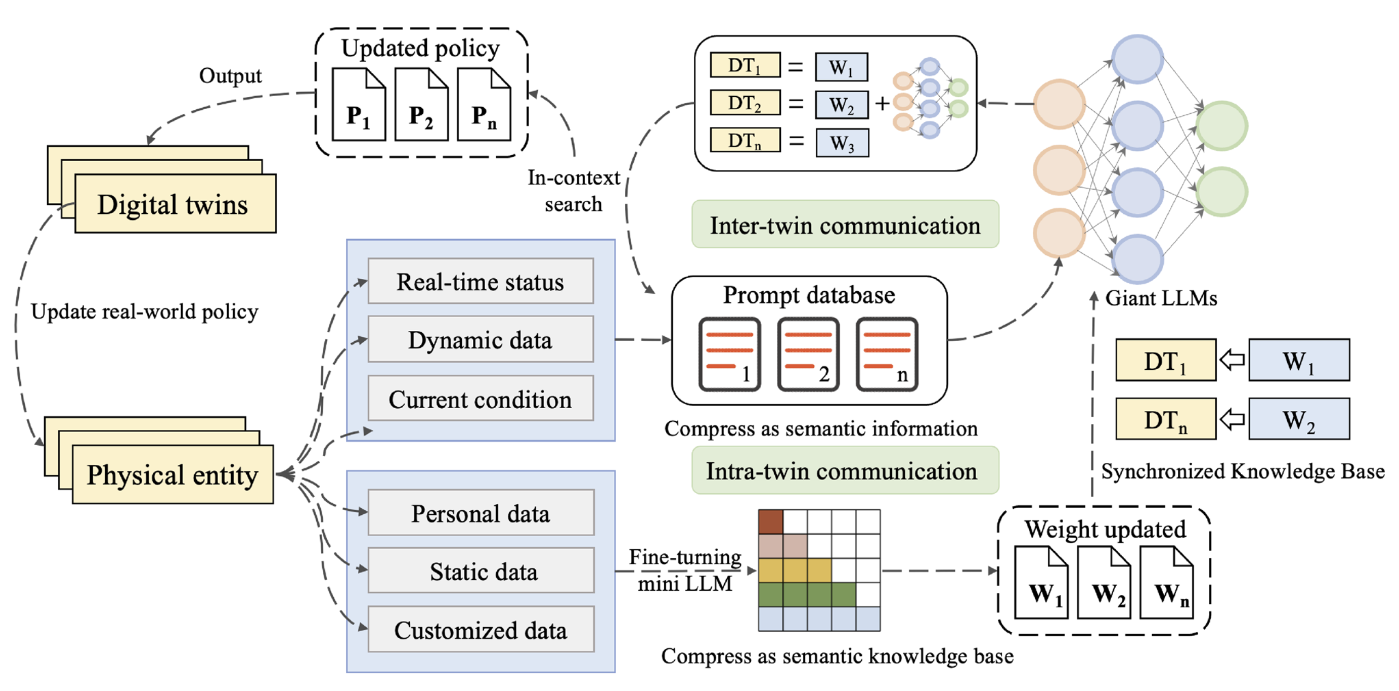}
    \caption{The LLM-Twin framework \cite{hong2024llm}.}
    \label{fig:enter-label}
\end{figure}

The authors in \cite{hong2024llm} proposed an innovative concept for using LLMs to manage and configure the cloud network on-the-fly based on variations in user demands, indicating continuous adaptability and support, illustrated in Fig. \ref{fig:enter-label}. LLM-Twin framework entails a sophisticated representation of the real network through a semantic representation. This semantic network essentially captures the intricate network component relations, connection protocols, and performance metrics, allowing LLM to mimic the network knowledge and behavior. Therefore, this network would enable detailed analysis and optimization, allowing the LLM-Twin framework to properly understand and generate solutions to complex queries, empowering advanced troubleshooting and complex network configuration decisions. 
Experimental studies to evaluate LLM-Twin performance evolve around the performance of the model in optimizing communication and computation in Digital Twin Networks (DTNs). The objective was to compare the performance of LLM-Twin against traditional Federated Learning (FL)-based DTNs. As part of the experimental setup, the framework was deployed on a high-performance server and used the LLAMA 7B model for fine-tuning. A case study of a smart home DTN was used, where data from physical entities was used to generate prompts for the LLM. The evaluation metrics include computational consumption, communication efficiency, and response times. The results illustrated that LLM-Twin outperforms FL-based DTNs in terms of computation time, requiring less processing power and better communication efficiency. 

The LLM-Twin idea proposed in \cite{hong2024llm} summarizes the state of different existing approaches for continuous network configuration using LLMs over the cloud. The main limitation that still persists in such academic efforts evolves around the need for continuously monitoring the LLM performance while considering the need to constantly improve the fine-tuning and prompt engineering process to account for enough knowledge for the model to adapt. Furthermore, the complex nature of LLMs might result in hallucinations, especially when the complexity and variation in context increases. Additionally, the increased computation demand for loading such models in memory and preparing fine-tuning is immense, thus quantization techniques need to be studied to bring processing close to end users and ensure increased security and privacy. 

\subsubsection{Infrastructure-as-code reparation \& security mitigations}
Automated cloud resource management is enabled via Infrastructure-as-Code (IaC) \cite{rahman2019systematic} solutions, but misconfigurations in IaC can lead to security vulnerabilities. Existing IaC scanning tools identify these issues but require manual intervention for repairs. LLMs, like GPT-4, can automate the repair process by generating corrected IaC code after misconfigurations are detected, utilizing contextual input and previous examples of proper code \cite{srivatsa2024survey}. This can reduce manual effort, speed up remediation, and ensure better resource allocation. However, challenges such as LLM-generated hallucinations—incorrect fixes that pass initial checks but fail deeper validation—remain. To address this, a two-pass approach can refine the LLM's suggestions, though manual verification still plays a critical role.

\begin{table*}[!t]
\scriptsize 
\centering
\caption{Summary of Related Works on LLM for Cloud-based NSM}
\begin{tabular}{|p{1.3cm}|p{2.3cm}|p{1.5cm}|p{3.2cm}|p{2cm}|p{1.4cm}|p{1.6cm}|}
\hline
\textbf{Taxonomy} & \textbf{NSM Task} & \textbf{Authors} & \textbf{Main Contributions} & \textbf{LLM Solution} & \textbf{Dataset Used} & \textbf{Evaluation Metrics} \\
\hline
\multirow{8}{*}{\parbox{1.5cm}{Network Monitoring \& Reporting}}
 & \multirow{4}{*}{\parbox{1.8cm}{Root cause analysis}} & Geol \textit{et al.} (2024) \cite{goel2024x} & Contextual information of the software development lifecycle to LLM & GPT-4 & Private by Microsoft & BLEU, METEOR, ROUGE, Bert Score, and NUBIA\\
\cline{3-7}
& & Zhang \textit{et al.} (2024) \cite{zhang2024lm} & The mitigation of the hallucination problem in LLM (LM-PACE) & GPT-4, Text-DaVinci-003, GPT-3.5-Turbo& Private by Microsoft& Reliability diagrams, ECE scores\\
\cline{3-7}
&& Roy \textit{et al.} (2024) \cite{roy2024exploring} & Chain-of-thoughts and RAG with contextual information passed to ReAct LLM & GPT-4 & Private data of incidents & C-BLEU, S-BLUE, ROUGEL, METEOR, BertS\\
\cline{2-7}
& \multirow{3}{*}{Anomaly detection} & Yu \textit{et al.} (2024) \cite{yu2024monitorassistant} & MonitorAssistant framework for model configuration and knowledge inheritance & GPT-4 & - & - \\
\cline{3-7}
& & Patil \textit{et al.} (2024) \cite{patil2024leveraging} & LLM fine-tuning on network traffic logs, security alerts, and incident reports & GPT-3.x & -& -\\
\cline{3-7}
&& Ott \textit{et al.} (2021) \cite{ott2021robust} & The use of sentence-level embeddings to identify ground truth and synthetic anomalies & GPT-2, XL, BERT & CloudLab OpenStack log dataset & precision, recall, f1-score\\
\cline{2-7}
& Vulnerability Detection \& Analysis & Cao \textit{et al.} (2024) \cite{cao2024llm}& Detection and analysis agents to automate vulnerability detection using RAG and CWE data & GPT-4 & Juliet and D2A datasets & Accuracy, recall, and f1-score \\
\hline
\multirow{5}{*}{\parbox{1.5cm}{AI-Powered Network Planning}} & \multirow{3}{*}{\parbox{1.8cm}{Intent-based network management}} & Mekrache \textit{et al.} (2024) \cite{mekrache2024intent} & Few-shot LLM learning for intent decomposition and translation through contextual data & LLama 13B, Mistral 7B, Llama 13B & private data & Rating score and creation time\\
\cline{3-7}
&& Panchal \textit{et al.} (2024) \cite{panchal2024simplifying} & Automated network services orchestration & - & - & - \\
\cline{3-7}
&& Dandoush \textit{et al.} (2024) \cite{dandoush2024large} & Dynamic network slicing in cloud environments & - & - & -\\
\cline{2-7}
& \multirow{2}{*}{\parbox{1.8cm}{Cloud incident mitigation}} & Ahmed \textit{et al.} (2023) \cite{ahmed2023recommending} & Automated mitigation of cloud incidents through LLM fine-tuning & RoBERTa, CodeBERT, GPT-3.x & Private by Microsoft & BLEU-4, ROUGE-L, METEOR, BERTScore, BLEURT, and NUBIA\\
\cline{3-7}
&& Hamadanian \textit{et al.} (2023) \cite{hamadanian2023holistic} & LLM for automated incident response through hypothesis generation, test, and reassessment & - & - & - \\
\cline{2-7}
& Security enhancement & Rigaki \textit{et al.} (2024) \cite{rigaki2024hackphyr} & A Fine-tuned 7 billion parameter LLM as a red-team agent for network security & GPT-4, GPT-3.5-Turbo & Locally created in NetSecGame environment & Win rate and reward\\
\hline
\multirow{3}{*}{\parbox{1.5cm}{Network Deployment and Distribution}} & \multirow{1}{*}{\parbox{1.8cm}{Intent-based network deployment}} & Dzeparoska \textit{et al.} (2023) \cite{dzeparoska2023llm} & An LLM-based pipeline for automated deployment of intent-based network & GPT-3.5, GPT-4 & A scenario using OpenStack in the SAVI testbed & Execution time \& number of policies\\ 
\cline{2-7}
& \multirow{1}{*}{\parbox{1.8cm}{Dynamic service placement}} & Rao \textit{et al.} (2024) \cite{rao2024eco} & LLM-based  code generation for automated task placement & GPT-4 & Video Analytics dataset & Total placement decisions \& total incorrect placements\\
\cline{2-7}
& \multirow{1}{*}{\parbox{1.8cm}{Adaptive load-balancing}} & Mcauley \textit{et al.} (2023) \cite{mcauley2023adaptive} & LLM for traffic routing, resource allocation, and real-time workload distribution & - & - & -\\
\hline
\multirow{5}{*}{\parbox{1.5cm}{Continuous Network Support}} & \multirow{1}{*}{\parbox{1.8cm}{Network topology design \& Optimization}} & Hong \textit{et al.} (2024) \cite{hong2024llm} & The use of LLM for on-the-fly cloud network configuration based on user demands & LLAMA 7B & Simulated smart home digital twin environment & Computational consumption, communication efficiency, and response times\\ 
\cline{2-7}
& \multirow{2}{*}{\parbox{1.8cm}{Infrastructure-as-code reparation \& security}} & Low \textit{et al.} (2024) \cite{low2024repairing} & The use of security scanners to identify misconfigurations followed by LLM fixed and prompt engineering & GPT-3.5 \& GPT-4 & Terragoat \& KaiMonkey & Fix rate\\
\cline{3-7}
&& Sarda \textit{et al.} (2023) \cite{sarda2023adarma}& ADARMA: a pipeline to address runtime anomaly detection and auto-remediation in microservices architectures using LLMs& - & Simulated environment using Chaos Mesh and Robot-shop in Kubernetes environment & F1-score, r2-score, MSE, precision, recall, functional and average correctness\\
\hline

\end{tabular}
\label{tab:cloudnetwork}
\end{table*}

In \cite{low2024repairing}, researchers explored the use of LLMs, specifically GPT-3.5 and GPT-4, to repair IaC in HashiCorp Configuration Language (HCL) for Terraform. The methodology starts with using security scanners to identify misconfigurations in the code, followed by dividing the IaC code into smaller blocks. Each block is associated with misconfiguration information, which is then input into the LLM via a structured prompt. The prompt design is critical for the LLM’s success, providing detailed context about the code and the specific misconfiguration, as well as additional human-provided context when needed. After the first pass of repairs, unresolved issues are flagged, and the second pass allows the LLM to request further details from the developer to refine the code. The experiments used vulnerable Terraform codebases, Terragoat, and KaiMonkey, and showed that GPT-4 outperformed GPT-3.5, fixing 18-34\% more misconfigurations in the first pass, with even greater improvements in the second pass (up to 57\%). Despite the promising results, some fixes were invalid or hallucinated, accounting for 20.4\% of the outputs, emphasizing the need for further validation. The paper also highlights that while LLMs provide significant assistance in IaC repair, challenges like hallucinations and incomplete fixes persist. Future work should focus on refining the prompt design, reducing hallucinations, and improving the overall robustness of LLM-assisted IaC repair tools.

\begin{figure}
    \centering
    \includegraphics[width=.72\linewidth]{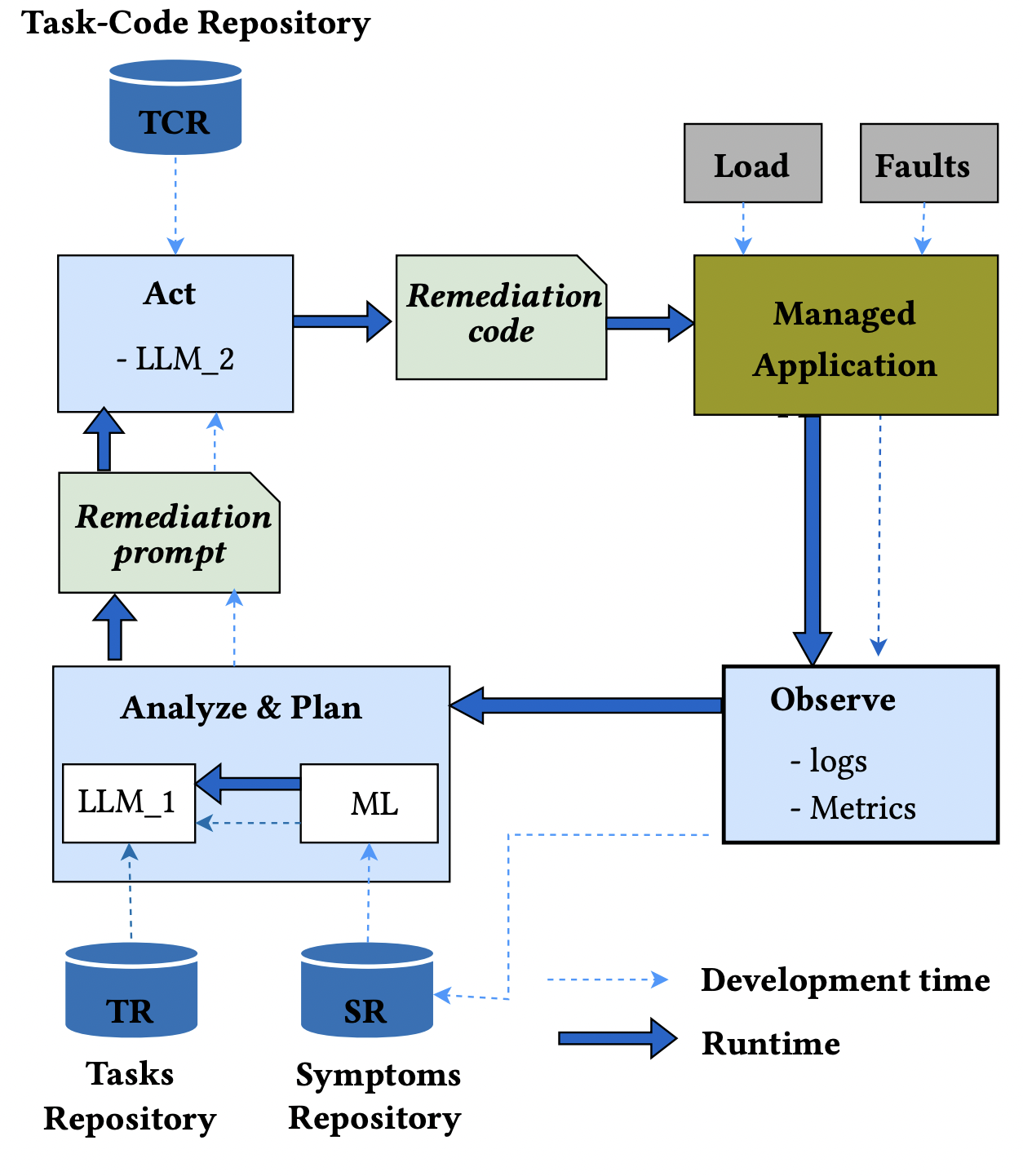}
    \caption{Overview of ADARMA framework \cite{sarda2023adarma}.}
    \label{fig:adarma}
\end{figure}

In the same context, Sarda \textit{et al.} \cite{sarda2023adarma} proposed ADARMA, a pipeline to address runtime anomaly detection and auto-remediation in microservice architectures using LLMs. The proposed system first detects anomalies in microservices using monitoring tools and then utilizes LLMs to identify the root cause and generate countermeasures. 
The novelty of the proposed work centers around the use of prompt engineering to tailor the LLM's output, which is fine-tuned with few-shot learning to improve the generation of actionable countermeasures. This process helps in continually improving the LLM process when new contexts are introduced to the model, thus the main reason this work is categorized under the use of LLM for continuous network support. Fig. \ref{fig:adarma} presents the proposed architecture in \cite{sarda2023adarma}. 
The environment setup is described as an injection of anomalies in Kubernetes environments. Functional correctness and average correctness metrics are used to measure the LLM's performance for evaluating the generated code. Preliminary results showed that the LLM-generated playbooks, when fine-tuned, achieve a 95.83\% correctness rate in automating remediation tasks, outperforming zero-shot or one-shot learning approaches. The system utilized classification, regression, and LLM models to achieve a 96\% accuracy in detecting anomalies and performing root-cause analysis. 

Existing LLM approaches for IaC repair address challenges like refining prompt engineering for accurate code generation and ensuring the correctness of fixes in complex environments. Additionally, scalability issues arise when handling large or diverse IaC configurations. Future work should focus on expanding datasets for fine-tuning, improving model robustness, and addressing the limitations of fully autonomous remediation to enhance practical applicability in real-world scenarios, especially when the provided context continuously evolves over time.

Table \ref{tab:cloudnetwork} provides a summary of related works on LLM for cloud-based NSM.



\section{LLM for Fog/Edge-based NSM}\label{sec:fogedge}
Fog and edge networks are becoming increasingly crucial as the demand for low-latency and high-efficiency data processing grows. These decentralized systems bring computation and data storage closer to users, improving responsiveness while easing the load on central servers. LLMs are emerging as powerful tools for optimizing both fog and edge networks, offering solutions for various challenges such as resource allocation, security, and fault management, as illustrated in Figure \ref{fog}. 
By leveraging LLMs, these networks can enhance performance, streamline operations, and ensure robust security, ultimately improving user experiences in a data-driven world. 
The upcoming subsections analyze the literature regarding the use of LLM for managing fog and edge networks, focusing on network monitoring and reporting, AI-powered network planning, network deployment and distribution, and continuous network support, as summarized in Table \ref{tab:edgetaxonomy}.


\subsection{LLM for Network Monitoring and Reporting}
Monitoring and reporting in edge and fog networks involves continuously tracking the health and performance of devices, as well as overseeing the availability of resources and the execution of services. It also includes real-time detection of potential issues, along with reporting these issues, identifying their causes, and analyzing resource consumption. However, the inherent complexity and diversity of networking tasks make effective monitoring and reporting a challenging endeavor. In this context, LLMs have been proposed as promising solutions for the efficient management and optimization of network systems, particularly in the dynamic and distributed environments of edge and fog networks. 

\subsubsection{LLM for network optimization}
LLMs have emerged as a promising solution to optimizing network performance, especially within edge computing environments. For example, the authors in \cite{wu2024netllm} introduced the NetLLM model to address various networking challenges, focusing on three specific tasks: viewport prediction, adaptive bitrate streaming, and cluster job scheduling. NetLLM, built on the Llama2-7B architecture, demonstrated superior performance and robust generalization capabilities, effectively processing multimodal data through an encoder module tailored to different tasks. This approach not only generates task-specific responses but also reduces the costs associated with fine-tuning LLMs and mitigates the issue of LLM hallucination by employing various networking heads, thereby facilitating output generation with fewer iterations.
Another case of edge network optimization is presented in \cite{rao2024eco}, where the authors proposed an LLM-based solution called ECO-LLM. This solution leverages GAI techniques to automatically adjust the placement of microservices, either at the edge or in the cloud, based on factors such as latency, workload, and task performance. This dynamic adjustment reduces operational costs and minimizes the need for manual configuration interventions.

\subsubsection{Security of LLM in Edge Networks}
Another concern when monitoring networks is the security of deployed LLMs. Deploying LLMs on edge devices is preferable because it improves data privacy by keeping sensitive information on the device. It is also important to hide the architecture of LLM to protect its intellectual property and prevent misuse. Fingerprinting methods are often used to safeguard AI models. The authors in \cite{nazari2024llm} proposed a non-intrusive fingerprinting technique that does not require physical access to the device for data collection. This method gathers detailed system information while the LLM is running and uses a trained classifier to accurately identify the LLM's architecture, focusing on RAM usage as a key feature.
Sarbi \textit{et al.} \cite{sarabi2023llm} proposed an LLM framework that directly learns from IoT raw text data to generate embeddings that characterize networked devices and their underlying hardware/software. With the help of their framework, the nodes can make sense of raw Internet scan information. The framework is also open-source and available on GitHub.

\subsection{LLM for AI-powered Network Planning}
Network planning for edge networks focuses on designing systems for efficient data processing to enhance performance and overall QoS. In this subsection, we review the literature on the application of LLMs in tackling the challenges of network planning in edge environments, specifically regarding optimized solutions for LLM deployment and data generation.

\begin{figure*}
\centerline{\includegraphics[width=7.2in]{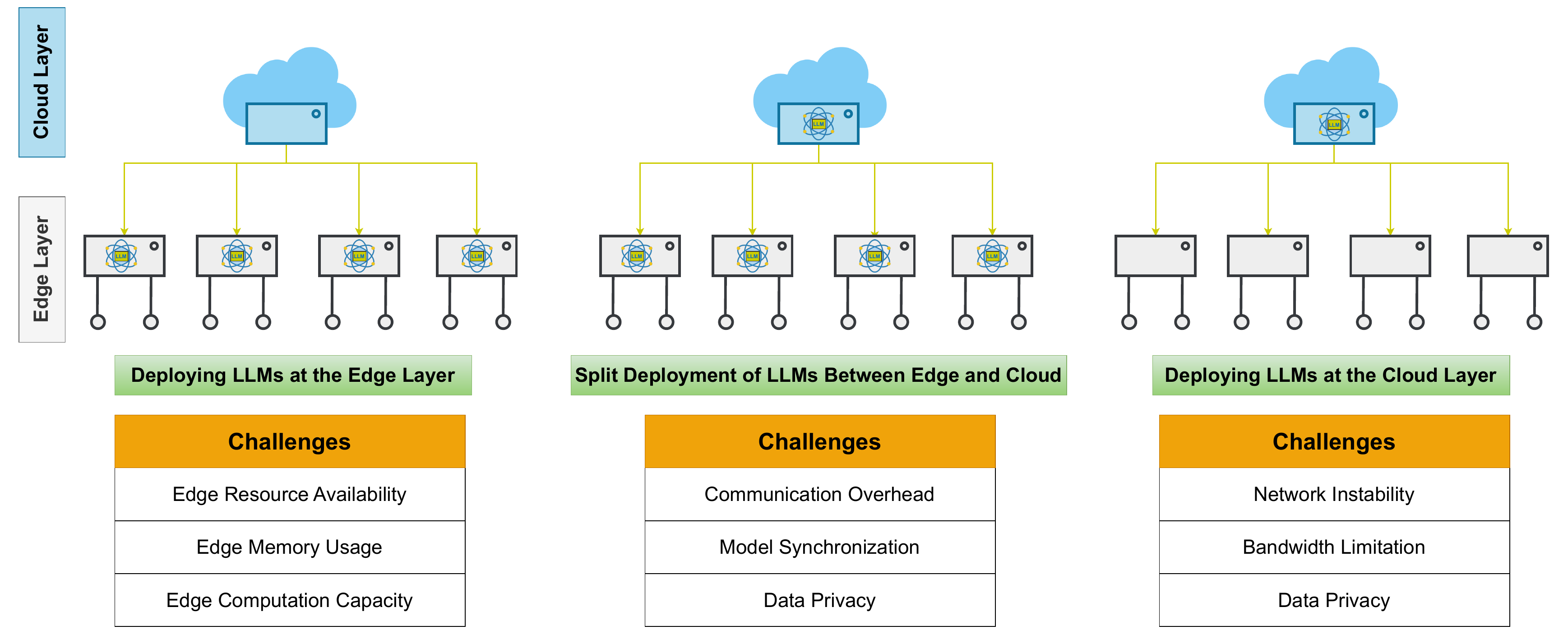}}
\caption{LLM deployment in edge networks.}
\label{edge_llm_deploy}
\end{figure*}
\subsubsection{Optimization of LLM deployment}
Deploying LLMs in resource-constrained environments, such as edge networks, presents a persistent challenge. Fig. \ref{edge_llm_deploy} illustrates different scenarios for deploying LLM in edge networks, namely: deploying LLM at the edge layer, deploying LLM at the cloud layer, and splitting the LLM between the edge and cloud layers. Each approach comes with its own set of challenges. In the following, we discuss some proposed solutions for each approach.

Shifting the processing of LLM tasks from centralized servers to edge devices is a key strategy to address challenges posed by the increasing number of user queries, which can affect QoS. However, certain constraints, such as limited memory and computational resources, hinder this transition to edge device deployment.
The authors in \cite{dhar2024empirical} addressed these limitations by evaluating the LLaMa-2 7B model on various edge devices using INT4 quantization and proposed several solutions. From a hardware standpoint, they suggested adding extra memory and resources and ensuring sufficient memory bandwidth. From a software perspective, they recommended enhancing memory efficiency and optimizing resource utilization through techniques such as model parallelism, sparse weight matrix encoding, shared storage, and synchronization with neighboring devices. 
In a similar context regarding the optimization of LLM deployment, the authors in \cite{zhang2024edge} suggested enhancing inference throughput through model quantization.  This technique involves storing the weights and activations of LLMs in lower bit precision, which reduces memory consumption on edge servers. Furthermore, they recommended a batching scheduling strategy to enable the simultaneous processing of requests from multiple users.
Instead of significantly increasing resources for LLM deployment, the authors in \cite{yao2024gkt} suggested a framework that focuses on knowledge transfer to use smaller models on edge devices. In this setup, an LLM acts as a teacher, generating guidance prompts to help a smaller language model, which serves as the student and completes responses based on these prompts. This approach uses the strengths of larger models to improve the efficiency and performance of smaller ones, allowing for quicker responses without needing the two models to have the same vocabulary.
Another strategy to enhance the deployment of LLMs and improve QoS is the implementation of vector database caching to store LLM request results at edge servers \cite{yao2024velo}. In this approach, the database maintains a cache of results from specific LLM requests to optimize the QoS in a cloud-edge collaborative environment. When a new request is received, a framework utilizing MARL determines whether to obtain the response from the cloud or to return it directly from the cached data at the edge. This method effectively reduces both response times and costs related to similar requests.
However, this cloud-edge collaborative approach may face issues due to unstable network connections. To address this challenge, the authors in \cite{zhang2024edgeshard} proposed a resource allocation strategy that employs collaborative edge computing by distributing LLM model shards across edge devices and cloud servers, taking into account the limitation of resources of edge devices. The proposed approach consists of three main steps: Profiling, which assesses the capabilities of both the devices and the LLM model; Inference Task Scheduling, which mathematically determines which devices should be involved and how to partition the LLM model; and Collaborative Inference, which carries out inference through sequential and pipeline parallel processing.
Likewise, the authors in \cite{khoshsirat2024decentralized} proposed a collaborative framework for distributing LLM layers among devices to enable parallel inference, utilizing a scheduling mechanism to select devices. The key difference from earlier work is their focus on incorporating renewable energy, connecting the devices to sustainable energy sources to support reliable and continuous inference. This approach not only makes the solution environmentally friendly but also fosters sustainable LLM inference.

\subsubsection{Data generation}
\begin{figure}
\centerline{\includegraphics[width=\linewidth]{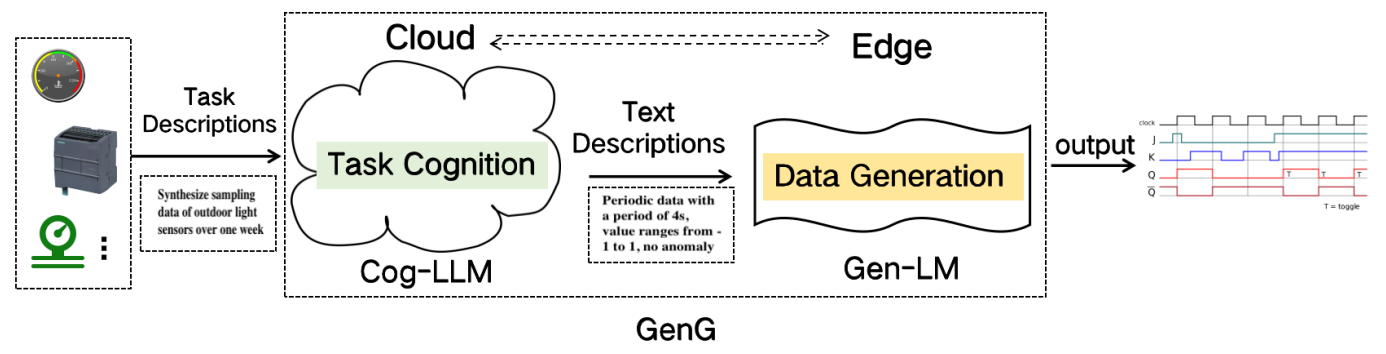}}
\caption{Overview of GenG framework \cite{zhou2024geng}.}
\label{geng}
\end{figure}

LLMs provide a promising solution for data generation, effectively addressing the challenges related to the high costs of obtaining high-quality real-world data, thus mitigating the problem of insufficient training data. The authors in \cite{zhou2024geng} proposed a two-step approach that integrates LLMs and Transformers to generate time-series data. As shown in Fig. \ref{geng}, the first component, Cog-LLM, employs an iterative self-training approach to fine-tune the LLM model by self-selecting training data from a substantial collection of unlabeled documents. The second component, Gen-LM, generates time series data based on the outputs of Cog-LLM. The connection between these components is supported by a Diffusion Model, emphasizing the significance of transfer learning within this framework.

\subsection{LLM for Network Deployment and Distribution}
Edge network and fog network deployment and distribution aim to position computational resources closer to end users to reduce latency, ensure data privacy, and optimize QoS. As these networks become increasingly critical for applications like IoT and autonomous vehicle systems, the need for efficient management of network configurations grows. In this context, LLMs have emerged as a promising solution to optimize the processes involved in network deployment and configuration management, offering enhanced automation, adaptability, and efficiency.

\begin{table*}[!t]
\scriptsize 
\centering
\caption{Summary of Related Works on LLM for Fog/Edge-Based NSM}
\begin{tabular}{|p{1.5cm}|p{1.8cm}|p{1.5cm}|p{2.0cm}|p{2.0cm}|p{1.4cm}|p{1.4cm}|p{2.7cm}|}
\hline
\textbf{Taxonomy} & \textbf{NSM Task} & \textbf{Authors} & \textbf{Main Contributions} & \textbf{Problem Addressed} & \textbf{LLM Solution} & \textbf{Dataset Used} & \textbf{Evaluation Metrics} \\
\hline
\multirow{4}{*}{\parbox{1.5cm}{Network\\ Monitoring\\ and Reporting}} 
& \multirow{2}{*}{\parbox{2.0cm}{LLM for Network\\ Optimization}} 
& Wu et al. (2024) \cite{wu2024netllm} & LLM for efficient networking tasks & Modality gap, delays, adaptation costs & Llama2-7B & Multimodal data & MAE, QoE, Job time \\ \cline{3-8}
& & Rao et al. (2024) \cite{rao2024eco} & Microservice placement optimization & Latency and placement issues & ChatGPT & Public traffic datasets & Placement decisions, Incorrect placements, Latency \\ \cline{2-8}
& \multirow{2}{*}{\parbox{2.6cm}{Security of LLM in\\ Edge Networks}} 
& Nazari et al. (2024) \cite{nazari2024llm} & Protect LLMs using fingerprinting & Security risks for LLMs & MiniLM, T5, RoBERTa & Tegrastats logs & Accuracy \\ \hline
\multirow{5}{*}{\parbox{1.5cm}{AI-powered Network Planning}} & \multirow{3}{*}{\parbox{1.8cm}{Optimization of LLM Deployment}} & Dhar \textit{et al.} (2024) \cite{dhar2024empirical} & LLM optimization for edge devices & Resource, memory constraints & LLama-2 7B & CPU, memory usage data & Latency, CPU \& memory usage \\ \cline{3-8}
& & Zhang \textit{et al.} (2024) \cite{zhang2024edge} & Collaborative edge LLM resource allocation & Privacy, latency, cloud
reliance & BLOOM variants, OPT-13B & Simulation of user request arrivals & Perplexity differential, Complexity reduction \\ \cline{3-8}
& & Yao \textit{et al.} (2024) \cite{yao2024gkt} & Knowledge transfer for efficient LLMs & Requirement of extensive resources and fine-tuning & Flan, Bloom, Llama2 variants & GSM8K, CSQA & Accuracy, running time \\ \cline{3-8}

& & Yao \textit{et al.} (2024) \cite{yao2024velo} & LLM caching for optimized QoS & Delays and high costs associated with cloud-based LLMs & Greedy variants & OASST1, Generated data using Qwen14b & Weight Ratio, Reward Value,
Request Completion \\ \cline{3-8}

& & Zhang \textit{et al.} (2024) \cite{zhang2024edgeshard} & LLM partitioning boosts inference efficiency & Latency, bandwidth costs, and privacy concerns & Llama2 variants & WikiText-2 dataset &  Latency, Throughput \\ \cline{3-8}

& & Khoshsirat \textit{et al.} (2024) \cite{khoshsirat2024decentralized} & Collaborative and energy-efficient LLM inference& LLM deployment on edges & Customized LLM & - & Energy Consumption, Processing Time, Job Throughput\\

\cline{2-8}
& \multirow{2}{*}{\parbox{1.8cm}{Data Generation}} & Zhou \textit{et al.} (2024) \cite{zhou2024geng} & LLM-based time-series generation framework & High costs of obtaining high-quality real-world data & LLaMA -7B, BERT & English portion of the Clueweb corpus  & GPT score, Human score, Fidelity, Usefulness \\ 
\hline
\multirow{9}{*}{\parbox{1.5cm}{Network Deployment and Distribution}} & \multirow{3}{*}{\parbox{1.8cm}{LLM for Network Deployment}} & Mekrache \textit{et al.} (2024) \cite{mekrache2024llm} & LLM-based intent translation for networking & Challenges in defining IBN intents &  Mistral 7B, LLaMA variants & EURECOM 5G facility, Generated NSD data & Cosine similarity, User Feedback \\ \cline{3-8}

& & Kwon \textit{et al.} (2023) \cite{kwon2023exploring} & LLMs automate edge-cloud infrastructure management & Automating infrastructure deployment & ChatGPT, Bard & 58 cases related to Ansible Playbook Repair & objective and subjective evaluation metrics \\ \cline{3-8}

& & Feng \textit{et al.} (2024) \cite{feng2024optimizing} & LLMs optimize microservice deployment & Microservice deployment latency & ChatGPT gpt-3.5-turbo-16k & Question-answer pairs & Code Correctness, Logical Soundness, Completeness, Time Complexity, Cyclomatic Complexity \\ 
\cline{2-8}
& \multirow{4}{*}{\parbox{1.8cm}{LLM for Model Interpretability in Edge Networks}} & Wang \textit{et al.} (2024) \cite{wang2024characterizing} & LLM generates explanations for model differences & Understanding differences between edge and base models & GPT-4o & ImageNet1K, CIFAR10, MIT Indoor Scenes, COCO& Accuracy, Size, FLOPS \\ 
\hline
\multirow{5}{*}{\parbox{1.5cm}{Continuous Network Support}} & \multirow{3}{*}{\parbox{1.8cm}{LLM Optimization}} & Yu \textit{et al.} (2024) \cite{yu2024edge} & LLM for  memory optimization, redundancy reduction & 
Computation overhead, memory usage, scheduling  &  LLaMA-7B & MMLU dataset
WikiText-2 dataset & Accuracy, layer-wise unified compression perplexity  \\ \cline{3-8}
& & Jin \textit{et al.} (2024) \cite{jin2024llm} & Knowledge pruning for LLM compression & LLM redundancy and computational cost & CLIP & UCI\_HAR, Opportunity, PAMAP2, WISDM, C-MAPSS & F1-score, RMSE, scoring function \\ \cline{3-8}
& & Zagar \textit{et al.} (2024) \cite{zagar2024dynamic} & Dynamic inference for decentralized LLM execution & Data privacy, trust, and cloud of LLMs costs hinder their adoption & Llama2, Gemma, Phi-2 & - & Inference speed, time to first token \\ 
\cline{2-8}
& \multirow{2}{*}{\parbox{1.8cm}{Edge Computing Security}} & Hasan \textit{et al.} (2024) \cite{hasan2024distributed} & Scalable, real-time cybersecurity for edge & Increased devices complicate real-time threat detection & GPT & Popular threat libraries & Accuracy \\ 
\cline{2-8}
& \multirow{2}{*}{\parbox{1.8cm}{Fault Tolerance Management}} & Fang \textit{et al.} (2024) \cite{fang2024large} & Efficient fault diagnosis based on LLM & Challenge of optimizing network fault tolerance & GPT4, ChatGLM3, LLaMA & Generated random traffic flows  & Transmission speed, Network throughput \\

\hline
\end{tabular}
\label{tab:edgetaxonomy}
\end{table*}
\subsubsection{LLM for network deployment}
Deploying and managing network configurations is essential for ensuring the seamless performance of networks, and LLMs can greatly enhance the efficiency and effectiveness of these processes. One key aspect of network deployment involves converting user intents into structured formats. The authors in \cite{mekrache2024llm} proposed an LLM-based intent translation system that enables users to articulate their intents in natural language, which the system then converts into Network Service Descriptors (NSDs). This approach helps users overcome the challenges of defining intents for IBN without the need for formal technical models, which often require significant expertise and effort. 
Another important application of LLMs in edge-cloud systems is the automation of infrastructure deployment and management. For instance, in \cite{kwon2023exploring}, the authors presented a solution based on LLM architectures like ChatGPT to enhance the quality of Ansible scripts used for deploying and managing infrastructure in Edge-Cloud systems by automatically suggesting code fixes for identified bugs.
Similarly, in \cite{feng2024optimizing}, the authors introduced the use of LLMs in conjunction with a RAG database and CoT techniques as a decision-making tool for microservice deployment in edge networks. Their approach aims to optimize microservice deployment latency by automatically generating and refining code to address challenges such as latency, complex request handling, and network node scaling.
\subsubsection{LLM for model interpretability in edge networks}
Model interpretability is crucial for optimizing DL models in edge networks and ensuring their effective deployment, with LLMs playing a significant role in enhancing this interpretability process.
In this context, the authors in \cite{wang2024characterizing} introduced XDELTA, an innovative explainable AI tool that clarifies the differences between high-accuracy base models and their computationally efficient edge counterparts. By utilizing an LLM layer, XDELTA generates detailed and readable explanations, highlighting the disparities in feature representation and accuracy. This capability is crucial for edge network deployment, as it helps practitioners understand how to optimize and tailor edge models for specific applications, ultimately enhancing performance and facilitating more effective real-world implementations. However, existing methods often fail to provide interpretable relative differences between model pairs, as their outcomes are too specific to individual inputs and do not generalize effectively across diverse datasets.

\subsection{LLM for Continuous Network Support}
Continuous network support is essential for ensuring service availability and achieving high QoS in edge and fog environments. This subsection examines the important role that LLMs can play in managing fault tolerance, as well as improving real-time threat detection and response, along with strategies for optimizing LLMs in resource-constrained settings.

\subsubsection{LLM optimization}
LLMs demand high computational resources and memory, and the knowledge they contain is often characterized by significant redundancy. Tuning LLMs in environments with restricted computation and memory resources represents a considerable challenge. The authors in \cite{yu2024edge} presented a new framework called EDGE-LLM, which addresses these issues by reducing redundancy through compression techniques and alleviating memory constraints by using skip connections to shorten backpropagation depth. They further enhanced performance with a voting mechanism that combines predictions from all exit layers. Additionally, they recommended hardware optimization with a scheduling module designed for efficient scheduling and offloading strategies to improve LLM inference throughput.
An alternative approach for compressing LLMs and eliminating redundancies is the Knowledge Pruning method proposed in \cite{jin2024llm}. The authors advocated for removing unnecessary knowledge and distilling only the relevant information into the target neural network model. This method achieves good results in time series classification and regression tasks, all while maintaining low computational costs.
In the context of fog networks, Zagar \textit{et al.} \cite{zagar2024dynamic} showed the potential of a dynamic inference framework to enhance LLM task execution across edge, fog, and cloud environments. By placing LLM execution closer to users' devices in a decentralized fog setup, their framework named SpeziLLM addresses key privacy, trust, and cost concerns associated with cloud solutions while keeping important functionalities near the data source.

\subsubsection{Edge computing security}
Ensuring security in edge networks is crucial, especially with the increasing number of edge devices, which complicates real-time threat detection. The authors in \cite{hasan2024distributed} tackled the issue of cybersecurity for edge devices by proposing a scalable and flexible solution that utilizes ML models to analyze local data, such as system logs, in real time and allows for immediate response actions. This approach helps maintain data privacy, reduces latency, and enhances responsiveness. Additionally, the solution includes a centralized LLM server that represents the threat intelligence system that guides the edge server. The LLM server was constantly trained to stay updated on new threats and attack scenarios, serving as a main resource for information on security risks.
Furthermore, at the edge layer, LLM can serve as an IoT honeypot to attract and analyze malicious activity targeting IoT devices. The authors in \cite{yosifova2024application} explored how LLMs can help defend against cyber threats by deceiving attackers. The model was adapted to respond to an attacker's attempts to breach the system by simulating console access and mimicking the behavior of a real Linux terminal. In this way, the attacker is deceived into believing they have successfully breached the system when, in fact, the LLM is simply role-playing the victim to mislead the attacker. Another similar contribution was made in \cite{sladic2024llm}, where the authors published a fine-tuned open-source framework
that allows LLM to run as a Linux shell. The name of the framework is ShelLM, and it was published on GitHub.

\subsubsection{Fault tolerance management}
Network fault tolerance is crucial for ensuring continuous service availability and maintaining an acceptable QoS. The authors in \cite{fang2024large} leveraged LLMs to autonomously handle the entire process of detecting, diagnosing, and recovering network faults. Their proposed solution employs various LLM variants as automated schedulers to oversee existing fault tolerance models, utilizing fault indicators and current network conditions, such as congestion and link interruptions.



\section{Challenges, Open Issues, and Future Directions}\label{sec:challenges}
The integration of LLMs into communication NSM offers transformative potential across various network types such as mobile networks, vehicular networks, cloud-based networks, and fog/edge-based networks. While these capabilities promise unprecedented automation and optimization, the opportunities come with unique challenges that need to be addressed to unlock their full potential. This section casts light on the critical challenges in LLM-empowered communication NSM, focusing on unresolved open issues and their implications. Then potential future research directions will be highlighted.

\subsection{Challenges and Open Issues}
\textbf{1) LLM adaptability and interoperability in communication networks:} The application of LLMs to domain-specific tasks in communication networks demands a nuanced understanding of network-specific data, as discussed in Section II.B. Communication networks generate diverse data according to the type of network, including system logs, configuration intents, road maps, sensor metrics, and routing information. While these datasets emanate from different sources with unique features, structures, and attributes, LLMs must integrate and contextualize this data while retaining interoperability across multiple domains. For instance, a mobile network domain-specific LLM pre-trained on 5G network data (BGP routing tables and packet capture files) demonstrates improvements in IP routing analysis \cite{10583947}. However, challenges arise in adapting this same mobile network domain-specific LLM to vehicular networks or cloud environments due to different data features and structures. The cost involved in building domain-specific LLMs from scratch is enormous, considering data collection, model pre-training, and fine-tuning. Therefore, coming up with adaptable and interoperable LLMs that could suffice in various communication network domains remains an open issue worthy of investigation. 

In terms of fine-tuning, adapting LLMs for vehicular or fog networks will involve additional tasks to capture domain-specific constraints such as vehicle mobility or fog device limitations. This is so because these networks essentially deal with image and video data files with or without text-based data \cite{lan2024traj,xu2024drivegpt4}. It remains an open issue as to how communication domain-specific LLMs can integrate effective fine-tuning approaches that will achieve reasonable results regardless of the specific communication network and fine-tuning data type. 

Moreover, the adaptability and interoperability of LLMs for communication NSM hinge on their ability to handle contextual constraints effectively. These constraints, shaped by the unique characteristics of each network environment, impact the applicability and success of LLMs. Whether managing cloud-based networks or vehicular networks, contextual constraints dictate how LLMs process data, generate insights, and deliver results across diverse and dynamic scenarios. Communication networks exhibit time-sensitive patterns that LLMs must learn and adapt to. Mobile networks face varying traffic loads during peak and off-peak hours, while vehicular networks see shifts in activity based on daily commuting schedules. An LLM managing both networks must adapt its resource allocation during morning rush hours to balance increased mobile data demand with vehicular network coordination for reduced congestion. Also, different network domains employ diverse communication protocols and standards, creating challenges for LLMs to integrate seamlessly. For example, a multi-domain LLM managing vehicular networks (using DSRC or C-V2X technologies) and mobile networks (using 4G/5G protocols) must align decision-making across these protocols for tasks such as handovers. Ensuring the communication network-based LLMs capture all these contexts remains a major hurdle.

\textbf{2) Computational resource demand and energy efficiency:} Generally, LLMs require substantial computational resources for training, fine-tuning, and inference, and communication network-specific LLMs are no exception. This poses a crucial challenge for their deployment in resource-constrained environments such as fog/edge-based networks. Notably, user devices are a main component of any communication network. However, these devices are battery-limited and may offload their most computation-intensive tasks, such as LLM pre-training, fine-tuning, and inference to edge, fog, cloud-based systems, or even vehicular networks. However, computational resources are not unlimited, and user devices may incur huge costs related to computational resources and energy costs. Pre-training LLMs demands significant energy and computational resources, often exceeding the capabilities of mobile or vehicular network devices or components. Most domain-specific LLMs in communication networks require multiple distributed GPU setups for pre-training and fine-tuning due to the fact that they use general-purpose LLMs as base models. Mobile-LLaMA \cite{10583947}, ConnectGPT \cite{10588835}, IoV-BERT-IDS \cite{fu2024iov}, and MistralBSM \cite{hamhoum2024mistralbsm} utilize LlaMA 2, GPT-4, BERT, and Mistral-7B as their base models, respectively. This means that pre-training would require extensive hardware, which may be impractical for real-time NSM.

LLM deployment must align with the United Nation's sustainable development goals on energy \cite{SOROOSHIAN2024142272}. Computation resources such as data centers consume massive amounts of electricity from the power grid or alternative sources. This high level of power consumption has adverse effects on the environment. Techniques such as LoRA that introduce low-rank matrices significantly reduce the computational and memory overhead associated with model training and inference, making it a crucial enabler for sustainable LLM deployment for communication NSM \cite{wu2024dlora,cai2024edge}. However, it is not widely adopted yet in LLM-empowered communication NSM tasks, which is an open issue for research study.

\textbf{3) Real-time inference speed:} Real-time responsiveness is a critical requirement for communication NSM, particularly in latency-sensitive applications like uRLLC scenarios (autonomous driving, factory automation). Many network applications require sub-millisecond or low-millisecond responses to function effectively. Autonomous vehicles rely on real-time traffic data and routing suggestions to avoid accidents, where delays could be catastrophic. A cloud-based NSM system managing mobile devices in a densely populated area during peak hours must process data at an immense scale. Communication network domain-specific LLMs contain thousands to billions of parameters, which require significant computational power to perform inference. This complexity leads to slower processing times, as larger models provide better accuracy but at the cost of higher latency. 

While data centers with GPUs and TPUs can support rapid inference, edge devices and fog nodes often lack such resources, making real-time performance difficult. For instance, deploying an LLM on an RSU in a vehicular network faces constraints in processing power and memory \cite{liu2024resource}. Real-time scenarios require rapid data ingestion, preprocessing, and feeding into the LLM. In such scenarios, slow preprocessing pipelines could become a bottleneck. Still on data processing, LLMs are often optimized for batch processing, where multiple queries are processed together to maximize efficiency. Real-time applications, however, require immediate processing of individual queries to achieve high real-time inference speed.

To improve inference speed, researchers have used quantization methods to reduce the precision of model weights (e.g., from 32-bit floating point to 8-bit integers) \cite{lang2024comprehensive}. Quantization decreases computational requirements without significantly affecting accuracy. Specifically, quantized LLMs have been deployed for edge-based IoT networks to reduce inference latency \cite{shen2024agile}. Although quantization results in faster inference, there is still the challenge of how to achieve a potential trade-off in acceptable accuracy levels. 

Lastly, allowing partial or approximate results to be generated quickly for less critical tasks while refining outputs in the background of more critical tasks will contribute to fast inference speed, especially for LLMs customized for all kinds of communication networks. However, in dealing with heterogeneous network domain characteristics, asynchronous processing makes it difficult to decide which application has the highest priority for task execution. Taking vehicular networks as an example, where sensitive tasks such as dynamic routing and path planning require substantial image and video data, even prioritizing these tasks may record lower inference speed compared to other cloud-based or edge-based tasks that require only textual data.

\textbf{4) Evaluation metrics and benchmark complexity:} Evaluating LLMs for NSM applications lacks standardization, particularly for non-traditional metrics. 
Unlike traditional AI models that may have well-defined metrics, LLMs must be assessed across multiple dimensions, often involving non-standard, domain-specific metrics. These complexities create barriers to fair comparisons, robust benchmarking, and effective optimization for communication networks.   

Each network domain introduces unique requirements and metrics, making it difficult to design generalized benchmarks. A mobile network LLM used for optimizing power control must be evaluated on average power consumption and service quality \cite{zhou2024largelanguagemodelllmenabled}; a vehicular network LLM must prioritize accuracy and precision in routing decisions, especially in dynamic traffic conditions \cite{10588835}; an edge-based LLM for shifting inference services to edge devices should be assessed on LLM inference latency and memory utilization \cite{dhar2024empirical}. 
LLM performance often extends beyond traditional AI metrics such as precision and recall, introducing non-standard and domain-specific evaluation needs. In critical NSM tasks such as anomaly detection, LLMs must justify their outputs using distinct metrics like interpretability and confidence score. Additionally, the absence of standard benchmarks for LLMs in communication NSM limits comparability between models and systems. Current benchmarks such as Bilingual Evaluation Understudy (BLEU) and Recall-Oriented Understudy for Gisting Evaluation (ROUGE) scores \cite{xu2024drivegpt4} are not tailored to the unique requirements of communication NSM tasks like resource allocation and anomaly detection. 

In summary, developing benchmarks that work across mobile, vehicular, cloud, and edge networks is crucial but challenging due to differing priorities and data structures. Incorporating metrics for bias, fairness, and explainability in LLMs customized for communication NSM tasks is an open issue to be explored.

\textbf{5) Security and privacy:} Security and privacy are critical concerns in deploying LLMs for communication NSM. As LLMs increasingly handle sensitive network data and perform tasks like anomaly detection, resource allocation, and intent-based configuration, they become targets for cyber threats and must adhere to strict data privacy regulations. Addressing these issues is essential for building trust, ensuring reliability, and safeguarding user and network data. In \cite{khowaja2024pathway}, the authors explored the
security vulnerabilities associated with fine-tuning LLMs in 6G
networks, in particular, the membership inference attack. Adversarial attacks exploit vulnerabilities in LLMs by injecting carefully crafted inputs to manipulate their outputs. These attacks can disrupt critical NSM functions, leading to incorrect decisions or compromised security. For instance, an attacker might inject subtle perturbations into network configuration commands interpreted by an LLM, leading to incorrect configurations and network disruptions. How to make LLMs robust against adversarial examples, especially in dynamic environments like mobile and vehicular networks, remains an open issue.

LLMs process large amounts of potentially sensitive data, including user activity logs, network performance metrics, and intent configurations. Improper handling of this data can lead to significant privacy violations. The pre-training and fine-tuning LLMs often necessitate access to extensive datasets, which may include sensitive user and network information. While it is uncommon to directly access network data from its source (mainly network operators), optimizing network performance often requires the release of partial datasets for tasks such as analysis, forecasting, and resource or QoS adjustments. However, even these partial datasets are vulnerable to cyberattacks, posing substantial risks to data security and privacy. In fog/edge networks, where data is processed closer to the source, the risk of exposure increases due to weaker security protocols on edge devices. Ensuring compliance with privacy regulations like GDPR and CCPA \cite{biswas2023guardrails} while maintaining model performance is an open issue.

As communication networks scale, securing LLMs across distributed environments becomes more challenging. This includes ensuring consistent security policies across cloud, edge, and fog deployments. Distributed environments introduce vulnerabilities due to diverse hardware and software platforms. Ensuring that all nodes in a fog network adhere to the same security protocols is difficult, especially in resource-constrained environments. for instance, in a smart city, an LLM managing both mobile and edge networks must protect against threats across a variety of devices, from smartphones to IoT sensors. Developing scalable and lightweight security frameworks for LLM deployments in distributed networks is an open issue.

\subsection{Future Research Directions}
Significant research opportunities exist in enhancing the capabilities of LLMs for communication NSM. The following future research directions provide a holistic roadmap for advancing LLM-empowered NSM in communication networks:

\textbf{1) Developing cross-domain adaptive LLM architectures:} The adaptability of LLMs across diverse communication networks is a crucial research area for extensive study. Creating cross-domain adaptive LLM architectures requires addressing the unique challenges posed by each domain, including differences in data types, communication protocols, and operational constraints. Future LLM architectures should incorporate modular design principles, enabling plug-and-play functionality for communication network domain-specific adaptations. Modular architectures can provide a foundation for cross-domain adaptability via a \textit{mixture-of-experts} design \cite{xiao2024configurable}. By designing LLMs with independent modules tailored for specific tasks or network domains, the system can dynamically load and execute domain-relevant modules while maintaining a shared global model. For instance, a modular LLM could use a vehicular module optimized for routing and traffic management and an edge module focused on energy-efficient task scheduling. When operating in hybrid environments, the system could integrate these modules seamlessly for multi-domain operations.

Techniques such as meta-learning and transfer learning can be extended to LLMs for adapting knowledge from one domain (e.g., mobile networks) to another (e.g., vehicular or edge networks) \cite{kim2024cross}. Developing unified, multi-domain LLMs that leverage shared knowledge across different network contexts while preserving domain-specific nuances is an important avenue for exploration. Researchers can investigate hybrid fine-tuning methods that combine parameter-efficient fine-tuning techniques such as LoRA and prompt tuning with dynamic task embeddings to address domain-specific constraints without retraining the entire model. Additionally, designing datasets and pre-training strategies that integrate communication protocols and data types will enhance cross-domain utility.

\textbf{2) Efficient resource allocation strategies for LLM workloads:} Future work must focus on optimizing computational resource allocation to meet the high demands of LLMs in resource-constrained environments like edge and fog-based networks. Since end devices in communication networks are battery-limited, it would be more practical to deploy lightweight LLMs that can run on these devices. End devices or edge/fog nodes can process the initial layers of LLM computations locally, and offload the remaining LLM computation tasks to edge computing servers \cite{liu2024resource}. Research into lightweight model architectures, such as sparsely activated models or token-efficient transformers, can reduce computational overhead. Distributed computing frameworks that balance tasks across heterogeneous devices, including federated learning paradigms, can be explored to offload LLM workloads while ensuring data privacy and energy efficiency.

Moreover, green AI initiatives should be integrated into the design of LLMs for communication NSM. Techniques such as quantization-aware training \cite{shen2024edgeqat}, low-rank approximations, and knowledge distillation can significantly lower energy consumption during model training and inference. Future studies should focus on optimizing the trade-off between resource efficiency and model accuracy, especially for latency-sensitive tasks.

\textbf{3) Enhancing real-time inference capabilities:} Real-time inference is a cornerstone of many applications in communication NSM, especially for latency-sensitive tasks. The ability of LLMs to provide instantaneous or near-instantaneous responses is vital to maintaining operational efficiency and safety in such scenarios. Improving the real-time inference capabilities of LLMs for NSM is critical for their adoption in communication networks. Future research should focus on designing asynchronous inference pipelines that prioritize high-urgency tasks dynamically based on contexts, such as vehicle routing in a vehicular network or anomaly detection in a mobile network. Techniques like adaptive precision and progressive inference can allow LLMs to provide approximate results within stringent latency constraints and refine outputs over time for non-critical tasks.

Additionally, model compression methods, such as pruning and mixed-precision computation, must be tailored for specific NSM applications to achieve the desired trade-off between speed, accuracy, and cost \cite{tang2024hobbit}. Researchers can also explore neuromorphic computing and hardware acceleration for running inference workloads. Hardware accelerators, such as GPUs, TPUs, and specialized inference chips, are critical for achieving low-latency performance in real-time applications. Neuromorphic computing, inspired by the human brain's architecture, offers promise for event-driven, low-power real-time inference.

\textbf{4) Standardizing evaluation metrics and benchmarks:} The lack of standardization in evaluating LLMs for NSM applications requires immediate attention. Future work should focus on developing comprehensive benchmarking suites tailored to communication networks, incorporating diverse metrics like energy efficiency, inference speed, accuracy, explainability, and robustness \cite{hodak2023benchmarking}. Future research can explore the possibility of developing multi-dimensional evaluation frameworks that integrate domain-specific metrics such as packet delivery ratio (for mobile networks), routing accuracy (for vehicular networks), and edge computation time (for edge-based networks) with traditional metrics like precision and recall. Additionally, standardized simulation environments for communication NSM, akin to OpenAI's Gym for RL implementations, can facilitate consistent LLM evaluation across various network domains. These environments hold promise to allow LLMs to be assessed under varying network conditions regardless of the domain-specific evaluation criteria.

\textbf{5) Strengthening security and privacy mechanisms:} Ensuring robust security and privacy of LLMs in communication NSM remains a vital research direction. Future work should focus on developing adversarial training techniques and robustness evaluations to defend LLMs against cyber threats such as data poisoning, adversarial inputs, and membership inference attacks \cite{khowaja2024pathway}. 

Secure federated learning frameworks tailored for communication NSM tasks can enable collaborative model training across distributed nodes while preserving data privacy. For instance, in fog-based networks, lightweight LLMs can be deployed on the fog nodes, which work as clients to learn the model and upload the client and model weights to be aggregated on the central server. 

Incorporating holomorphic encryption or secure multi-party computation during model pre-training and fine-tuning can protect sensitive network data. Research into lightweight cryptographic protocols for LLMs in resource-constrained environments will also be critical. Moreover, designing security standards and protocols for LLMs operating across distributed environments, particularly in multi-domain deployments, represents a promising area for future exploration. Furthermore, integrating responsible AI principles into LLMs for communication NSM is essential for building systems that are fair, transparent, secure, and sustainable. Future research may focus on creating frameworks and methodologies that align LLM-driven NSM solutions with ethical and societal values. By addressing these dimensions, researchers can ensure that LLMs not only enhance network performance but also uphold trust and accountability in their deployment.

\section{Conclusion}\label{sec:conclusion}
The recent advancements in LLMs for understanding complex patterns and generating human-like content have reinforced their remarkable potential in enhancing communication NSM. This survey has provided a comprehensive exploration of LLMs in enhancing NSM across diverse communication network domains, including mobile networks and IoT technologies, vehicular networks, cloud-based networks, and fog/edge-based networks. We first presented LLM fundamentals in terms of the transformer architecture and the function of each of its components, general-purpose and domain-specific LLMs, model pre-training and fine-tuning, and a summary on customizing LLMs for communication NSM. Then, we established a comprehensive taxonomy, namely, LLM for network monitoring and reporting, LLM for AI-powered network planning, LLM for network deployment and distribution, and LLM for continuous network support, to demonstrate the versatility and applicability of LLMs in automating complex NSM tasks. Under each taxonomy, several related works were reviewed to identify their innovative points in bridging the gap between advanced AI capabilities and their integration into various communication network domains. Despite their advantages, the adoption of LLMs for NSM is not without challenges. Therefore, we identify critical challenges and open issues associated with LLM-empowered communication NSM. Finally, future research directions that hinge on creating adaptable, robust, and energy-efficient LLM solutions tailored for communication NSM were discussed. By leveraging the strengths of LLMs, researchers and practitioners can unlock new opportunities for innovation, paving the way for smarter, more resilient network ecosystems. 

\bibliographystyle{IEEEtran}
\bibliography{output}

\begin{IEEEbiography}[{\includegraphics[width=1in,height=1.25in,clip,keepaspectratio]{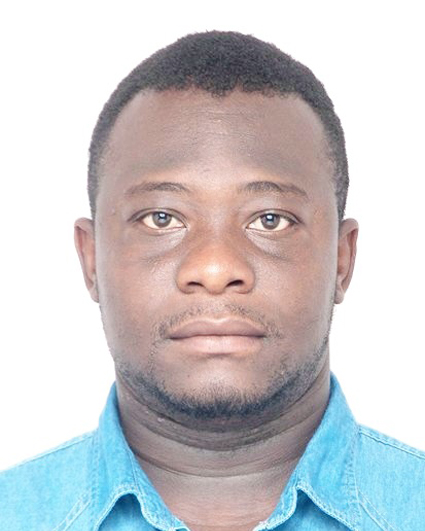}}]{Gordon Owusu Boateng} received his Bachelor's degree in Telecommunications Engineering from the Kwame Nkrumah University of Science and Technology (KNUST), Kumasi-Ghana, in 2014. He received his M.Eng. degree and Ph.D. degree in Computer Science and Technology from the University of Electronic Science and Technology of China (UESTC), in 2019 and 2023, respectively. He is currently a Postdoctoral Fellow at Khalifa University, Abu Dhabi, UAE. 
From 2014 to 2016, he worked under sub-contracts for Ericsson (Ghana) and TIGO (Ghana). Till now, Gordon has co-authored over 35 scientific journal and conference papers. 
His research interests include 5G/6G wireless networks, blockchain, reinforcement learning, vehicular networks, and large language models.
\end{IEEEbiography}

\begin{IEEEbiography}[{\includegraphics[width=1in,height=1.25in,clip,keepaspectratio]{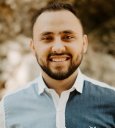}}]{Hani Sami} is currently a Postdoctoral Fellow at École de technologie supérieur, Montréal, QC, Canada. He received his Ph.D. in Information Systems Engineering from Concordia University. He also received his M.Sc. in Computer Science from the American University of Beirut  and completed his B.S. and worked as Research Assistant at the Lebanese American University. The topics of his research are Fog Computing, Vehicular Fog Computing, Reinforcement Learning, Reward Shaping, and Blockchain. He is a reviewer of several prestigious conferences and journals.
\end{IEEEbiography}

\begin{IEEEbiography}[{\includegraphics[width=1in,height=1.25in,clip,keepaspectratio]{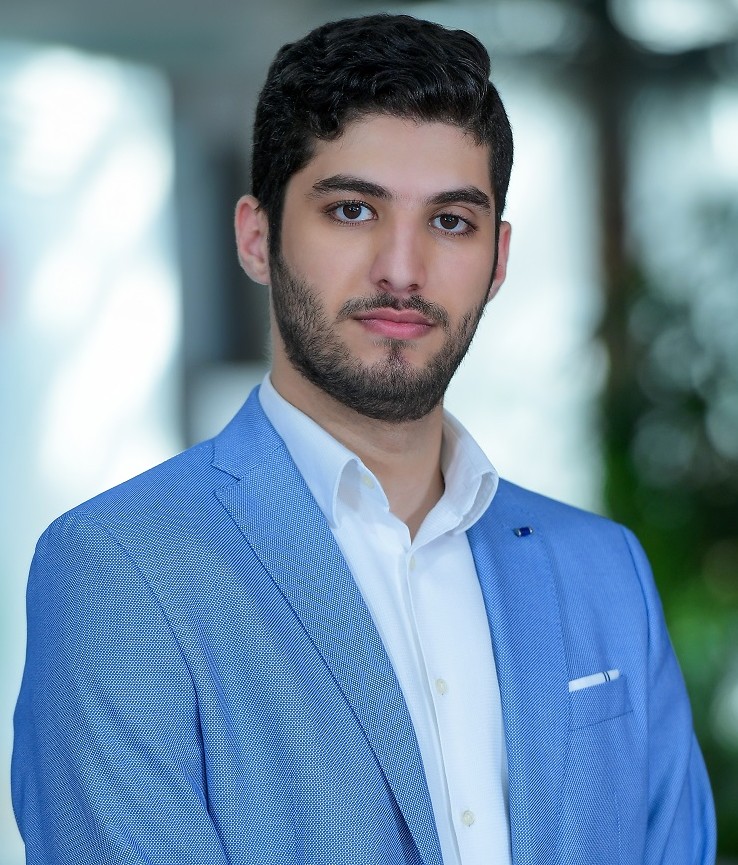}}]{Ahmed Alagha} is currently a Postdoctoral Fellow at Concordia University, Montréal, QC, Canada. He received his Ph.D. in Information Systems Engineering from Concordia University, and the B.Sc. and M.Sc. degrees in Electrical and Computer Engineering from Khalifa University, Abu Dhabi, UAE, where he also worked as a Research Associate. He is a recipient of the prestigious FRQNT doctoral and postdoctoral awards. His research interests include multiagent systems, deep reinforcement learning, vision-language models, IoT, sensing technologies, crowd sensing and sourcing. 
\end{IEEEbiography}

\begin{IEEEbiography}[{\includegraphics[width=1in,height=1.25in,clip,keepaspectratio]{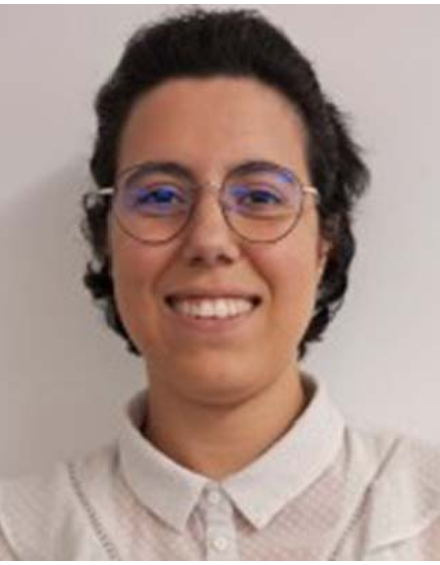}}]{Hanae Elmekki} received an engineering diploma in Telecommunications and Information Technology from the National Institute of Postal and Telecommunication (INPT) in Rabat, Morocco, in 2014. She is currently a Ph.D. candidate at the Concordia Institute for Information Systems Engineering (CIISE), Concordia University, Montreal, QC, Canada, having fast-tracked to the program from an M.Sc. in Quality Systems Engineering at the same institution. Her current research interests include artificial intelligence, reinforcement learning, cardiovascular diseases, robotics, federated learning, and cloud/edge computing.
\end{IEEEbiography}

\begin{IEEEbiography}[{\includegraphics[width=1in,height=1.25in,clip,keepaspectratio]{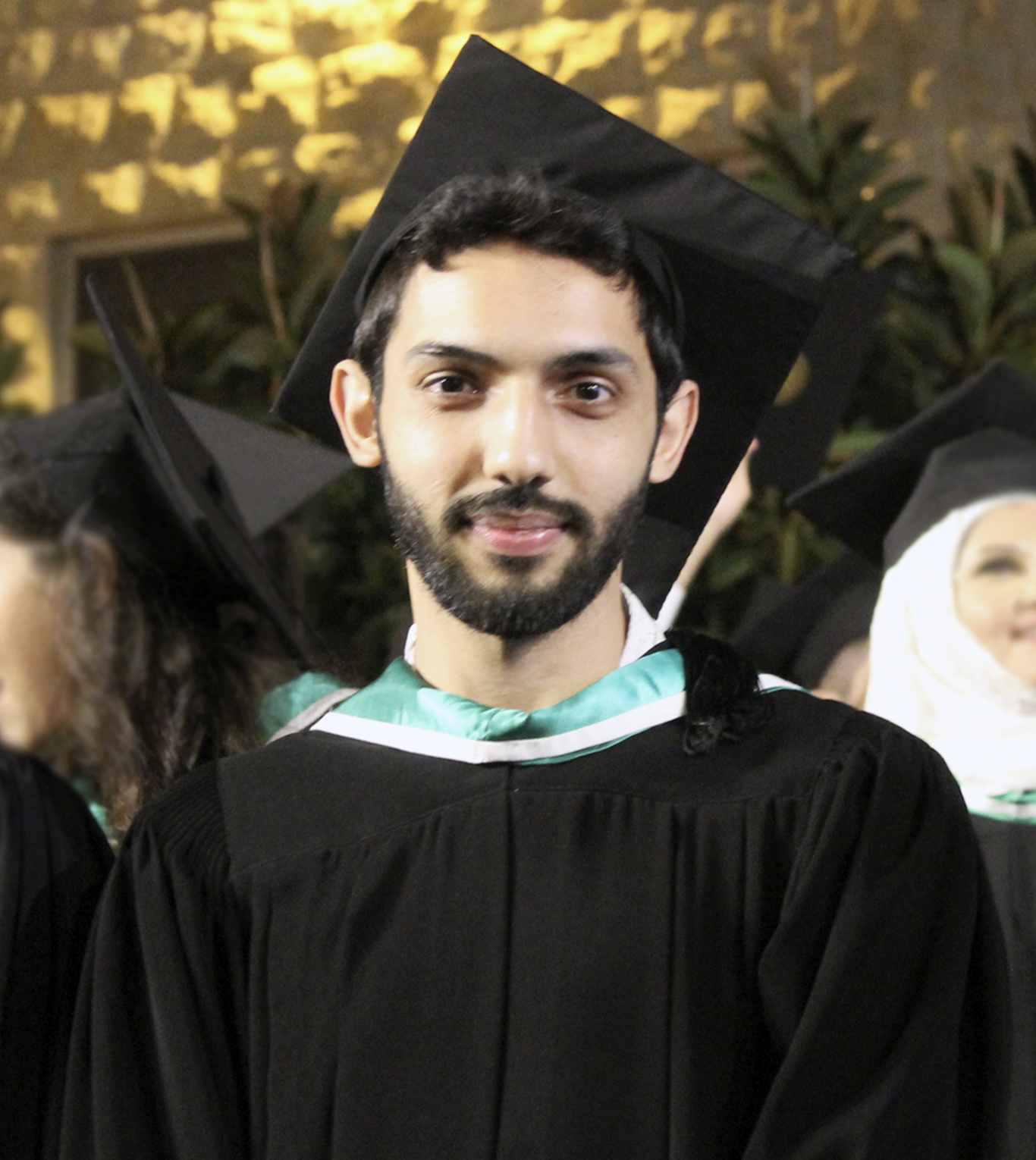}}]{Ahmad Hammoud} received the B.S. degree in business computing from Lebanese University in 2016, and the M.S. degree in computer science from Lebanese American University in 2019. He is currently pursuing a Ph.D. degree with the Electrical Engineering Department, École de Technologie Supérieure. His current research interests include Metaverse, cloud and fog federations, federated learning, the Internet of Vehicles, game theory, Blockchain, artificial intelligence, and security.
\end{IEEEbiography}

\begin{IEEEbiography}[{\includegraphics[width=1.1in,height=1.3in,clip,keepaspectratio]{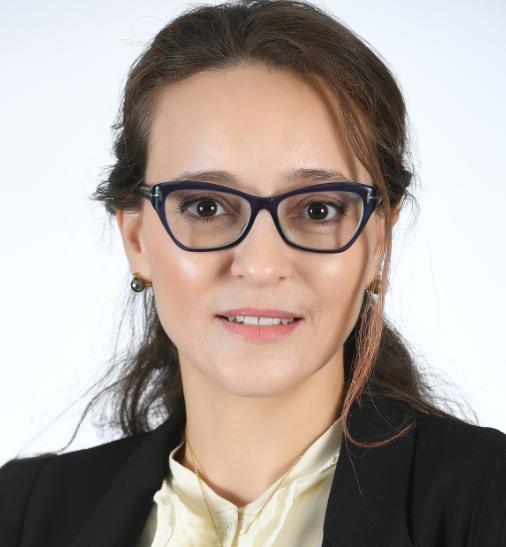}}] {Rabeb Mizouni} is an Associate Professor in the Department of Electrical Engineering and Computer Science at Khalifa University, Abu Dhabi, United Arab Emirates. She got her M.Sc. and Ph.D. in Electrical and Computer Engineering from Concordia University, Montreal, Canada in 2002 and 2007 respectively. Currently, she is interested in the deployment of context aware mobile applications, crowd sensing, Artificial Intelligence, IoT and Blockchain. Dr. Mizouni is currently an Associate Editor for IEEE Internet of Things magazine.
\end{IEEEbiography}

\begin{IEEEbiography}[{\includegraphics[width=1.1in,height=1.3in,clip,keepaspectratio]{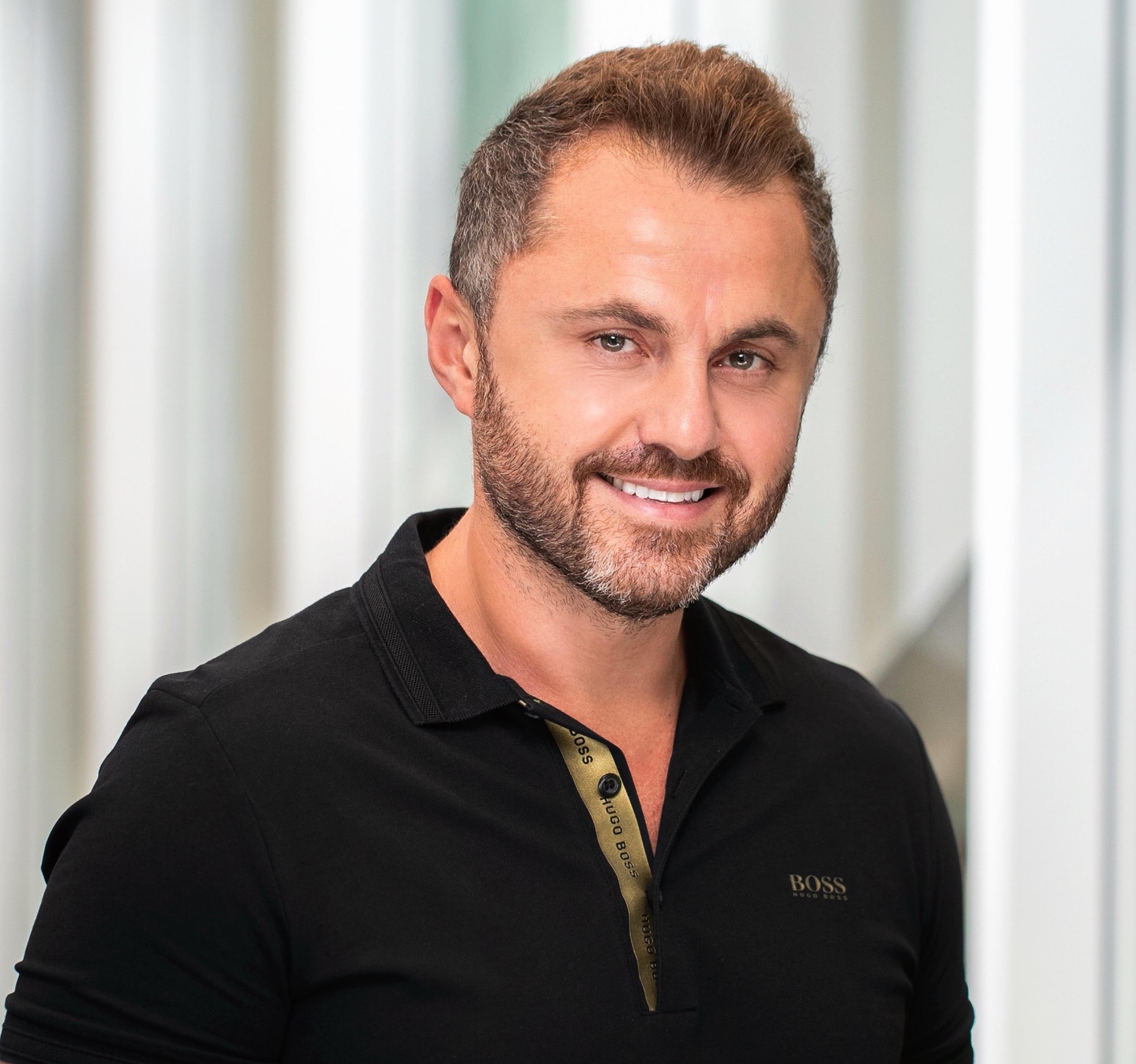}}]{Azzam Mourad} received his M.Sc. in CS from Laval University, Canada (2003) and Ph.D. in ECE from Concordia University, Canada (2008). He is currently a Professor of Computer Science and Founding Director of the Cyber Security Systems and Applied AI Research Center with the Lebanese American University, Visiting Professor of Computer Science with New York University Abu Dhabi and Affiliate Professor with the Software Engineering and IT Department, Ecole de Technologie Superieure (ETS), Montreal, Canada. His research interests include Cyber Security, Federated Machine Learning, Network and Service Optimization and Management targeting IoT and IoV, Cloud/Fog/Edge Computing, and Vehicular and Mobile Networks. He has served/serves as an associate editor for IEEE Transactions on Services Computing, IEEE Transactions on Network and Service Management, IEEE Network, IEEE Open Journal of the Communications Society, IET Quantum Communication, and IEEE Communications Letters, the General Chair of IWCMC2020-2022, the General Co-Chair of WiMob2016, and the Track Chair, a TPC member, and a reviewer for several prestigious journals and conferences. He is an IEEE senior member.
\end{IEEEbiography}

\begin{IEEEbiography}[{\includegraphics[width=1.1in,height=1.3in,clip,keepaspectratio]{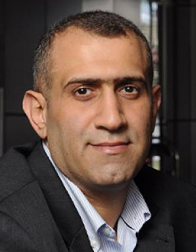}}]{Hadi Otrok} received his Ph.D. in ECE from Concordia University. He holds an Associate Professor position in the department of Electrical Engineering and Computer Science (EECS) at Khalifa University. Also, he is an Affiliate Associate Professor in the Concordia Institute for Information Systems Engineering at Concordia University, Montreal, Canada, and an Affiliate Associate Professor in the Electrical department at Ecole de Technologie Superieure (ETS), Montreal, Canada. His research interests include the domain of blockchain, reinforcement learning, crowd sensing and sourcing, ad hoc networks, and cloud security. He co-chaired several committees at various IEEE conferences. He is also an Associate Editor at IEEE Transactions on Network and Service Management (TNSM), Ad-hoc networks (Elsevier), and IEEE Network. He also served from 2015 to 2019 as an Associate Editor at IEEE Communications Letters.
\end{IEEEbiography}

\begin{IEEEbiography}[{\includegraphics[width=1.1in,height=1.3in,clip,keepaspectratio]{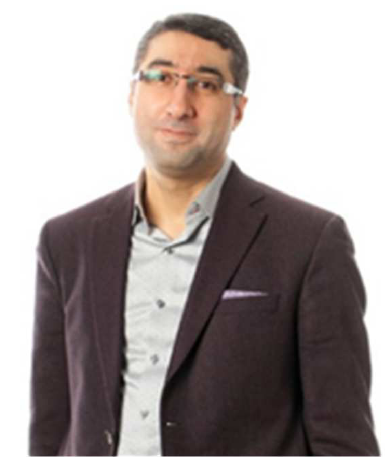}}] {Jamal Bentahar} received the Ph.D. degree in computer science and software engineering from Laval University, Canada, in 2005. He is a Professor with Concordia Institute for Information Systems Engineering, Concordia University, Canada. From 2005 to 2006, he was a Postdoctoral Fellow with Laval University, and then NSERC Postdoctoral Fellow at Simon Fraser University, Canada. He was an NSERC Co-Chair for Discovery Grant for Computer Science (2016-2018). He is a visiting professor at Khalifa University of Science and Technology. His research interests include the areas of computational logics, reinforcement learning, multi-agent systems, service computing, game theory, and software engineering.
\end{IEEEbiography}

\begin{IEEEbiography}[{\includegraphics[width=1.1in,height=1.3in,clip,keepaspectratio]{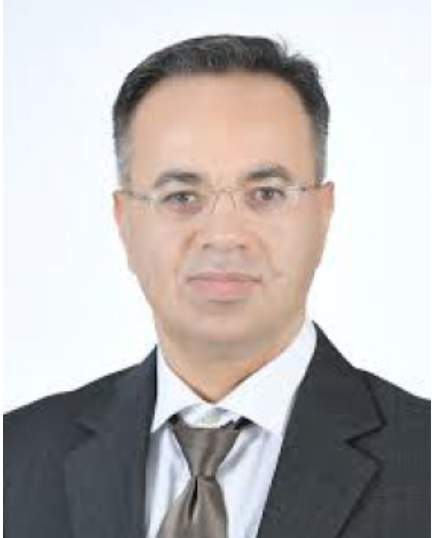}}] {Sami Muhaidat} received his Ph.D. in Electrical and Computer Engineering from the University of Waterloo, Ontario, in 2006. From 2007 to 2008, he was an NSERC Postdoctoral Fellow in the Department of Electrical and Computer Engineering at the University of Toronto, Canada. From 2008 to 2012, he served as an Assistant Professor in the School of Engineering Science at Simon Fraser University, British Columbia, Canada. Currently, he is a Professor and the Associate Dean for Research in the College of Computing and Mathematical Sciences at Khalifa University. He is also an Adjunct Professor at Carleton University, Ontario, Canada. Sami’s research interests include advanced digital signal processing techniques for wireless communications, intelligent surfaces, machine learning for communications, optical communications, and massive multiple-access techniques. He has served in various editorial roles, including as Area Editor for the IEEE Transactions on Communications, Guest Editor for the IEEE Network special issue on "Native Artificial Intelligence in Integrated Terrestrial and Non-Terrestrial Networks in 6G," and Guest Editor for the IEEE Open Journal of Vehicular Technology (OJVT) special issue on "Recent Advances in Security and Privacy for 6G Networks." Additionally, he has held positions as Senior Editor and Editor for IEEE Communications Letters, Editor for the IEEE Transactions on Communications, and Associate Editor for the IEEE Transactions on Vehicular Technology.
\end{IEEEbiography}

\begin{IEEEbiography} [{\includegraphics[width=1.1in,height=1.3in,clip,keepaspectratio]{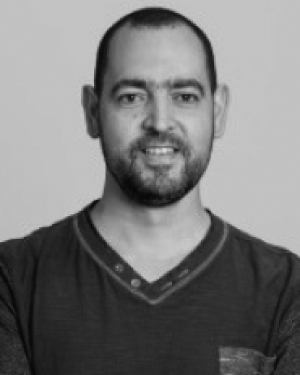}}]{Chamseddine Talhi} Received the Ph.D. degree in computer science from Laval University, Quebec, QC, Canada, in 2007. He is a Professor with the Department of Software Engineering and IT, ÉTS, University of Quebec, Montreal, QC, Canada. He is leading a research group that investigates smartphone, embedded systems, and IoT security. His research interests include cloud security and secure sharing of embedded systems.
\end{IEEEbiography}

\begin{IEEEbiography}[{\includegraphics[width=1.1in,height=1.3in,clip,keepaspectratio]{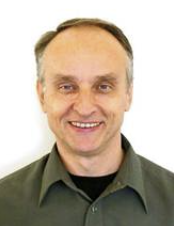}}] {Zbigniew Dziong} received the Ph.D. degree from the Warsaw University of Technology, Poland, where he also worked as an Assistant Professor. From 1987 to 1997, he was with INRS-Telecommunications, Montreal, QC, Canada. From 1997 to 2003, he worked with Bell Labs, Holmdel, NJ, USA. Since 2003, he has been with the École de Technologie Supérieure (University of Quebec), Montreal, as a Full Professor. He is an expert in the domain of performance, control, protocol, architecture, and resource management for data, wireless, and optical networks. He has participated in research projects for many leading telecommunication companies, including Bell Labs, Nortel, Ericsson, and France Telecom. He won the prestigious STENTOR Research Award (1993, Canada) for collaborative research. His monograph ATM Network Resource Management (McGraw Hill, 1997) has been used in several universities for graduate courses.
\end{IEEEbiography}

\begin{IEEEbiography}[{\includegraphics[width=1.1in,height=1.3in,clip,keepaspectratio]{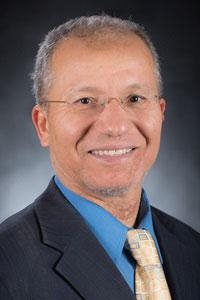}}]{Mohsen Guizani} received the BS (with distinction), MS and PhD degrees in Electrical and Computer engineering from Syracuse University, Syracuse, NY, USA in 1985, 1987 and 1990, respectively. He is currently a Professor of Machine Learning and the Associate Provost at Mohamed Bin Zayed University of Artificial Intelligence (MBZUAI), Abu Dhabi, UAE. Previously, he worked in different institutions in the USA. His research interests include applied machine learning and artificial intelligence, Internet of Things (IoT), intelligent systems, smart city, and cybersecurity. He was elevated to IEEE Fellow in 2009 and was listed as a Clarivate Analytics Highly Cited Researcher in Computer Science in 2019, 2020 and 2021. Dr. Guizani has won several research awards including the ‘‘2015 IEEE Communications Society Best Survey Paper Award’’, the Best ComSoc Journal Paper Award in 2021 as well five Best Paper Awards from ICC and Globecom Conferences. He is the author of ten books and more than 800 publications. He is also the recipient of the 2017 IEEE Communications Society Wireless Technical Committee (WTC) Recognition Award, the 2018 AdHoc Technical Committee Recognition Award, and the 2019 IEEE Communications and Information Security Technical Recognition (CISTC) Award. He served as the Editor in-Chief of IEEE Network and is currently serving on the Editorial Boards of many IEEE Transactions and Magazines. He was the Chair of the IEEE Communications Society Wireless Technical Committee and the Chair of the TAOS Technical Committee. He served as the IEEE Computer Society Distinguished Speaker and is currently the IEEE ComSoc Distinguished Lecturer.
\end{IEEEbiography}

\end{document}